\documentclass[pra,onecolumn,unsortedaddress,floatfix]{revtex4}
\usepackage{amssymb,amsmath}

\usepackage{graphicx}
\usepackage[unicode=true,
  colorlinks=true,
  linkcolor=blue]{hyperref}

\hypersetup{breaklinks=true, pdfborder={0 0 0}}
\usepackage{float}
\usepackage[caption=false]{subfig}
\usepackage[outercaption]{sidecap}

\usepackage{fancyhdr}
\usepackage{placeins}

\makeatletter
\renewcommand{\fnum@figure}{\textbf{FIG.~\thefigure}}
\makeatother

\begin{document}

\hspace{5.2in} \mbox{}

\title{A silicon qubit platform for {\em in situ} single molecule structure determination}

\author{V. S. Perunicic}
\email{vpe@unimelb.edu.au}
\affiliation{School of Physics, The University of Melbourne, Victoria 3010, Australia.}
\author{M. Usman}
\affiliation{School of Physics, The University of Melbourne, Victoria 3010, Australia.}
\affiliation{School of Computing and Information Systems, The University of Melbourne, Victoria 3010, Australia.}
\author{C. D. Hill}
\affiliation{School of Physics, The University of Melbourne, Victoria 3010, Australia.}
\affiliation{School of Mathematics and Statistics, The University of Melbourne, Victoria 3010, Australia.}
\author{L. C. L. Hollenberg}
\email{lloydch@unimelb.edu.au}
\affiliation{Centre for Quantum Computation and Communication Technology, School of Physics, The University of Melbourne, Victoria 3010, Australia.}

\date{\today}

\begin{abstract}{\bf
Imaging individual conformational instances of generic, inhomogeneous, transient or intrinsically disordered protein systems at the single molecule level \textit{in situ} is one of the notable challenges in structural biology. Present techniques access averaged structural information by measuring over large ensembles of proteins in nearly uniform conformational states in synthetic environments. This poses significant implications for diagnostics and drug design which require a detailed understanding of subtle conformational changes, small molecule interactions and ligand dynamics. Here we tackle the problem by designing a single molecule imaging platform technology embracing the advantages silicon-based spin qubits offer in terms of quantum coherence and established fabrication pathways. We demonstrate through detailed simulation, that this platform enables scalable atomic-level structure-determination of individual molecular systems in native environments. The approach is particularly well suited to the high-value lipid-membrane context, and as such its experimental implementation could have far-reaching consequences for the understanding of membrane biology and drug development.
}\end{abstract}

\pacs{}
\maketitle
\linespread{1}

\section{Introduction}
\label{Sc:Introduction}

The function of biomolecules such as proteins is intrinsically linked to their atomic structure, which in turn depends on highly complex, crowded, heterogeneous native cellular environments \cite{Dhar2010,Ellis2001, Elcock2010,Benton2012,Feig2017}. Current structure reconstruction techniques, X-ray crystallography \cite{Chapman2011,martin2016serial,Shi2014}, cryo-EM  \cite{Carroni2016,Li2013} and NMR spectroscopy \cite{Kay2016,Jiang2017,Ashbrook2018,Renault4863}, are exceptionally powerful, however, they have access to only average structural information of proteins in artificial environments and require large ensembles of proteins in nearly identical conformational states. Unlike crystalline physical systems, proteins have complex sets of metastable conformation states that depend on a multitude of variables, including the presence of various other molecules in their local environment \cite{Phillips2009}. There is an abundant and essential class of molecules which do not possess well-defined conformation states \cite{Sickmeier2007,uversky2016dancing}, constantly moving through a continuous conformation space, which itself is parameterised by countless environment variables \cite{Karplus2005, Baldwin2009}. As a result, imaging 3D atomic sites of individual, arbitrary molecular systems is outside the scope of the existing structure reconstruction techniques, which impacts our understanding of the behaviour of proteins on the length scales fundamental to their function.

Due to the nanometre dimensions of molecular systems, a pathway for achieving single protein imaging can be found in sensors that are atomic in scale. Recently, the nitrogen vacancy (NV) centre in diamond \cite{Doherty13,JPrev14} has shown a great promise as a biocompatible, nanoscopic qubit-probe for room temperature magnetometry \cite{Balasubramanian08,Taylor11,Cole09,Kaufmann13} and, in particular, high resolution nuclear magnetic resonance (NMR) spectroscopy \cite {Zhao2011, Cai2013, Perunicic2014,Trifunovic2015, Ajoy2015, Kost2015, Laraoui2015, Lazariev2015, Wang2016,  Schmitt2017,Boss2017,Glenn2018, Zopes2018, Abobeih2019, cujia2021arxive}. However, imaging the atomic makeup of single molecular structures requires low temperatures, as a means of arresting atomic motion and capturing desired instances of protein conformation states. Even at low temperature, the spin coherence times of NV centres are relatively short while their  fixed quantisation poses further limitations for the application considered here. Furthermore, the single-qubit deterministic fabrication specifications required for practical development of the instrumentation technology at scale are lacking for NV centres. This specific context opens up the interesting prospect of using low temperature silicon qubits as bio-sensors. Silicon donor qubits are equipped with a combination of long coherence times \cite{Tyryshkin2011,Muhonen2014} and deterministic fabrication methods \cite{Morello2010,Pla2012,Pla2013,Mahapatra2011,Fuechsle2012, Usman2016} making them uniquely positioned for practical applications such as quantum computing, and in particular the sensing concept introduced here.

In this paper, we therefore outline a platform for single-molecule structure-determination based on qubits in silicon allowing a form of magnetic resonance imaging (MRI) of individual molecular structures at the atomic level. This work extends the idea of using a spin qubit as both detector and source of localized field gradients, initially proposed in the context of nitrogen-vacancy centres in diamond in \cite{Perunicic2016}. Here we seek to capture the technological advantages of silicon donor qubits, requiring full treatment of the qubit electronic wave function. We characterise the control mechanism over the quantum interaction between the qubit electronic wave-function and individual nuclei in the targeted molecular specimen, providing the foundation of MRI at the atomic level. To overcome the exceptional challenges of spatial-frequency encoding at the single molecule level, we introduce the nuclear spin storage protocol which enables qubit sensors to be used for 3D atomic level imaging through careful sampling and extraction of structural information from a molecular specimen. Donor placement is an important fabrication question addressed by determining the optimal donor depth based on the effects of silicon surface termination. Finally, we perform a simulated demonstration by imaging the atomic structure of macro-molecular system comprising an individual trans-membrane protein and its lipid bilayer environment.

\section{Results}
\subsection{Silicon qubit-based single molecule structure imaging system}
\label{Sc:Molecular_imaging}

In this section, we outline the physical setup and principles of the device function, dedicating the remainder of the paper to the development of quantum control and 3D sampling aspects of qubit-based molecular structure imaging in the context of a highly non-trivial test-system. We develop the platform using group-V donor qubits in silicon, taking phosphorus (Si:P) \cite{Morello2010,Pla2012,Pla2013,Mahapatra2011,Fuechsle2012} as an example, noting that the method is applicable to other spin qubit systems with similar coherence times and wave function extent.

\begin{figure}[!htb]%
\centering
 \includegraphics[width=1\linewidth]{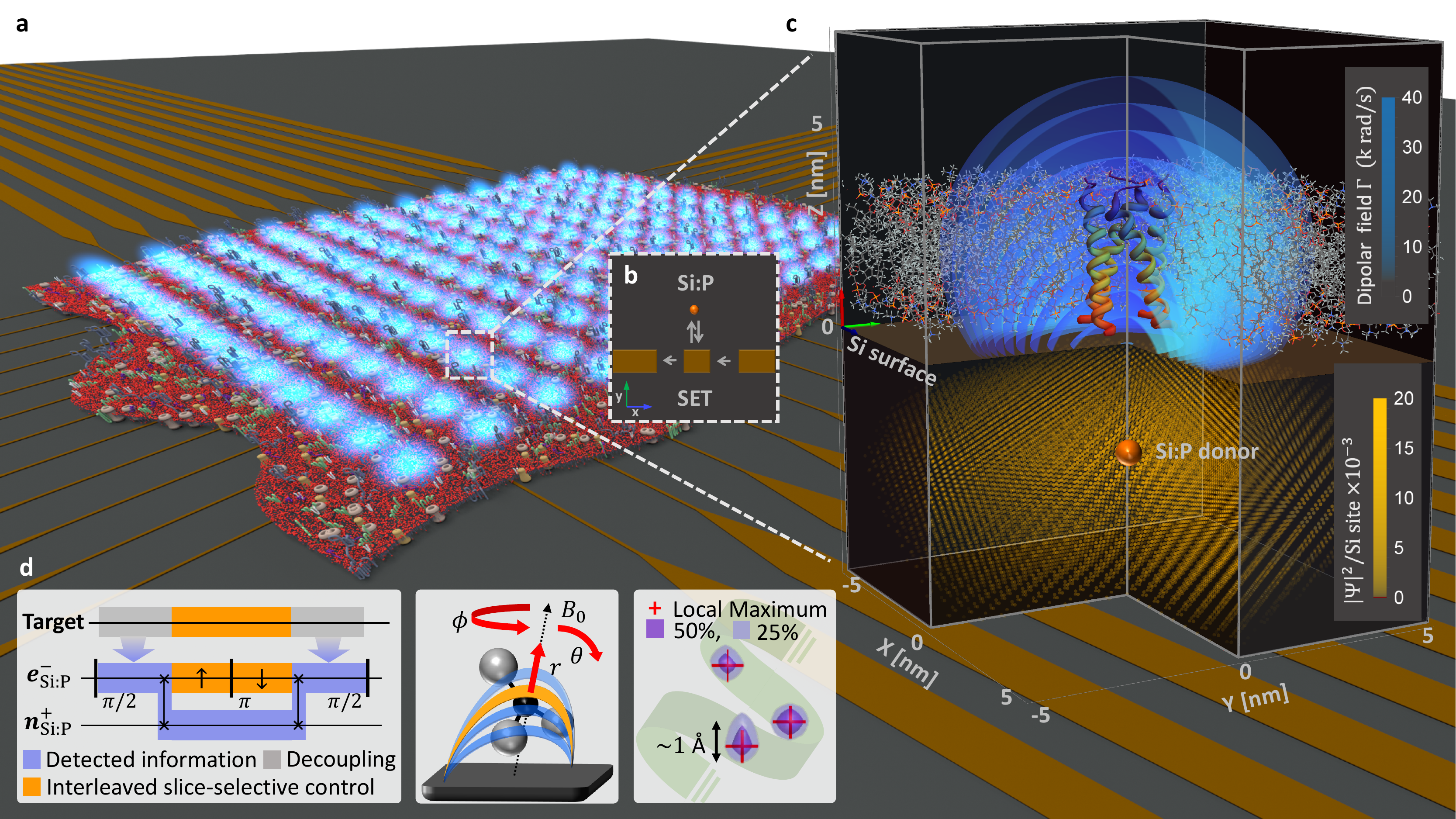}
 \vspace{-15pt}
  \caption{\textbf{Silicon qubit molecular imaging system.} 
   \textbf{a}) An array of noninteracting near-surface donor spin qubits in a silicon chip. Biological specimens, such as a native cell membrane, are positioned on the surface (red layer). The sample is snap frozen allowing each qubit to image the conformational instance of individual molecular structures that inhabit its sensing volume (blue). In-plane control structures based on standard STM fabrication techniques \cite{Fuechsle2012, Weber2014} (orange metallic leads).
   \textbf{b}) Local Single Electron Transition (SET) islands (top-down view) \cite{Morello2010,Fuechsle2012,Pla2012} facilitate read-out and initialisation of independent qubit sensors.
   \textbf{c}) Example of a single transmembrane-lipid system consisting of the M2 proton channel of influenza A virus \cite{Cady2010} in 1,2-Didecanoyl-sn-glycero-3-phosphocholine (DDPC) lipid bilayer (upper). The sample is positioned above the phosphorus donor qubit (Si:P) at depth of $4.75\,a_0$ ($\sim2.5\,\rm nm$) (orange sphere, lower) such that the sample is immersed into the qubit's magnetic dipole field (equipotential slices shown in blue) emerging from its electronic wave function (orange, lower). Water molecules suppressed for clarity, their positions can be directly determined as part of the imaging.
   \textbf{d}) The nuclear spin storage (NSS) detection protocol (high-level schematic, left). The dipole-dipole decoupling methods adapted to the qubit's gradient field environment (gray) and interleaved target nuclei control (orange), allow the qubit to probe (flow of quantum information, blue) the quantity of the specimen's nuclei present in the targeted equipotential slice of its dipole field. A dipole-adapted MRI protocol samples the specimen across a multitude of overlapping slices (middle). Measurement signals from individual slices inverted into the atomic-level 3D image of the targeted specimen's nuclear density (illustration, right).
  }\label{Fig:1}
   \vspace{15pt}
\end{figure}

In this initial work we consider a biological specimen such as a section of a cellular membrane is positioned on a silicon chip containing near-surface qubit sensors (Fig.\,\ref{Fig:1}\,a). Subsequently, the sample is snap frozen at a specific point, capturing the desired configuration of target molecular structures in vitrified ice \cite{Al-Amoudi2004, McDonald2007,Bellare2018}. The chip may contain a single sensing qubit \cite{Morello2010,Pla2012} or a grid arrangement of qubits \cite{Hill2015}. The donor qubits can be introduced using single ion implantation \cite{Morello2010,Pla2012,Pla2013}, and in the case of Si:P, they can also be deterministically fabricated by STM lithography \cite{Mahapatra2011,Fuechsle2012} and fully characterised by STM donor metrology \cite{Usman2016, Usman2017}, which is particularly advantageous for the imaging scheme. We assume the silicon surface is hydrogen-terminated, which is quite typical in such systems, noting that oxide termination would lead to reductions in the sensing volume. 
The nanoscopic size and fabrication process of the platform technology opens up a pathway towards scalability in the field of protein structure determination. This feature uniquely contrasts with conventional techniques where specimen imaging parallelisation is fundamentally limited by the relatively large physical size of the system required for a single imaging site.

The sensing signal is obtained by reading out the electron spin state of each qubit independently (Fig.\,\ref{Fig:1}\,b). As each qubit addresses a molecular-sized imaging volume, we focus on describing the protocol from the perspective of a single donor qubit (Fig.\,\ref{Fig:1}\,c). The dipole-dipole coupling between the donor electronic wave function and the molecular nuclear target spins species ($\rm ^{1}H, ^{13}C, ^{14}N$ etc.) extends above the silicon surface establishing a hemispherical sensing volume (Fig.\,\ref{Fig:1}\,a,\,c blue). The method leverages this dipole-dipole field to perform 3D imaging in a manner analogous to clinical Magnetic Resonance Imaging (MRI), with dipolar slices $\Gamma$ (blue lobes, Fig.\,\ref{Fig:1}\,c) defining distinct spatial ``lobes'' containing target nuclear spins with effectively the same coupling to the donor electron. 

The key enabler behind the qubit-based molecular imaging technique is a nuclear spin storage (NSS) protocol particularly focused on the silicon qubit context (Fig.\,\ref{Fig:1}\,d, left), which allows us to probe the molecular target's nuclear destiny at individual dipolar slices by reading out the state of the qubit. In order to generate spatio-frequency encoding across Angstrom distances, the NSS protocol successfully embeds dynamical dipole-dipole decoupling (such as CORY-24 sequence) into the gradient magnetic field produced by the donor wave function. Through careful interleaved and phase-matched control (Fig.\,\ref{Fig:1}\,d, left, orange), the donor's electronic spin resonantly couples only to the target nuclei in a chosen slice, while the coupling between the target nuclei is suppressed and the unwanted effects of the gradient filed canceled out. Information about the target's nuclear density in the probed slice is intermediately stored onto the state of the donor's nuclear spin (Fig.\,\ref{Fig:1}\,d, left, blue) in order to protect it against decoherence and enable efficient readout. Details of the NSS protocol are given in Section \ref{Sc:NSS_protocol}.

The steps of the 3D imaging procedure (Fig.\,\ref{Fig:1}\,d, middle) are summarised as follows. First, we choose a spin carrying nuclear species to image inside a target molecule specimen, i.e, $\rm ^{1}H, ^{13}C, ^{14}N$ etc. Next, we orient the donor's dipole field along a particular direction. The properties and control mechanisms of the dipole field are given in Section\ref{Sc:donors_dipolar_field}. We apply the NSS protocol to measure the nuclear density of the target across a series of dipole slices. The slices are measured sequentially for each of the multiple orientations of the donor's dipole field (Fig.\,\ref{Fig:1}\,d, middle, red arrows)  in order to sample the entire sensing volume (see Sections \ref{Sc:NSS_protocol} and  \ref{Sc:optimal_control_parameters}). In the final step, the measurement outcomes are inverted from dipolar slice-space to 3D space, enabling the nuclear density of the targeted molecular structure to be imaged in sub-Angstrom voxel resolution (Fig.\,\ref{Fig:1}\,d, right) (see Section \ref{Sc:protein-lipid_system}).
Sections that follow expand on each of the mentioned steps, detailing how silicon donor qubits can be used to image the nuclear spin density of the target molecule in 3D at the atomic level.

\subsection{Controlling the dipolar field produced by a spatially extensive donor wave function}
\label{Sc:donors_dipolar_field}

Traditional MRI uses a gradient magnetic field to correlate specific (usually planar) region of the sensing volume with a particular nuclear Zeeman splitting magnitude \cite{lauffer1996mri}, this concept is called spatial-frequency encoding. In contrast, we rely on the natural dipole-dipole interaction between the target nuclei and the qubit electron spin to provide spatial-frequency encoding. The feasibility of silicon-qubit based molecular imaging rests on the nontrivial properties of the dipolar field produced by the wave function describing the qubit's electron. In this section, we focus on the shape, magnitude and the spatial control of the dipole field, while the aspects of resolution and slice cross-sectional profile are addressed later.

Qubit probes such as the NV-centre in diamond have a highly confined (Angstrom scale) electronic wave function producing relatively strong point-dipole like fields with spatial inhomogeneity characterised by the magic-angle axis, which is ideal for 3D molecular imaging via frequency encoding \cite{Perunicic2016}. However, as noted NV centres have relatively limited quantum coherence times and the orientation of their dipole field is difficult to control due to the loss of readout contrast when the background magnetic field tilts away from the centre's quantisation axis. Donor qubits in silicon, on the other hand, have electronic wave functions on the nanometre scale \cite{Wellard2005}, an order of magnitude greater compared to the structural features of interest in molecular systems. It is, therefore, crucial to examine the behaviour of the above-surface dipole-dipole field produced by such delocalised donor wave functions. We use the experimentally benchmarked Tight-Binding (TB) formalism \cite{Usman2016,Usman2017} to compute a near-surface donor electronic probability density (see Method section: Tight binding formalism) and calculate its associated dipolar field from first principles (see Method section: Calculation of the Dipole-Dipole field). This formalism has been shown to reproduce the donor electron wave function to atomic precision when compared directly with STM imaging of sub-surface donors \cite{Usman2016,Usman2017}. The same approach can be used in experimental realisations of the method, where the donor TB wave function representation obtained from STM donor spatial metrology eliminates the need for the nontrivial characterisation of the dipolar field itself.

\begin{figure}[!htb]%
\centering
 \includegraphics[width=1\linewidth]{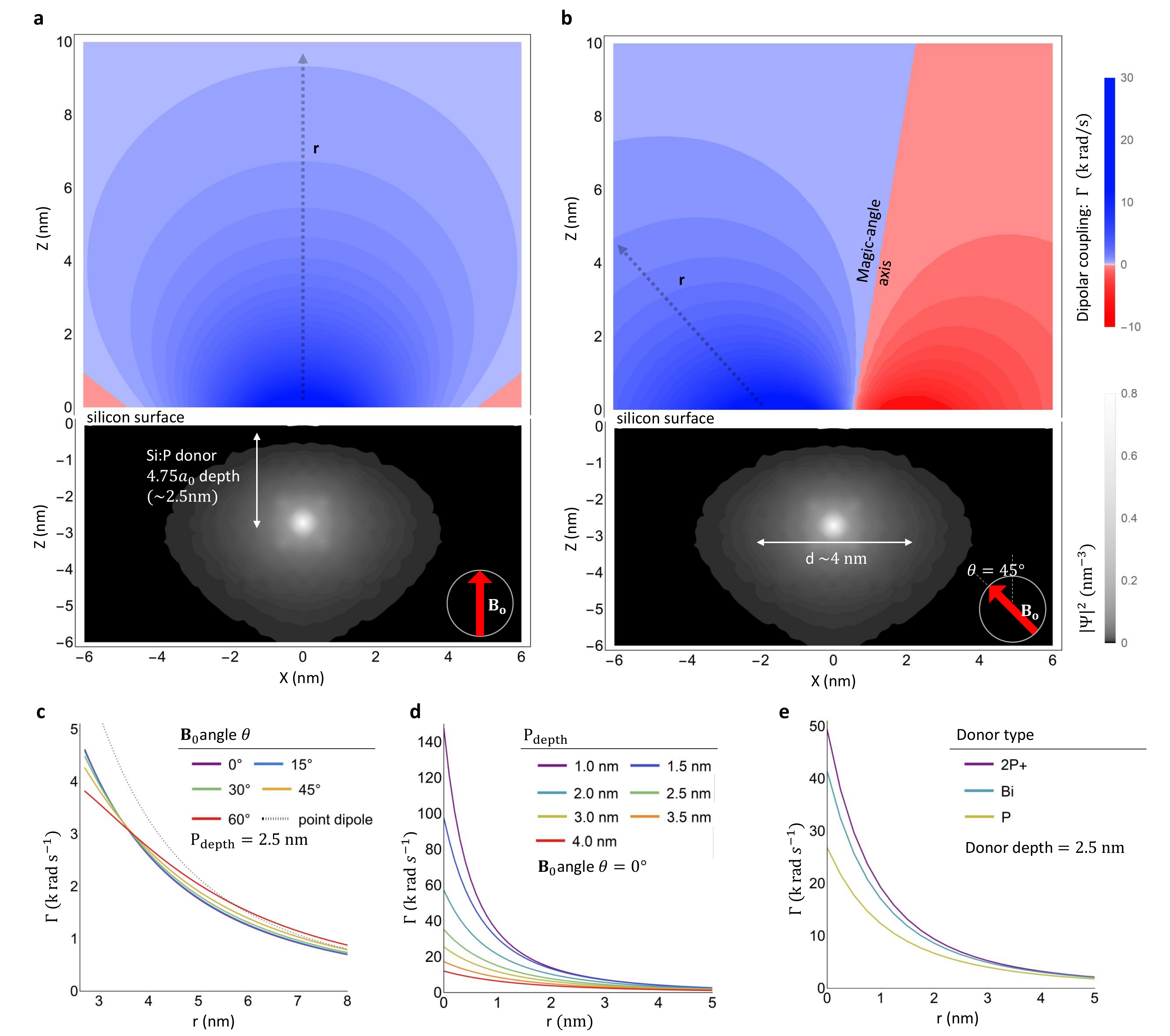}
 \vspace{-15pt}
  \caption{\textbf{Tight-binding characterisation of the spatial-frequency encoding of the dipole-dipole field produced by delocalised wave functions of donor electrons in silicon.}
  \textbf{a}) The cross-section of the ZZ component of the dipole-dipole field $\Gamma$ (upper) produced by the donor wave function with a diameter of $\sim4\rm\,nm$ of the P donor in Si at a depth of $\sim2.5\rm\,nm$ (lower) calculated using the NEMO tight-binding framework \cite{Usman2016,Usman2017} based on $\sim3$ million Si atoms (see Methods section: Tight binding formalism).
  \textbf{b}) Tilting the background magnetic field $\bf B_0$ allows for axis ($r$) control of the dipolar lobes and maintains the magic-angle axis. This behaviour is characteristic of a point-dipole, indicating suitability for molecular imaging. 
  \textbf{c}) Variations in the axial profile of $\Gamma(r)$ as a function of $\bf B_0$ orientation (solid, coloured) originate from the asymmetry of the near-surface wave function. It is important to account for these variations in the context of 3D imaging, however, they are collectively relatively close to the point-dipole field (dashed, grey) particularly for more distant targets. 
  \textbf{d}) Dipole filed coupling $\Gamma$ as a function of $r$ for a range of donor depths. Donor depth is a potent way of controlling the gradient of the dipolar field, especially for closer targets as illustrated here. 
  \textbf{e}) The near-surface gradient can be further engineered by decreasing the size of the wave function using bismuth donors (blue) or an ionised neighbouring pair of phosphorus donors (2P+) (purple).
  }\label{Fig:2}
   \vspace{15pt}
\end{figure}

From the TB simulations, we find that the secular ZZ component of the dipolar field (Fig.\,\ref{Fig:2}\,a upper) emerging from the donor probability density (Fig.\,\ref{Fig:2}\,a lower) effectively retains the same shape as a point-dipole, characterised by the curved lobes and magic-angle axis (for details see Appendix \ref{Si_Sc:Effects_of_spatial_delocalisation}: Effects of spatial delocalisation). Furthermore, the characteristic dependence of a point-like dipole field on the background magnetic field $\bf B_0$ orientation is maintained. This is depicted in Fig.\,\ref{Fig:2}\,b), whereupon introduction of a tilt in $\bf B_0$ ($\theta=45^\circ)$, the dipole coupling field produced by the donor orients along its direction. These key properties enable the spatial-frequency encoding to be controlled through the co-latitude and longitude ($\theta, \phi$) orientations of $\bf B_0$. 

To facilitate 3D imaging, we introduce a custom spatial-frequency encoding formalism to uniquely address individual dipole slices in the presence of varying $\bf B_0$ orientation. A dipole slice is the lobe-shaped surface on which the target nuclei experience the same coupling $\Gamma$ to the donor. In 3D space, we specify this surface by three parameters $(r, \theta, \phi)$, defined as follows. The parameter $r$ is the distance between the point on substrate surface above the donor and the given slice, along the direction of $\bf B_0$ (dashed line, Fig.\,\ref{Fig:2}\,a\,and\,b). While, parameters $\theta$ and $\phi$ are the orientations of the $\bf B_0$, as noted before. Therefore, a slice specified by single parameter $\Gamma$ in frequency space, is specified by a vector ($r, \theta, \phi$) in 3D space. The vector ($r, \theta, \phi$) is used to address of the entire lobe-shaped surface, produced by the donor wave function under $\bf B_0$ (note, it should not to be confused with the spherical polar coordinate notation for a single point in space).

To achieve Angstrom level resolution, it is essential to characterise the dependence of the dipolar field on the co-latitude direction of the background field $\theta$ (Fig.\,\ref{Fig:2}\,c). We note that the dipole field remains relatively close to that of a point-dipole (Fig.\,\ref{Fig:2}\,c dotted) under a range of $\bf B_0$ tilt ($\theta\leq60^\circ)$ sufficient for 3D molecular imaging.
The dipole field can be engineered by varying the donor depth, as indicated by Fig.\,\ref{Fig:2}\,d (see Appendix \ref{Si_Sc:Engineering_the_gradient_by_donor_placement}: Engineering the gradient by donor placement). Note, we discuss the optimal depth later. The dipole field can also be produced using donors with more confined wave functions, such as bismuth (Fig.\,\ref{Fig:2}\,e blue) or an ionised pair of phosphorus atoms 2P+ (Fig.\,\ref{Fig:2}\,e purple) (see Appendix \ref{Si_SC:Engineering_the_gradient_field_through_confinement}: Engineering the gradient field through confinement).

\subsection{Nuclear spin storage protocol}
\label{Sc:NSS_protocol}

\begin{figure}[!htb]%
\centering
 \includegraphics[width=1\linewidth]{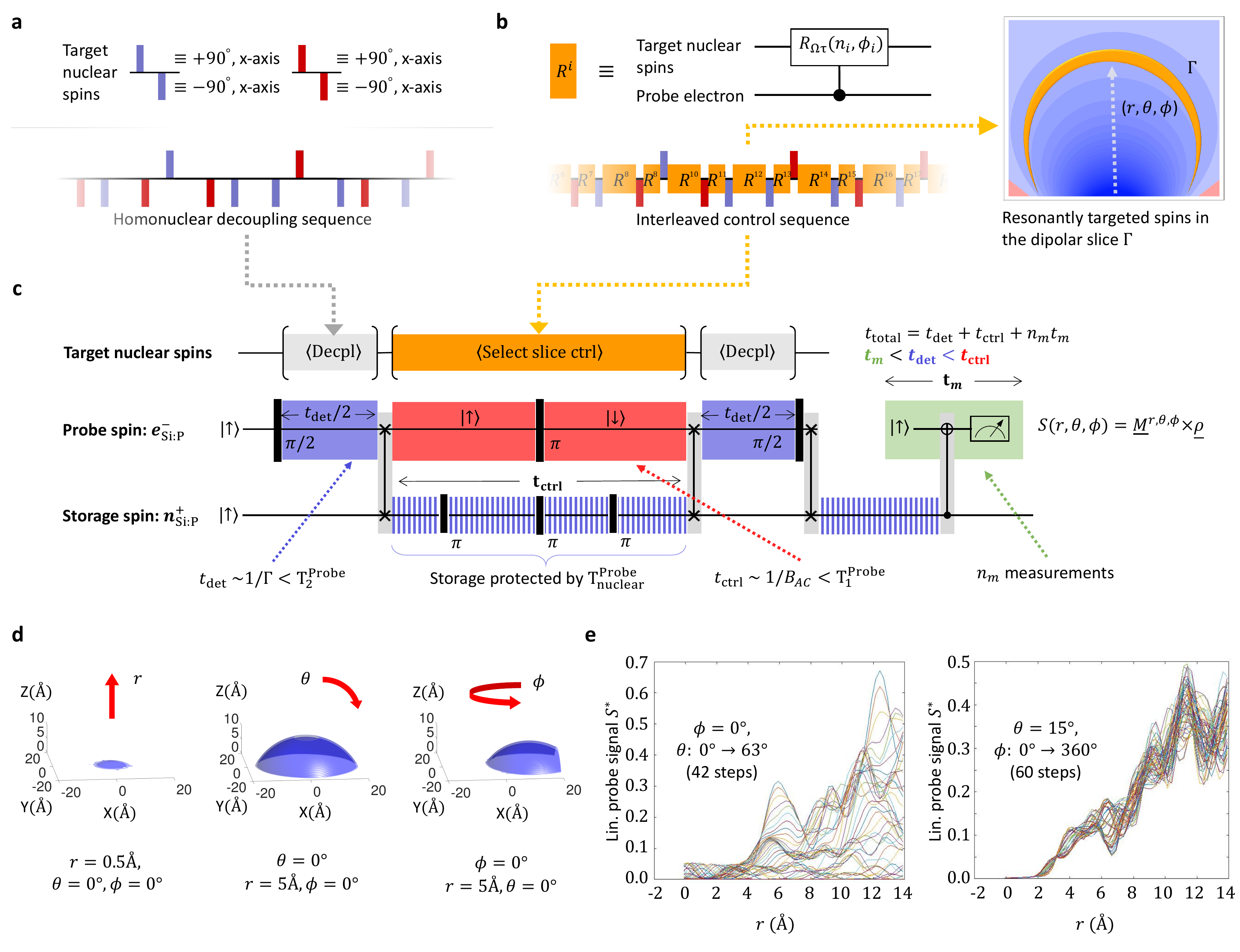}
 \vspace{-15pt}
  \caption{\textbf{Schematic of the Nuclear Spin Storage (NSS) detection protocol.}
  \textbf{a}) Continuous application of the homonuclear decoupling sequence, CORY-24 \cite{Cory91} comprising of $\pi/2$ pulses (red/blue), ensures that nuclear target spins have sufficiently long coherence times.
  \textbf{b}) Interleaving \cite{Perunicic2016} control rotation pulses (orange) into the decoupling sequence enables dipolar slices in the sensing volume to be individually addressed.
  \textbf{c}) The structure of the NSS detection protocol features three distinct segments.The detection segment of length $t_{\rm det}\sim1/\Gamma_i$ limited by $T_2^{\rm probe}$ (purple). Nested in between is the control segment of length $t_{\rm ctrl}\sim1/\gamma B_{\rm AC}$ limited by the relatively longer $T_1^{\rm probe}$ timescale (orange), during which the nuclear storage is used to protect the information about the target magnetisation collected by the first half of the detection segment. Lastly, the signal is obtained by measuring the state of the electron (green) which allows for $n_m$ projective measurements to be collected in a relatively short time $t_m\ll t_{\rm det}+t_{\rm ctrl}$, without repeating the entire detection protocol.
  \textbf{d}) The sensing volume is sampled in the ``criss-cross" MRI fashion consisting of slice radius sweeps ($r$) repeated for an array of $\bf B_0$ orientations ($\theta, \phi$).
  \textbf{e}) Example of the signal profiles over $r$ for a selection of $\theta$ and $\phi$ orientations (right/left). The signal peaks encode the number of nuclear target spins present in a given slice enabling the atomic-level image of the sensing volume to be formed in 3D.
  }\label{Fig:3}
   \vspace{15pt}
\end{figure}

The nuclear spin storage (NSS) detection protocol (depicted in entirety in Fig.\,\ref{Fig:3}\,c.) optimally leverages the long longitudinal ($T_1^{\rm probe}$), transverse ($T_2^{\rm probe}$) and nuclear transverse ($T_{\rm nuclear}^{\rm probe}$) coherence times of the group-V donor qubits to extract the information about the number of molecular target spins in a given slice ($r, \theta, \phi$).  The NSS protocol requires several layers of careful quantum control over both the qubit and the target nuclear spins.

Collectively, the targeted nuclei are subjected to strong homonuclear decoupling \cite{Cory91} with spectrally broad pulses resonant with the background Zeeman splitting $\omega_0=\gamma_tB_0$, (Figs.\,\ref{Fig:3}\,a\,and\,c top register - denotes numerous nuclear spins). Decoupling suppresses the dipole-dipole interactions between the nuclear spins, narrowing their spectral broadening thus enabling the spatial-frequency encoding to be facilitated. Simultaneously with target nuclear decoupling, the NSS protocol features three segments that perform different functions. It starts by the detection segment (Fig.\,\ref{Fig:3}\,c mid register, blue) where the probe's electron spin is allowed to evolve for a time period $t_{\rm det}/2$ after such that its phase encodes the information about the state of the current molecular target magnetisation. This information is stored on the probe's nuclear state protected by its long transverse coherence time $T_{\rm nuclear}^{\rm Probe}$ (Fig.\,\ref{Fig:3}\,c, bottom register, dashed blue).

In the middle of the detection segment is the control segment of duration $t_{\rm ctrl}$  (Fig.\,\ref{Fig:3}\,c, mid register, red) where the electron spin remains protected by the relatively long $T_1^{\rm Probe}$. A dipolar slice of choice is addressed resonantly using by application of interleaving \cite{Perunicic2016} (Fig.\,\ref{Fig:3}\,c, top register, orange) which ensures that only the spins in that dipolar slice are control rotated based on the state of the probe spin by the phase $\Omega t_{\rm ctrl}$, where is the control Rabi frequency $\Omega=\gamma_tB_{\rm AC}$ (Fig.\,\ref{Fig:3}\,b).  The target nuclei experience gradient Zeeman splitting, $\omega=\gamma_tB_0+\Gamma$, while the nuclear quantum storage is subject to contact hyperfine, both require strategic refocusing by $\pi$-pulses. Details can be found in the Methods section: Details of the NSS protocol.
At the end of the control segment, the phase is brought back from storage onto the probe electron spin, followed by the second half of the detection segment (Fig.\,\ref{Fig:3}\,c, mid register, blue, right). The states from the first and the second half of the detection segment interfere with the resulting state encoding only the relative change in magnetisation of the target molecule, which is by construction proportional to the number of nuclear spins present at the probed slice.

In the final segment, the resulting phase is projected onto population difference and placed in nuclear storage, enabling measurement statistics ($n_m$ electron spin projections) to be acquired independently, without the need to repeat the entire sequence (Fig.\,\ref{Fig:3}\,c, green).

Detailed numerical analysis of the NSS protocol and illustration of its behaviour can be found in the Appendix \ref{Si_Sc:Testing-spatial-frequency-encoding-and-stability-of-NSS-detection protocol}: Testing spatial-frequency encoding and stability of NSS detection protocol. The signal produced by a single target nucleus with coupling $k$ to the probe and in the spectral vicinity if the given slice $\Gamma$ ($k\approx\Gamma$) can be captured sufficiently well for our purposes by the following relations:
\begin{align} \label{eq:Signal_from_sigle_nucleus}
S(\Gamma,k)=\left(1-\cos\left(ak/2\,t_{\rm det}\right)\right) \frac{(\gamma_{\rm t}B_{\rm AC})^2}{\Omega_{\rm k}^2 + (2\pi/T_{\rm \rho\_target})^2} \sin^2(\Omega_{\rm k}/2\,t_{\rm ctrl}),
\end{align}
\begin{align} \label{eq:fine_Rabi_rate}
\Omega_{\rm k}&=(a(\Gamma-k))^2 + (\gamma_{\rm t}B_{\rm AC})^2,
\end{align}
where $a\sim0.3$ is an effective scale factor of the coupling between the probe and the target nuclear spins dependent on the specific decoupling sequence used, and $T_{\rm \rho\_target}$ is the effective coherence time of the target nuclear spins under NSS protocol, which we estimate can reach up to $100\,\rm ms$ (extensive analysis is featured in Appendix \ref{Si_Sc:Target_coherence_under_NSS_detection_protocol}: Target coherence under NSS detection protocol).

The net signal $S_{\rm net}(\Gamma)$ detected from a slice $\Gamma$ that contains $n_{\rm net}$ nuclear spins (of the same nuclear species) is the product of the individual signals $S_i$, assuming each nucleus $i$ is in a thermally mixed state:
\begin{eqnarray} \label{eq:Net_Signal_from_all_nuclei}
S_{\rm net}(\Gamma)=\frac{1}{2}e^{-t_{\rm det}/T_{\rm 2\_ probe}}\prod_{i=1}^{n_{\rm net}}S(\Gamma,k_i),
\end{eqnarray}
where $S(\Gamma,k_i)$ is given by Eq.\,\ref{eq:Signal_from_sigle_nucleus} and $k_i$ is the coupling to the $i^\text{th}$ nuclear spin. Therefore, linearisation:
\begin{eqnarray} \label{eq:scaled_Net_Signal_from_all_nuclei}
L_{\rm net}=-\log(2S_{\rm net}),
\end{eqnarray}
ensures that the processed signal is positive definite and directly proportional to the number of target nuclear spins present in the slice $\Gamma$.

\subsection{Optimal control parameters for 3D sampling}
\label{Sc:optimal_control_parameters}

The resolution and the net experimental running time intrinsically depend on the real space cross-sectional profile of dipolar slices and the total number of the slices $n_{\rm slices}$ used to probe the sensing volume. To account for the nonlinear, nonplanar dipole-dipole field, we develop dynamic control over parameters $t_{\rm det}$, $B_{\rm AC}$ and $t_{\rm ctrl}$ as the function of slice frequency $\Gamma$. We follow by outlining a procedure for 3D sampling of the sensing volume by slice $\Gamma$ parameters $r$, $\theta$ and $\phi$.

The amplitude of the signal response per target nucleus is kept uniform by varying the detection time segment with respect to slice frequency $\Gamma$, as follows:
\begin{eqnarray} \label{eq: varing_t_det}
 t_{\rm det}= \frac{\pi\rho_{\rm det}}{a\Gamma},
\end{eqnarray}
where $\rho_{\rm det}\in[0,1]$ is the response per target nucleus ratio ($\rho_{\rm det}\approx0.2$ for dense molecular samples). For details see Appendix \ref{Si_Sc:Optimal_detection_segment_length}: Optimal detection segment length.

Similarly, the width of the slice cross-sectional profile is kept uniform by varying the strength of the resonant interleaving field $B_{\rm AC}$ with respect to the dipolar gradient:
\begin{eqnarray} \label{eq: varing_B_AC}
 B_{\rm AC}(\Gamma)= \frac{B^{\rm AC}_0}{\partial \Gamma/\partial r(r=0)} \frac{\partial \Gamma}{\partial r},
\end{eqnarray}
where, $B^{\rm AC}_0$ is the interleaving field strength used to address surface slices ($r=0$). This enables for the control segment $t_{\rm ctrl}$ to be further optimised:
\begin{eqnarray} \label{eq: varing_t_cntl}
 t_{\rm ctrl}(\Gamma)= \frac{\pi \rho_{\rm ctrl}}{\gamma_t B_{\rm AC}(\Gamma)},
\end{eqnarray}
where the factor $\rho_{\rm ctrl}$ is the nuclear target spin control rotation ratio. The detailed description is illustrated in Appendix \ref{Si_Sc:Controlling_the_cross-sectional_slice_profile_in_real_space}: Controlling the cross-sectional slice profile in real space.

There are a multitude of ways to probe the sensing volume, for demonstration purposes we adopt uniform sampling over variables ($r, \theta,\phi$) (Fig.\,\ref{Fig:3}\,d) which define each slice of the dipolar field in the sensing volume. The NSS protocol is performed over the set of slices (combinations of $r, \theta,\phi$) that spans the entire sensing volume, thus producing signal profiles proportional to the number of target nuclei present in each slice (Fig.\,\ref{Fig:3}\,e). For the methodology of converting such a collection of signal profiles into a 3D image of the nuclear target density, see Method section: 3D density image processing.

\subsection{Surface termination and optimal donor depth}
\label{Sc:Surface_termination}

The hydrogen-terminated silicon (Si-H) has favourable physical thickness and fabrication techniques, however, it contains surface nuclear spins which, unlike molecular hydrogen, is subjected to contact hyperfine interaction with the qubit electron spin. The Si-H can become a source of decoherence as they follow profoundly different dynamics under decoupling sequences designed to address dipole-dipole coupling only. The contact hyperfine interaction creates the clustering of Si-H above the donor that effectively experience different Zeeman splitting. We perform extensive analysis over the number of spins in such clusters, their maximum hyperfine interaction and map their coherence when subjected to the NSS protocol - as a function of the donor depth and the strength of the decoupling pulses, see Appendix \ref{Silicon_surface_and_the_optimal_qubit_depth}: Silicon surface and the optimal qubit depth. The results enable us to find the optimal qubit depth, which for phosphorus donors in silicon is around $2-2.5\rm\,nm$. We note that recent studies based on STM fabrication and metrology of donor qubits have found phosphorus atoms at a similar depth range below the H-passivated surface \cite{Usman2016, Voisin2020}.

\subsection{Simulated 3D imaging of a single membrane protein-lipid system}
\label{Sc:protein-lipid_system}

We demonstrate the overall qubit-based imaging protocol by simulating the atomic-level structure determination of a single transmembrane protein-lipid molecular system, the M2 proton channel of the influenza virus \cite{Cady2010} embedded in the DDPC lipid bilayer(Fig.\,\ref{Fig:1}\,c). Initially, we use a relatively fast sectional imaging mode that focuses only on the part of the protein-lipid system using the $^{13}$C nuclear sub-density - as an example of the method's nuclear species selectivity. Finally, we demonstrate the extended imaging mode capable of atomic-level reconstruction of the whole system, i.e. imaging through the entire lipid membrane. We will selectively image the $^{14}$N nuclear sub-density in the extended imaging mode.

\begin{figure}[!htb]%
\centering
 \includegraphics[width=1\linewidth]{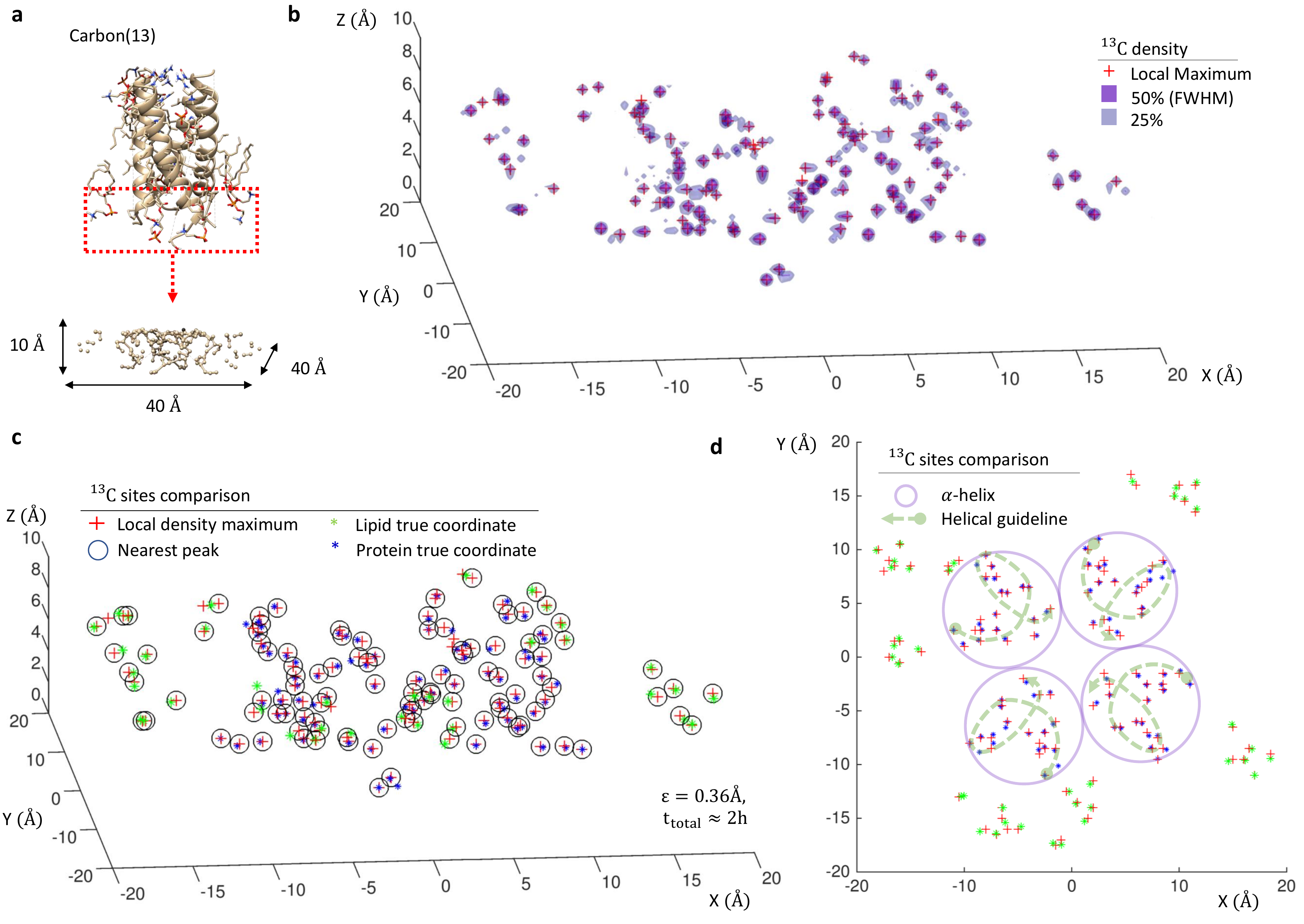}
 \vspace{-15pt}
  \caption{\textbf{Simulated qubit-based molecular imaging of the surface section of the protein-lipid system with the focus on $^{13}$C sub-density.}
  \textbf{a}) To facilitate a relatively fast imaging mode, the sensing volume is initially restricted to the $\rm 40\times40\times10\AA$ region of the external part of the M2 proton channel-lipid system (only neighbouring lipids shown, water molecules suppressed for clarity).
  \textbf{b}) The image of the nuclear sub-density (simulation), featuring 3D peaks (red +), the $50\%$ (FWHM) and $25\%$ contours. The imaging parameters reflect presently achievable experimental conditions for Si:P qubit systems: $B^{\rm AC}_0=0.5\rm\,\mu T$, $T_2^{\rm probe}=T_{\rm \rho\_target}=100\rm\,ms$, $r_{\rm max}=1.4\,\rm\AA$, $dr=0.25\,\rm\AA$, with $\theta_{\rm max}=63^\circ$, $d\theta=1.5^\circ$, $d\phi=6^\circ$, $n_m=1000$, $\rho_{\rm det}=\rho_{\rm ctrl}=0.2$. Experiment run time estimate $\sim 2$\,hours.
  \textbf{c}) Density peak locations are automatically paired (circled) with their nearest true atomic coordinates (marked with *). The average radial deviation from true coordinates is $\epsilon=0.36\,\rm\AA$.
  \textbf{d}) Top-down view (z-projection) of the density image shoves projections of $\alpha$-helix segments (circled in purple), with helical guidelines superimposed (green dashed).
  }\label{Fig:4}
   \vspace{15pt}
\end{figure}

In the sectional imaging mode, the sensing volume is restricted to the external (Fig.\,\ref{Fig:4}\,a, lower) $\rm 40\times40\times10\AA$ segment of the protein-lipid system. The imaging mode is optimal for identifying proteins, their conformation states as well as their near-surface reactants and ligands. It is relatively fast as the nuclei in the sensing volume possess stronger coupling to the probe, enabling for both shorter $t_{\rm ctrl}$ and $t_{\rm det}$ timescales. The 3D $^{13}$C nuclear sub-density (Fig.\,\ref{Fig:4}\,b) for the section of the molecular system can be attained within $\sim2$\,hours of signal acquisition under presently achievable experimental conditions (see Appendix \ref{Si_Sc:Carbon(13)_nuclear_density_imaging}: Carbon(13) nuclear density imaging).

Individual 3D density peaks feature prominently in the image, interpreting their maxima (Fig.\,\ref{Fig:4}\,c, +) as reconstructed atomic sites without optimisation (i.e. for bond lengths and angles) proves a close match to the true coordinates (Fig.\,\ref{Fig:4}\,c, *) with average deviation $\epsilon=0.36\,\rm\AA$ in line with the voxel grid spacing of $d=0.5\,\rm\AA$. The top-down view of the density image (Fig.\,\ref{Fig:4}\,d) provides a more intuitive insight into the protein's symmetric structure. Notice the four $\alpha$-helices (Fig.\,\ref{Fig:4}\,d, circled in purple), with their helical guidelines superimposed to aid visual interpretation (Fig.\,\ref{Fig:4}\,d, green dashed). Furthermore, the extensive overlapping of dipolar slices makes the imaging method relatively resistant to noise and quantum errors (see Appendix \ref{Si_Sc:Impacts_of_signal_noise_in_density_imaging}: Impacts of signal noise in density imaging). For comparative studies of sectional imaging mode applied to hydrogen and nitrogen sub-densities, see Appendix \ref{Si_Sc:Nitrogen(14)_density_imaging}: Nitrogen(14) density imaging and Appendix \ref{Si_Sc:Hydrogen_density_imaging}: Hydrogen density imaging.

\begin{figure}[!htb]%
\centering
 \includegraphics[width=1\linewidth]{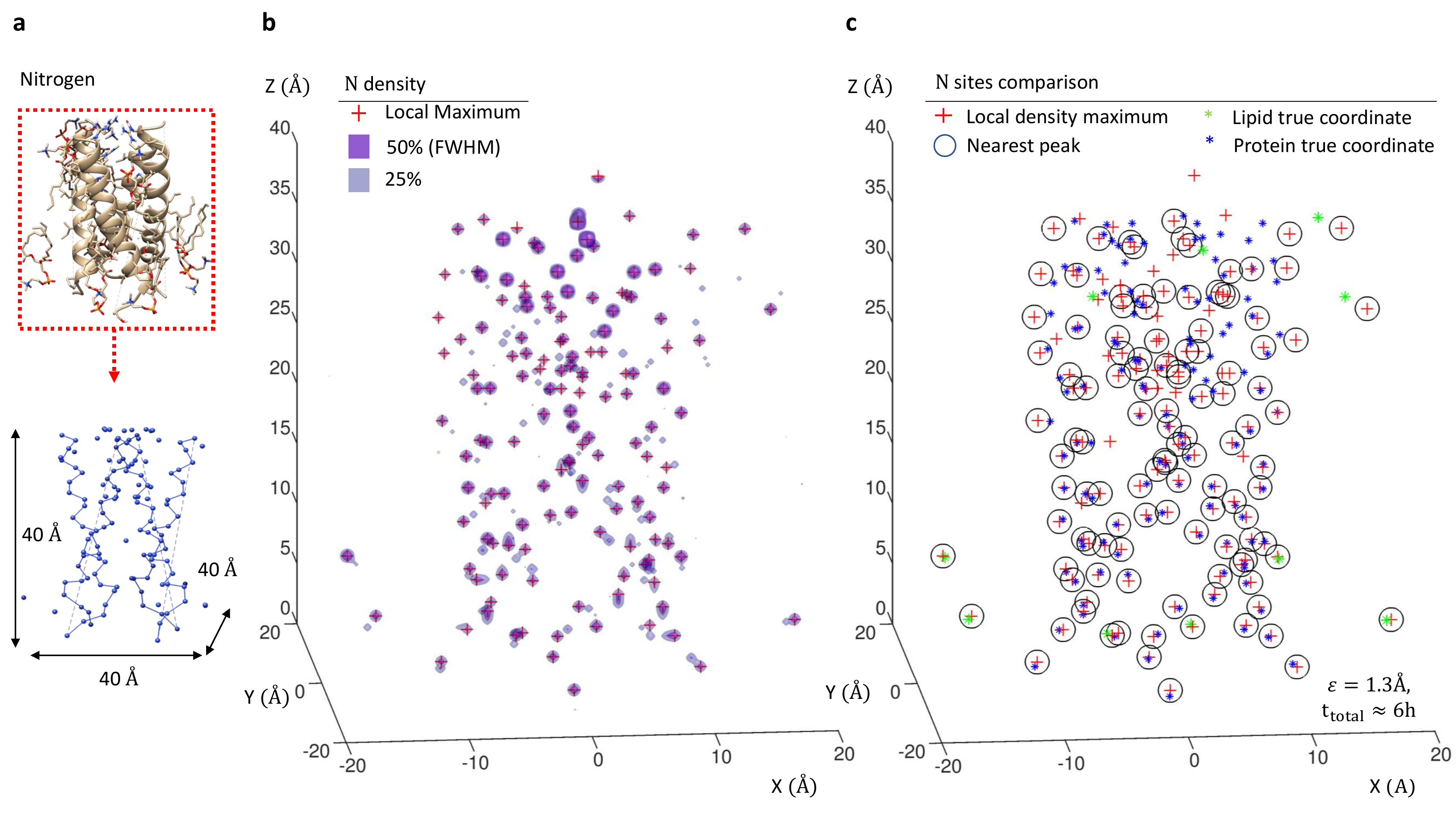}
 \vspace{-15pt}
  \caption{\textbf{Simulated qubit-based molecular imaging across the entire lipid membrane.}
  \textbf{a}) The imaging is illustrated for the $^{14}$N sub-density of M2 proton channel-lipid system in the sensing volume ($\rm 40\times40\times40\AA$).
  \textbf{b}) The full across-the-membrane image of the $^{14}$N density of the M2 proton channel-lipid system (simulation), featuring 3D peaks (red +), the $50\%$ (FWHM) and $25\%$ contours. The sampling parameters are: $d=0.75\,\rm\AA$, $B_0^{\rm AC}=1.25\rm\,\mu T$, $dr=0.5\,\rm\AA$, $r_{\rm max}=4\,\rm\AA$ $d\phi=5^\circ$, $\theta_{\rm max}=45^\circ$. Experimental run time estimate $\sim 6$\,hours.
  \textbf{c}) The average deviation between the density peaks (red +) and the true coordinates (marked with *) is $\epsilon=1.3\,\rm\AA$.
  }\label{Fig:5}
   \vspace{15pt}
\end{figure}

The extended imaging mode expands the sensing volume to the ($\rm 40\times40\times40\AA$) region encompassing the entire M2 proton channel and the neighbouring lipids (Fig.\,\ref{Fig:5}\,a). The nitrogen sub-density 3D image of the entire protein-lipid system (Fig.\,\ref{Fig:5}\,b) would experimentally take around $6$\,hours of signal acquisition. Here the average deviation from the true coordinates (Fig.\,\ref{Fig:5}\,c) is $\epsilon=1.3\,\rm\AA$. The lower parts of the molecule show close matching to the original coordinates, while the upper, being further away from the donor, take longer to sample allowing decoherence to lower the signal contrast and increase the deviation from true coordinates. For additional information see Appendix \ref{Si_Sc:Imaging_the_entire_transmembrane_protein-lipid system}: Imaging the entire transmembrane protein-lipid system.

\section{Discussion}
\label{Sc:Conclusion}

In this paper, we proposed a platform technology based on near-surface qubits in silicon for noninvasive \textit{in situ} structure determination of individual molecules, well-suited to lipid membrane-protein systems. In this qubit-based molecular imaging technique, each qubit functions as a probe site with a sensing volume above the silicon surface commensurate with the nanoscopic size of the targeted protein stature. The technique is enabled by the carefully devised NSS protocol that controls the effective quantum interactions between the qubit's electronic wave function and the nuclei in the sensing volume. This allows the spin qubits in silicon to facilitate MRI of the target protein's individual atomic sites.
Our platform technology leverages established atomic fabrication techniques developed for silicon-based quantum computing over the last decade. It opens up a unique avenue towards scale-up in the field of protein structure determination, which is at preset fundamentally limited by exclusive reliance on macroscopic equipment, allowing incumbent technologies to achieve up to a handful of imaging sites per research facility.

Using a single proton channel of the influenza virus as an archetypal example of a lipid membrane protein system, we showed through simulation that under achievable experimental conditions the molecular imaging protocol has the ability to directly image $^1$H, $^{13}$C and $^{14}$N nuclear densities of individual protein-lipid systems in sub-Angstrom resolution.
Experimental realisation of the qubit-based molecular imaging platform would provide noninvasive access into atomic structure of arbitrarily conformational instances of individual inhomogeneous molecular systems. This ability uniquely complements the present methods: X-ray crystallography, NMR and cryo-EM, that rely on indirect protein stature determination by averaging over conformationally uniform ensembles of molecules. By offering such an unprecedented level of insight into the function of proteins, the proposed platform technology model charts a course of significant implications to biomedical field.

\section*{Materials and Methods}
\label{Sc:Methods}

\subsection{Tight-binding description of the donor electron wave function}
\label{Methods_Sc:Tight_binding}

We investigate the effects of a donor's wave function spherical symmetry loss on the dipole field by utilising the Tight Binding (TB) description of the wave function \cite{Usman2016,Usman2017}. TB is a benchmarked numerical formalism that enables the entire wave function to be represented based on silicon lattice sites acting as continuous, local sources of the wave function following the Slater-orbital description. To reconstruct the complete wave function at any point, the orbital contributions from all of the lattice sites are summed up - for practical purposes, this approach is tractable because Slater-orbitals have an exponential drop off on the Angstrom scale.

TB formalism enables the wave function to be numerically modelled over a relatively large volume cell comprising of several million silicon atoms. That allows for the impact of the proximate silicon surface on the wave function shape to be comprehensively represented. Given the wave function description using TB has been thoroughly benchmarked experimentally via STM donor spatial metrology \cite{Usman2016,Usman2017, Usman2020}, it is an optimal choice for our purposes. Therefore, it is a viable pathway to calculating the realistic donor's dipole field to high precision suitable for applications in molecular imaging.

We summarise the modelling procedure as follows. The $sp^{3}d^{5}s^{*}$ TB Hamiltonian is optimised for silicon band structure \cite{Boykin2004} while the central-cell-correction model is used to include intrinsic strain and non-static dielectric screening effects \cite{Usman_JPS2015}. The donor atom is represented by a Coulomb-like potential that has been cut-off to a specific value and screened by a k-dependent dielectric function \cite{Nara1965}. Its nearest neighbour bond lengths (P-Si) have been adjusted to reflect the strain as provided by DFT calculations \cite{Overhof2004}. In accordance with the often-used experimental setups, the sample surface is set to have (001) orientation, consisting of dimer rows of Si atoms and follows the $2\times1$ surface reconstruction scheme. The impact of the surface strain due to the reconstruction is included in the tight-binding Hamiltonian by scaling the inter-atomic interaction energies in accordance with variations in the lattice bond lengths.

Hydrogen termination is highly effective in saturating the silicon surface dangling bonds. We model H passivation implicitly by shifting the energy of dangling bonds in the model, effectively neutralising their electrostatic potential~\cite{Lee2004}. The silicon volume cell around the P donor is chosen to be a square box of $40\,\rm nm^3$ (approximately 3 million atoms) with closed boundary conditions, which has been shown to be sufficient for wave function benchmarking \cite{Usman2017}. The real-space Hamiltonian is solved in parallel by the Lanczos algorithm to calculate the single particle energies and wave functions of the donor atom. The tight-binding Hamiltonian is implemented within the framework of NEMO-3D. For further details see Usman et al.\cite{Usman2017, Usman2021}.

The tight binding model provides a means to evaluate the full wave function at any point in space, however, to calculate the dipole-dipole coupling field (next Method section), it is sufficient to consider the probability density at the individual silicon and phosphorus atomic sites.

\subsection{Calculation of the Dipole-Dipole field}
\label{Methods_Sc:Dipole-Dipole_field}

The Si:P dipolar field responsible for the spatial-frequency encoding depends on the orientation of the background magnetic field $\bf B_0$. To construct the dipole-dipole coupling field above the silicon surface, we calculate the weighted sum of contributions from each lattice site:
\begin{align} \label{Methods_eq:wfn_sum}
&\Gamma(x,y,z,{\bf B_0})= \sum_{i}^{N_{\rm lattice}}|\Psi_i|^2D_{\rm point}^{\rm ZZ}(x-x_i,y-y_i,z-z_i,{\bf B_0}),
\end{align}
where $\Gamma$ is the net dipole coupling at the point $(x,y,z)$ above the Si surface for a given background magnetic field $\bf B_0$, $|\Psi_i|^2$ is donor electron's probability at the $i^{\rm th}$ site location $(x_i,y_i,z_i)$ inside a lattice consisting of $N_{\rm lattice}$ atoms describing the TB wave function. Here, $D_{\rm point}^{ZZ}(x-x_i,y-y_i,z-z_i,{\bf B_0})$ is the point-to-point dipole-dipole coupling term between the lattice site $i$ and the point $(x,y,z)$. For ${\bf r}=(x-x_i,y-y_i,z-z_i)$, $D_{\rm point}^{ZZ}$ is given by:
\begin{align} \label{Methods_eq:pointDD}
&D_{\rm point}^{ZZ}({\bf r, B_0})=\gamma_{\rm e} \gamma_{\rm n} \frac{\hbar\mu_0}{4\pi} \frac{1}{|{\bf r}|^3} \left(1-3(\hat{\bf B}_{\rm 0}\cdot\hat{\bf r})^2\right),
\end{align}
which has the shape characterised by the dipolar lobes (see Appendix \ref{Si_Sc:Effects_of_spatial_delocalisation}: Effects of spatial delocalisation) aligned according to $\bf B_0$. In principle, the probability density $|\Psi_i|^2$ can be calculated to a higher resolution by using additional coordinates in between the lattice sites, however this requires significant numerical resources while providing a relatively modest improvement to the gradient field accuracy. Furthermore, to optimise numerical calculations, we threshold the lattice sites by selecting those that satisfy $|\Psi_i|^2/|\Psi_{\rm max}|^2\geq10^{-4}$. We are thus able to reduce $N_{\rm lattice}$ from several million to around 50 thousand, while accounting for $98\%$ of the net probability density.

\subsection{Details of the NSS protocol}
\label{Methods_Sc:NSS_protocol}

We detail the internal function of the NSS protocol in the context of the phosphorus donor which possesses a spin-$1/2$ nuclear spin, other group-V elements or qubit systems equipped by highly coherent nuclear spins would follow an equivalent procedure. The outline of the (NSS) detection protocol is illustrated in Fig.\,\ref{Fig:3}\,c, and we will now describe its construction in detail here.
\\
\noindent
{\bf Electron spin phase accumulation during the first detection segment:} The detection segment of NSS protocol is responsible for the encoding of the state of the targeted nuclear density onto the phase of the probe. In contrast to \cite{Perunicic2016}, we separate it from the lengthy process of spectrally narrow control rotations interleaved into the dipole-dipole decoupling. This is achieved by splitting the spin echo sequence on the electron spin in \cite{Perunicic2016} and inserting the interleaved control section in the middle. We begin by preparing both the donor's electron and nuclear spins in the ``up'' state. Reading from the left (Fig.\,\ref{Fig:3}\,c), the pulse sequence applied to the probe spin starts off by a $\pi/2$ pulse equivalent to the echo sequence. During the course of the initial detection segment (purple) only the decoupling pulses are run on the molecular target nuclei. The decoupling pulses are fast relative to detection time $t_{\rm det}$ and their effect suppresses the dipole-dipole coupling between target spins. Collectively this renders the target's magnetisation effectively static, allowing the probe to acquire the phase associated with the entire nuclear spin environment.
\\ 
\noindent
{\bf Storing the electron spin state to the nuclear spin storage:} Instead of immediately interfering the phase acquired in the first detection segment with that of second, we swap the probe's electron spin state width the donor's nuclear spin state. Therefore the information about the molecular magnetisation is now stored in the state of the probes nuclear qubit, protected by minutes long $T_2$ of the donor's nuclear spin. The probe spin is now polarised ``up'' and the slice-selective control interleaved protocol (orange) is initiated for $t_{\rm ctrl}$ duration.
\\
\noindent
{\bf Select control during nuclear spin storage:} At a half mark ($t_{\rm ctrl}/2$) the probe spin is flipped. This is an important step, and it is useful to consider in detail. Decoupling pulses are resonant with the background Zeeman frequency ($\gamma B_0$), while the target nuclear spins near the probe are shifted away from it by the probe's dipolar field which is effectively comprised of many lobes each with a different coupling $\Gamma_i$. The spectral width of the decoupling pulses is orders of magnitude higher than any of the $\Gamma_i$, therefore the pulses will flip the nuclear spins without difficulty, however, errors in nuclear spin phase will accumulate on the timescale of order $\sim 1/\Gamma_i$. Since $t_{\rm ctrl}$ needs to be longer than that timescale (to allow for narrow spectral width of lobe slices), these errors will become significant, ultimately changing the apparent magnetisation of the entire molecule. By introducing the $\pi$ flip on the probe in the middle of the sequence, the coupling frequency changes sign, allowing for the gradient induced phase errors in the nuclear spins to gradually refocus. The interleaved protocol needs to be carefully phase-matched after the $\pi$ pulse since the fine control frequency changes from $\sim\gamma B_0+\Gamma_i$ to $\sim\gamma B_0-\Gamma_i$.
\\ 
\noindent
{\bf From nuclear spin storage to second detection segment:} By the end of the interleaved protocol, only the selected dipole lobe slice is targeted. During this period, the probe spin (red) has stayed in the polarised state, protected by seconds long $T_1^{\rm probe}$. In the meantime, the information stored (purple bars) on the P nuclear spin needs to be curated. There are multiple ways to do this - our choice in this instance is straightforward, featuring three $\pi$ pulses. The first and last are to cancel the hyperfine phase imparted by the probe electron spin on to the P nuclear spin during each of the two control segments (red). The middle $\pi$ pulse serves the role of the traditional echo pulse that causes interference between the two detection segments. At the end of $t_{\rm ctrl}$ the state is swapped back from storage to the probe spin, and allowed to interfere with the new phase acquired during the second detection segment (purple). This phase contains the information about the targeted nuclear spins located on slice $\Gamma_i$.
\\ 
\noindent
{\bf Electron spin measurement segment:} At the end of the second detection segment, the state is stored back on the donor's nuclear spin. Instead of repeating the entire sequence numerous times, we can interrogate the state in storage by re-initialising the electron spin and performing a control-not operation based of the nuclear spin. This process is relatively fast ($\rm\mu s$ time scale) and is repeated on its own $n_m$ times, therefore substantially improving the total time of the protocol.
\\ 
\noindent
{\bf The signal:} Thus far, the NSS detection protocol provides several advantages. The detection segment duration can be tailored independently to match the lobe being interrogated $t_{\rm det}\sim1/\Gamma_i$. The detection segments evolution is limited by $T_2^{\rm probe}$ and made distinct from the control segment. This allows the control segment (orange/red) to be longer, $t_{\rm ctrl}\sim1/B_{\rm AC}$, in order to achieve higher spectral resolution while being protected by a relatively lengthy $T_1^{\rm probe}$. Keeping the operations in the NSS protocol segmented enables us to capture the behaviour of the signal originating from individual target nuclei in a simple analytic form given by Eq.\,\ref{eq:Signal_from_sigle_nucleus} (note, carefully interleaved dipole-dipole decoupling underpins all the segments). It is crucial to note that the cumulative effect of the detection segments and $\pi$ pulses on donor nuclear spin produce an echo effect. This leaves the signal proportional only to the effective interaction rate between donor's electron spin and the target nucleus during the course of their integration $t_{\rm det}$, as captured by the first term in Eq.\,\ref{eq:Signal_from_sigle_nucleus}). The net phase accumulated from this integration on the donor's electron spin depends on the change in the target's effective magnetisation from the first to the second detection segments. In the NSS protocol, the change in magnetisation is effected by the control rotation of the targeted nuclear spin during the control segment, as captured by the second term in Eq.\,\ref{eq:Signal_from_sigle_nucleus}.

\subsection{3D density image processing}
\label{Methods_Sc:density_image}

A set of $\bf B_0$ orientates is selected, for $\theta: [0,\theta_{\rm max}]$, with spacing $d\theta$ and $\phi: [0,2\pi]$, with spacing $d\phi$. For each orientation the signal profile is recorded by sweeping across slice lobe radius $r: [0,r_{\rm max}]$ with the uniform spacing steps $dr$. Figure\,\ref{Fig:3}\,e depicts the signal profile as a function of $r$ for a selection of $\theta$ and $\phi$ orientations. All of the signal profiles are concatenated in an ordered vector $\underline{S}$ of length $n_{\rm slices}$. The sensing volume is represented in the discretised format of grid dimensions ($l,w,h$) with spacing $d$ containing $n_{\rm voxels}$ voxels such that $n_{\rm slices}\approx n_{\rm voxels}$. This discretisation format is used to define the unknown nuclear target spin density, which is stored in a vectorised form $\underline{\rho}$. To convert data from $\underline{S}$ to $\underline{\rho}$ we carefully construct a matrix transform $\underline{\underline{M}}$, by computing each slice in 3D space over the discretised sending volume, vectorising it and row-wise concatenating all of the slice vectors into a matrix $\underline{\underline{M}}$. It follows that (pseudo)inversion of the transform provides the 3D nuclear density voxel image $\underline{\rho}=\underline{\underline{M}}^{-1}\times\underline{S}$. This approach is robust however, it causes the numerical resources to scale quadratically with the number of voxels ($O(n_{\rm voxels}^2)=O(l^6)$, for section volume of side length $l$). Therefore, focusing on smaller volume segments is an efficient way to lower the computational requirements.

The inversion protocol used here is general and resource intensive limiting the voxel grid resolution, however, this is physically inconsequential. The gains can be made in developing more advanced sampling procedures. Uniform spacing in ($r, \theta,\phi$) undersamples radially more distant locations in 3D space, leading to density peak splitting in distant corners of the sensing volume. This can be fixed by inserting more slices in between radial directions for larger $r$, instead of a uniform sampling grid. Adaptive sampling can make even more significant advances. Here, the 3D conversion process could run in parallel to data acquisition, choosing slice parameters $(r, \theta, \phi, B_{\rm AC}, t_{\rm det}, t_{\rm ctrl}, n_m)$ iteratively on the run to minimise a chosen cost function, e.g. voxel entropy. Such an approach would optimise the sampling of unknown molecular structures during runtime.

\section*{Data Availability}
All relevant data are available from the authors on reasonable request.


\section*{Acknowledgements}
The authors wish to thank David Simpson, Julia McCoey and Alastair Stacey for useful discussions. L.C.L.H. acknowledges support of an ARC Laureate Fellowship (FL130100119). L.C.L.H. and M.U. acknowledge support from the Australian Research Council (ARC) funded Center for Quantum Computation and Communication Technology (CE170100012). The computational resources were provided by the National Computing Infrastructure (NCI) and Pawsey Supercomputing Center through National Computational Merit Allocation Scheme (NCMAS).

\section*{Author contributions}
V.S.P., C.D.H. and L.C.L.H. conceived the idea of using low-temperature Si qubits for bio-molecular imaging. V.S.P. and C.D.H. developed the NSS quantum control protocol and V.S.P. developed the 3D measurement protocol for the qubit-based molecular imaging. V.S.P., C.D.H., M.U. and L.C.L.H. contributed to the development of the theoretical framework of the protocols. M.U. designed and carried out tight binding wave function modelling. V.S.P. developed analytical and numerical Hamiltonian models and carried out simulations. V.S.P. analysed the results and wrote the manuscript in consultation with all the authors.

\section*{Additional information}

Competing financial interests: The authors declare no competing financial interests.

Correspondence and requests should be addressed to  V.S.P. (email: vpe@unimelb.edu.au) or to L.C.L.H. (email: lloydch@unimelb.edu.au).

\appendix

\FloatBarrier\section{Effects of spatial delocalisation}
\label{Si_Sc:Effects_of_spatial_delocalisation}

For describing a dipole field due to an NV-centre or other highly localised qubits, the point dipole description is a practical approximation when the spatial extent of the qubit's electronic wave function is comparable or smaller than Angstrom scale molecular features we are interested in imaging. Despite possessing numerous highly desirable characteristics (long coherence time, full spin control and deterministic fabrication in the case of Si:P), group-V donors in silicon have a relatively large wave function with Bohr radius on the nanometre scale~\cite{Usman2016}. The donor electron is therefore highly delocalised, spanning a region comparable in size to the target molecule itself. It is essential to quantify the dipole field group-V donors produce for 3D imaging of molecular structures using the protocol outlined in the main text of the paper. To this end, we consider the net dipole-dipole interaction $\Gamma(\vec{r})$ between a spatially delocalised electron with the probability distribution $|\Psi(\vec{r})|^2$ and a target nuclear spin assumed to be a point particle. It follows that $\Gamma(\vec{r})$ is a convolution of the donor's probability density $|\Psi(\vec{r})|^2$ and the point dipole-diploe field $D_{\rm point}(\vec{r})$, i.e. $\Gamma=|\Psi|^2 \ast D_{\rm point}$. A naive contemplation may lead us to conclude that the wave function will impart its large size and globular shape to $\Gamma(\vec{r})$ making the coupling field both more spherical and spatially extensive. Such an outcome would appear discouraging. Firstly, the larger size would result in diminished coupling gradient and therefore reduced resolution. Secondly, the increased spherical symmetry would diminish the spatial features associated with the dipolar magic-angle, therefore rendering the gradient slices effectively unresponsive to the orientation of the background magnetic field.

\begin{figure}[htb!]%
\centering
 \includegraphics[width=0.75\linewidth]{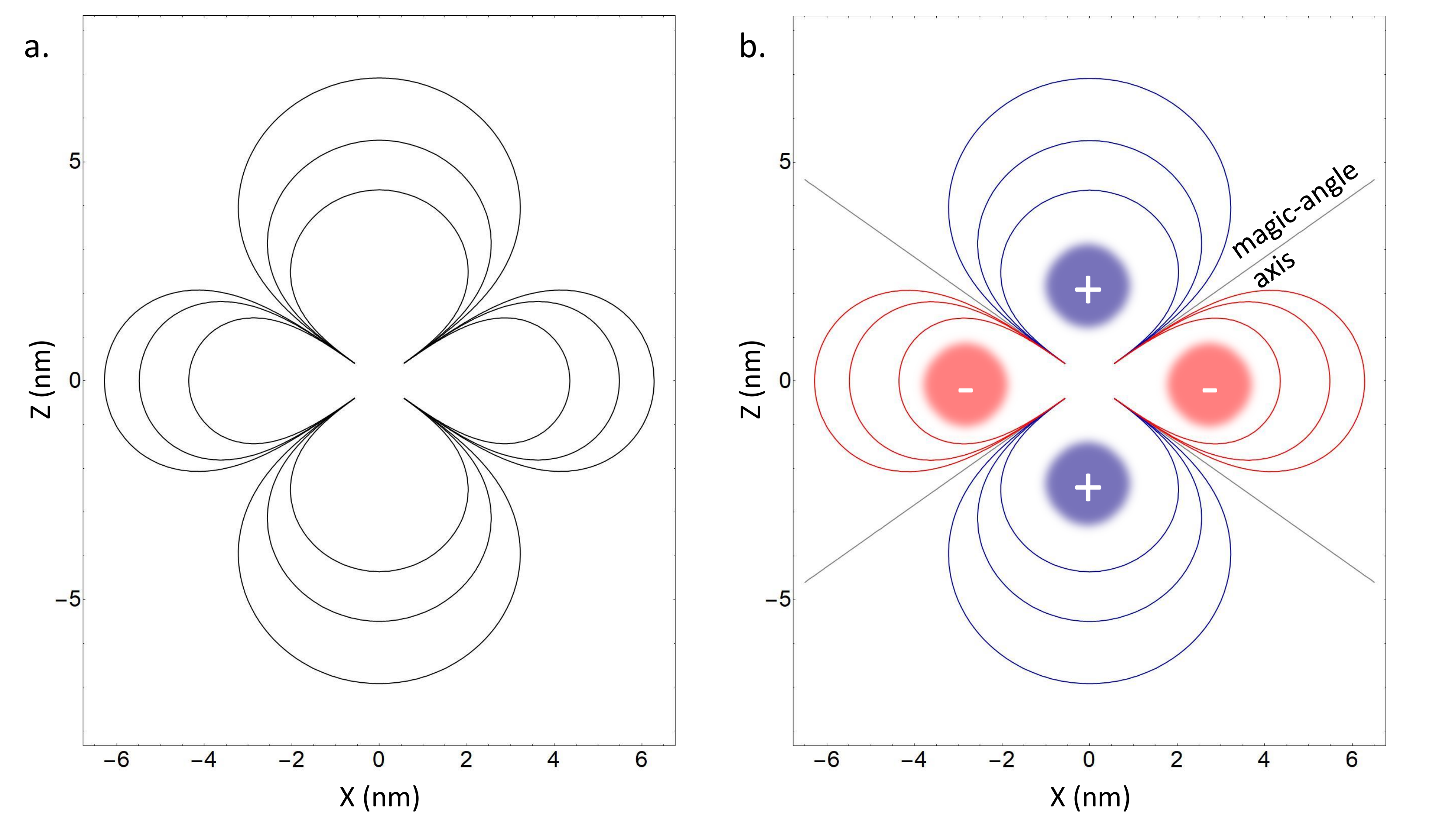}
 \vspace{-5pt}
  \caption{\textbf{The cross section of the ZZ dipole-dipole interaction terms $D_{\rm point}^{\rm ZZ}$}. 
  \textbf{a}) The contour lines showing the same coupling magnitude is a sufficient practical tool for interpreting the behaviour of coupled spin qubit systems. 
  \textbf{b}) The vertical and horizontal contours, although being equal in magnitude, carry an opposite sign. This characteristic caries little significance in experimental qubit control, however, it is essential to understanding dipole-dipole interactions between spatially delocalised spins.
   }\label{fig:SiP_dipole_explenaiton}
   \vspace{15pt}
\end{figure}

To investigate this question further, we focus on the characteristics of the dipole-dipole coupling function. Figure\,\ref{fig:SiP_dipole_explenaiton} depicts the lobe contour lines of the ZZ dipole-dipole coupling term $D_{\rm point}^{\rm ZZ}$ for point particles (i.e. before convolution with the wave function). It is worth emphasising that the shape of $D_{\rm point}^{\rm ZZ}$ follows the ``${\rm 8+\infty}$'' pattern, where the vertical set of lobes (``${\rm 8}$'') signify the positive coupling and the horizontal (``${\rm\infty}$'') negative, while the separation between the two follows the magic-angle cone of zero coupling (note: the allocation of a sign to each of the lobe pairs depends on magnetic moments of the interacting particles, however, dual polarity is always present). The sign of the coupling operates on the level of the interaction phase, forming an intrinsic part of the dipole-dipole coupling. The sign is, therefore, easy to neglect, particularly from the perspective of experimental quantum control of spin qubits.

However, the positive to negative transition has a crucial effect on the coupling that emerges from a delocalised wave function. In particular, it can ensure that the net coupling $\Gamma$ retains a similar shape as that of the point dipole-dipole $D_{\rm point}$, provided that the shape of the donor's probability density $|\Psi|^2$ is sufficiently symmetric. This can also be understood intuitively by looking at $D_{\rm point}^{\rm ZZ}$ as an example. If we visualise the magic-angle line between positive and negative regions, we can see that a sum of two closely spaced point dipole-dipole functions $(D_{\rm point}^{\rm ZZ}(\vec{r})+D_{\rm point}^{\rm ZZ}(\vec{r}+d\vec{r}))$ will, due to cancellation, result in a new function that also has similar a positive-negative division. As the net interaction at any particular point in space is the weighted sum of a point dipole $D_{\rm point}^{\rm ZZ}$ by $|\Psi|^2$, it follows that the net coupling can also resemble the shape of a point dipole-dipole interaction.

\FloatBarrier\section{Engineering the gradient by donor placement}
\label{Si_Sc:Engineering_the_gradient_by_donor_placement}

\begin{figure}[htbp!]%
\centering
 \includegraphics[width=0.75\linewidth]{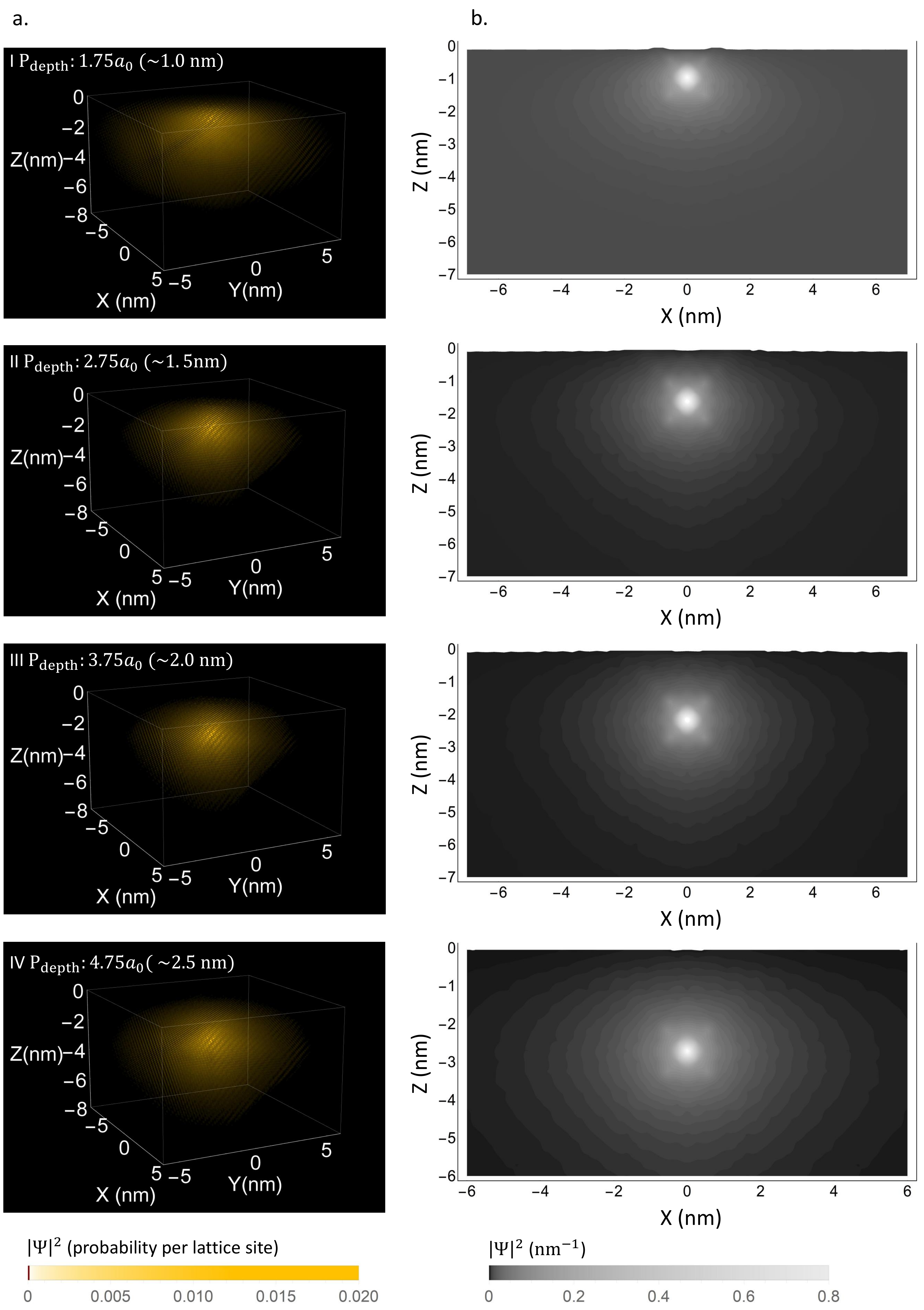}
 \vspace{-5pt}
  \caption{\textbf{Probability density of Si:P electronic wave function for donor depths ranging from $\mathbf{1}$\,nm to $\mathbf{2.5}$\,nm (I-IV).} 
  Representation in 3D (\textbf{a}) with the corresponding z-x cross-sections (\textbf{b}).
  }\label{fig:SiP_many_3D_wfns1}
   \vspace{15pt}
\end{figure}

\begin{figure}[htbp!]%
\centering
 \includegraphics[width=0.75\linewidth]{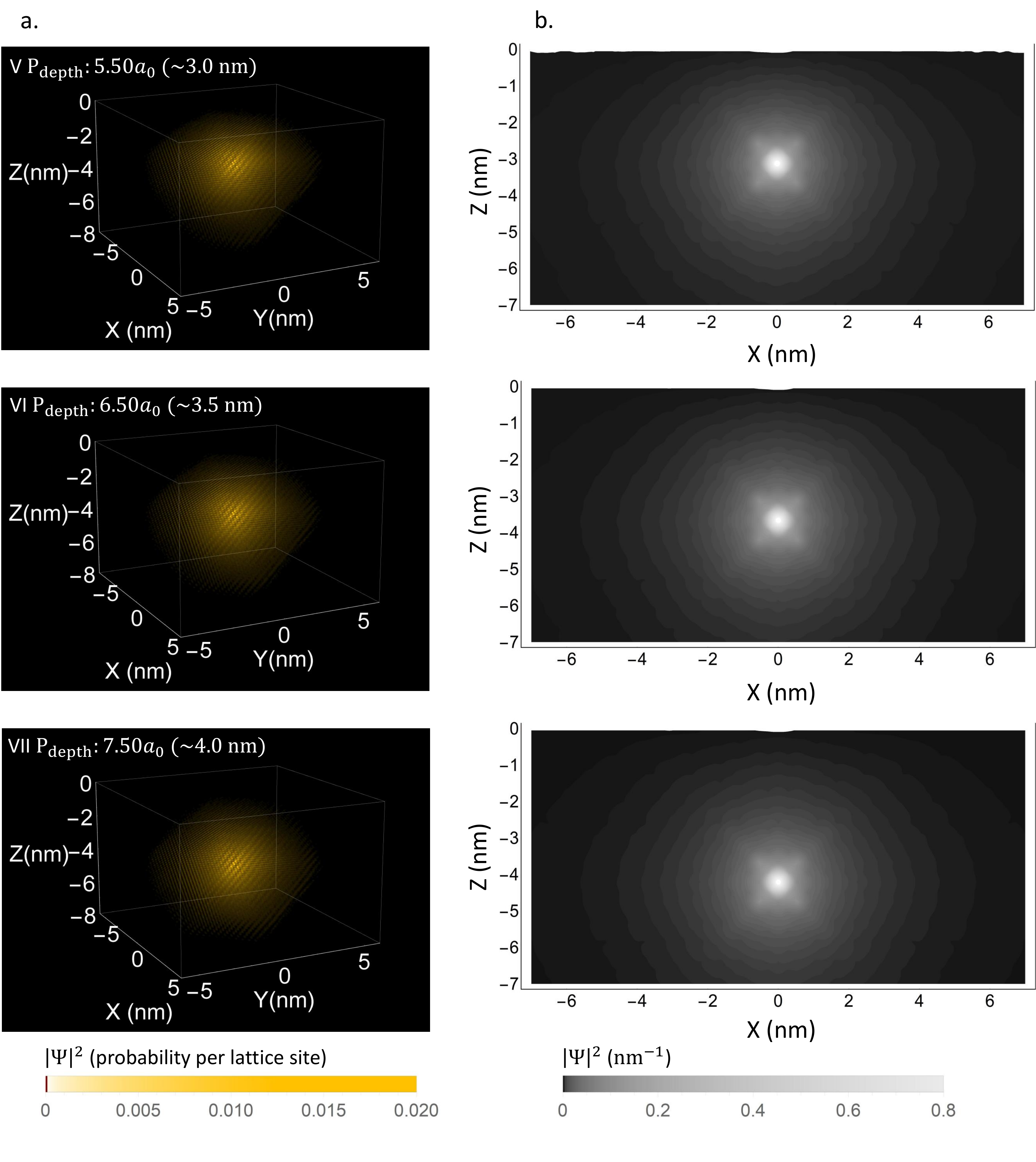}
 \vspace{-5pt}
  \caption{\textbf{Probability density of Si:P electronic wave function for donor depths ranging from $\mathbf{3}$\,nm to $\mathbf{4}$\,nm (IV-VII).} Representation in 3D (\textbf{a}) with the corresponding z-x cross-sections (\textbf{b}). Note the emerging trend where deeper donors (VII) maintain their spherical symmetry while shallow ones (see Fig.\,\ref{fig:SiP_many_3D_wfns1} case I) become vertically compressed leading to a probability distribution that is more localised in the z-direction and less in the x-y plane.
  }\label{fig:SiP_many_3D_wfns2}
   \vspace{15pt}
\end{figure}

\begin{figure}[htbp!]%
\centering
 \includegraphics[width=0.75\linewidth]{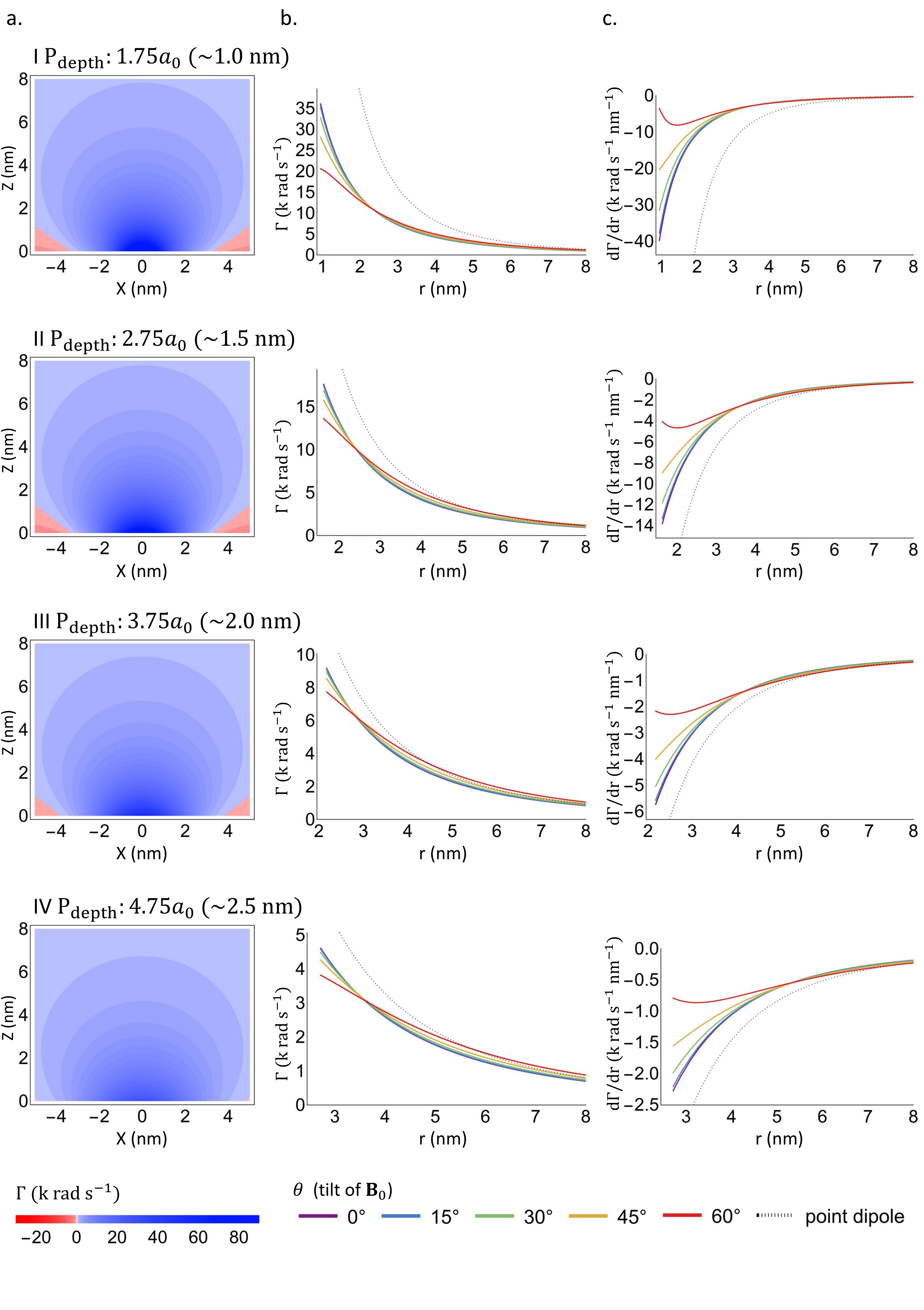}
 \vspace{-5pt}
  \caption{\textbf{The cross-sections (x,z) of the dipole-dipole field above the silicon surface for donors I to IV (Fig.\,\ref{fig:SiP_many_3D_wfns1})}. 
  The dipole field under the perpendicular ($\theta=0^\circ$) background magnetic field alignment (\textbf{a}). The linear cross-sections of dipole-dipole field under various tilt angles (\textbf{b}) and their respective gradients (\textbf{c}), all shown in comparison to the point dipole field (dotted line).
  }\label{fig:SiP_line_crosssection_for_many_3D_wfns1}
   \vspace{15pt}
\end{figure}

\begin{figure}[htbp!]%
\centering
 \includegraphics[width=0.75\linewidth]{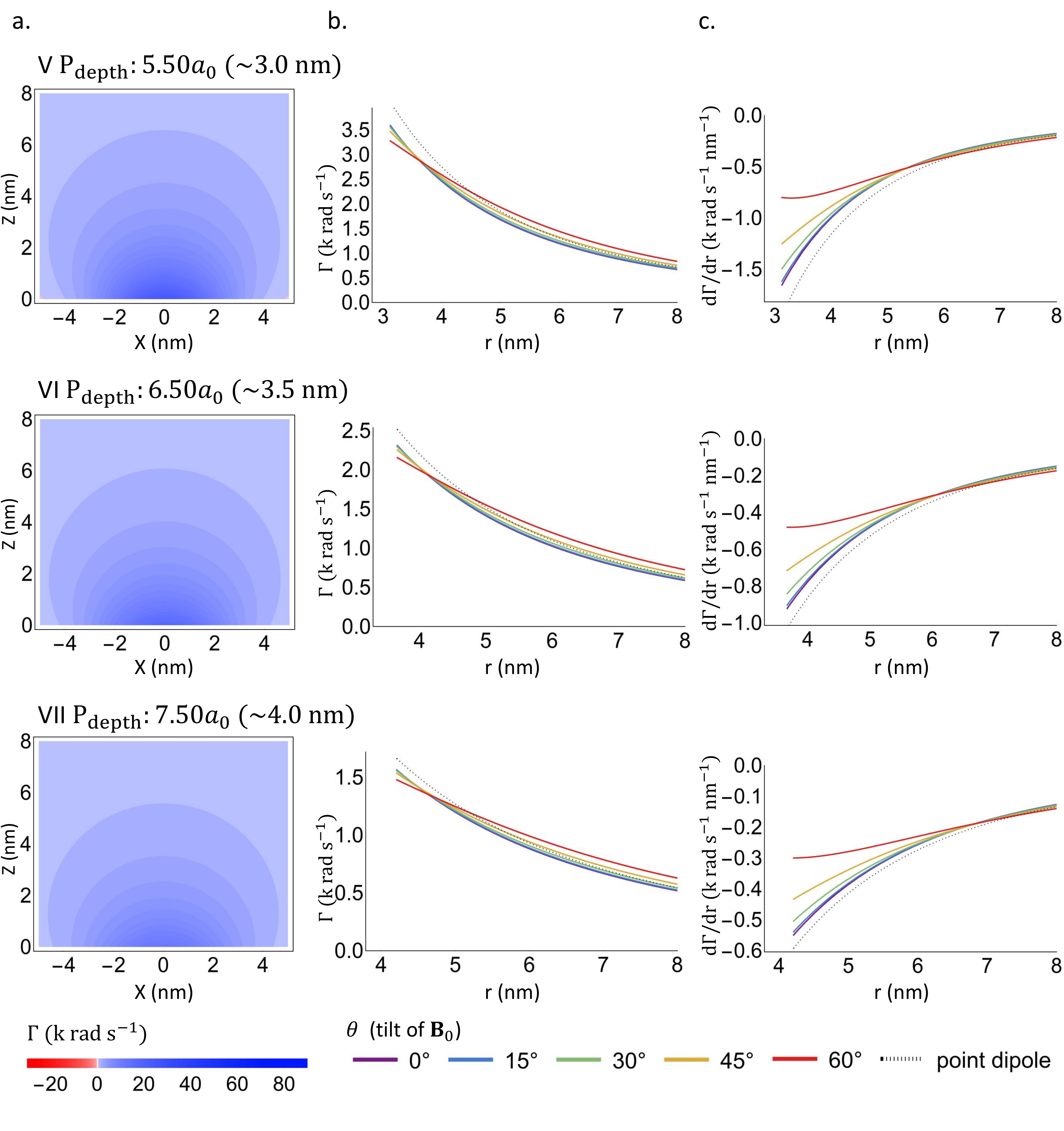}
 \vspace{-5pt}
  \caption{\textbf{The cross-sections (x,z) of the dipole-dipole field above the silicon surface for donors V to VII (Fig.\,\ref{fig:SiP_many_3D_wfns2}).}
  The dipole-dipole field under the perpendicular ($\theta=0^\circ$) background magnetic field alignment (\textbf{a}). The linear cross-sections of dipole-dipole field under various tilt angles (\textbf{b}) and their respective gradients (\textbf{c}), all shown in comparison to the point dipole field (dotted line).
  }\label{fig:SiP_line_crosssection_for_many_3D_wfns2}
   \vspace{15pt}
\end{figure}

This section quantitatively examines the dipole-dipole gradient field generated by a donor and how its strength  can be controlled by varying the physical depth of the donor. To understand the impact of the surface on the shape of the donor wave function and its associated dipole-dipole field, we take a comprehensive look at the range of different donor depths from $1\,\rm nm$ to $4\,\rm nm$. As seen in Figs.\,\ref{fig:SiP_many_3D_wfns1} and \,\ref{fig:SiP_many_3D_wfns2}, as the donor approaches the surface the probability density experiences gradual compression in the z-axis, changing its shape from spherical ($4\,\rm nm$) to elliptical ($1\,\rm nm$), thus leading to greater delocalisation in the lateral direction.
Figures\,\ref{fig:SiP_line_crosssection_for_many_3D_wfns1}\,a. and \ref{fig:SiP_line_crosssection_for_many_3D_wfns2}\,a. depicts the cross-section of the above-surface ZZ dipole-dipole field as a function of the donor depth, under the influence of a background field $\bf B_0$ orientated in a perpendicular direction. The presence of characteristic dipolar coupling features (lobes and magic-angle) persists even for near-surface donor wave functions (I) that suffer from a higher level of deformation. However, compared to a point dipole, the donor effective dipole fields appear relatively laterally broadened and vertically compressed (by $\sim 10-30\%$). This can be further seen in Figs.\,\ref{fig:SiP_line_crosssection_for_many_3D_wfns1}\,b. and \ref{fig:SiP_line_crosssection_for_many_3D_wfns2}\,b., for a multitude of magnetic field orientations (solid colours) and in comparison to a point dipole (dotted black).  The corresponding dipolar field gradients are shown in Figs.\,\ref{fig:SiP_line_crosssection_for_many_3D_wfns1}\,c. and \ref{fig:SiP_line_crosssection_for_many_3D_wfns2}\,c. To properly interpret Figs. \ref{fig:SiP_line_crosssection_for_many_3D_wfns1}\,b.,c. and \ref{fig:SiP_line_crosssection_for_many_3D_wfns2}\,b.,c., it necessary to note that the radial distance $r$ starts from different values to ensure appropriate comparison of $\bf B_0$ orientations. With this in mind, we recognise that the wave function deformation away from spherical symmetry (depth dependence going from VII to I) leaves a relatively modest imprint on the dipole filed gradient, up to the $\bf B_0$ tilt angle of $\theta\sim55^\circ$.

\begin{figure}[htbp!]%
\centering
 \includegraphics[width=0.75\linewidth]{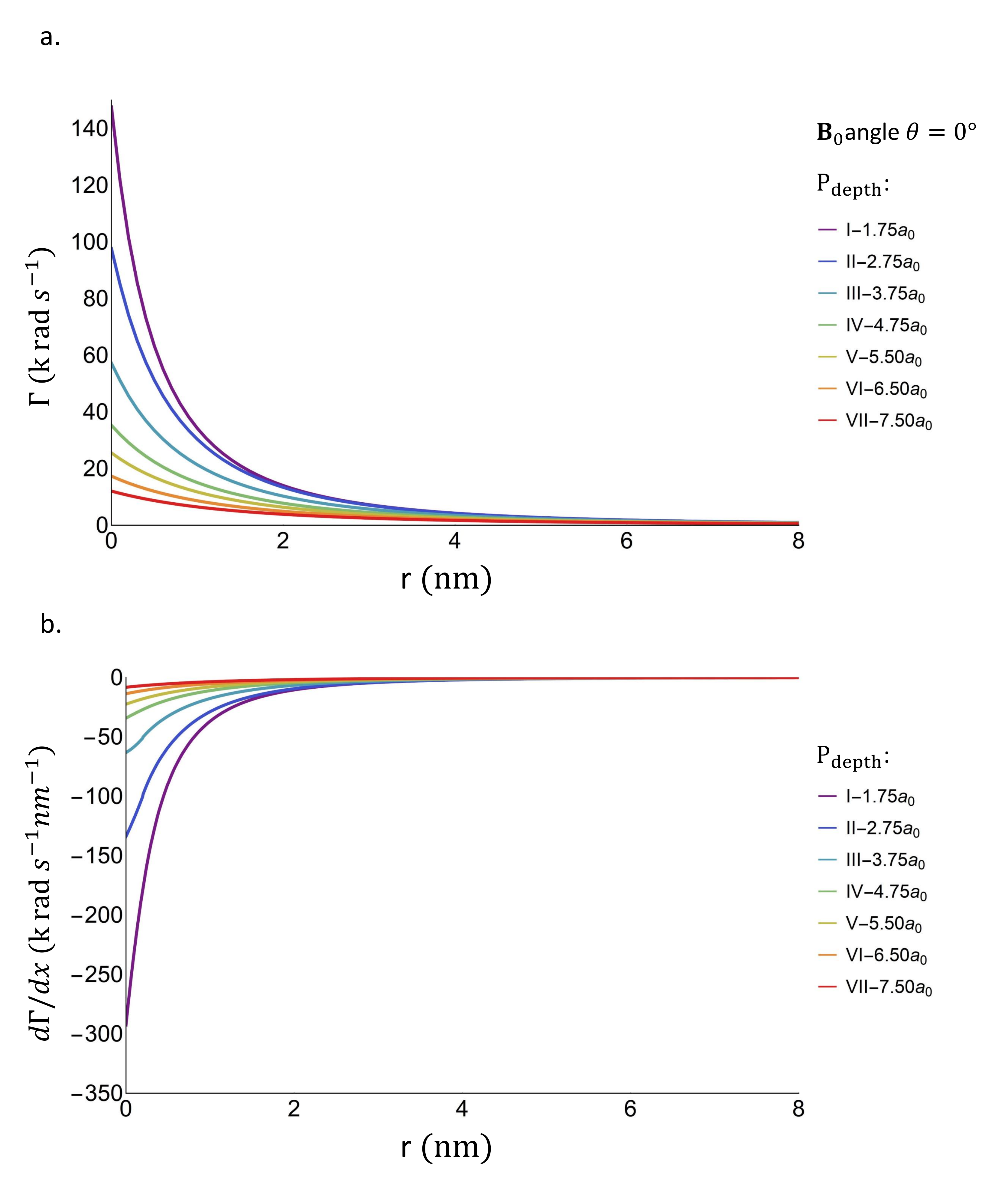}
 \vspace{-5pt}
  \caption{\textbf{The dipole-dipole field as a function of donor depth.}
  \textbf{a}) The dipolar field strength for a range of donor depths (I-VII).
  \textbf{b}) The gradient of the dipole fields. The field is calculated along the vertical vector $r$ that starts from the silicon surface (at the point directly above the donor) and is parallel to the background magnetic field ($\theta=0^\circ$).
  }\label{fig:SiP_wfns_on_one_graps_line_cross_sections}
   \vspace{15pt}
\end{figure}

Effects of the donor's depth on the dipolar field above the surface can be seen in Fig.\,\ref{fig:SiP_wfns_on_one_graps_line_cross_sections} a and b. Note, the $r$ vector was redefined to start from the surface along the perpendicular background magnetic field ($\theta=0^\circ$) thus allowing the strength and the gradient of different instances of donor depth to be appropriately contrasted on a single plot.  That provides an indicator of the extent to which spectral imaging can be efficiently done. By visual inspection, we can see that sensing becomes more challenging in the range of $r=6-8\rm\,nm$, irrespective of donor depth. We will later consider in more detail the dipole field gradient in the context of 3D imaging.

\FloatBarrier\section{Engineering the gradient field through confinement}
\label{Si_SC:Engineering_the_gradient_field_through_confinement}

\begin{figure}[htbp!]%
\centering
 \includegraphics[width=0.75\linewidth]{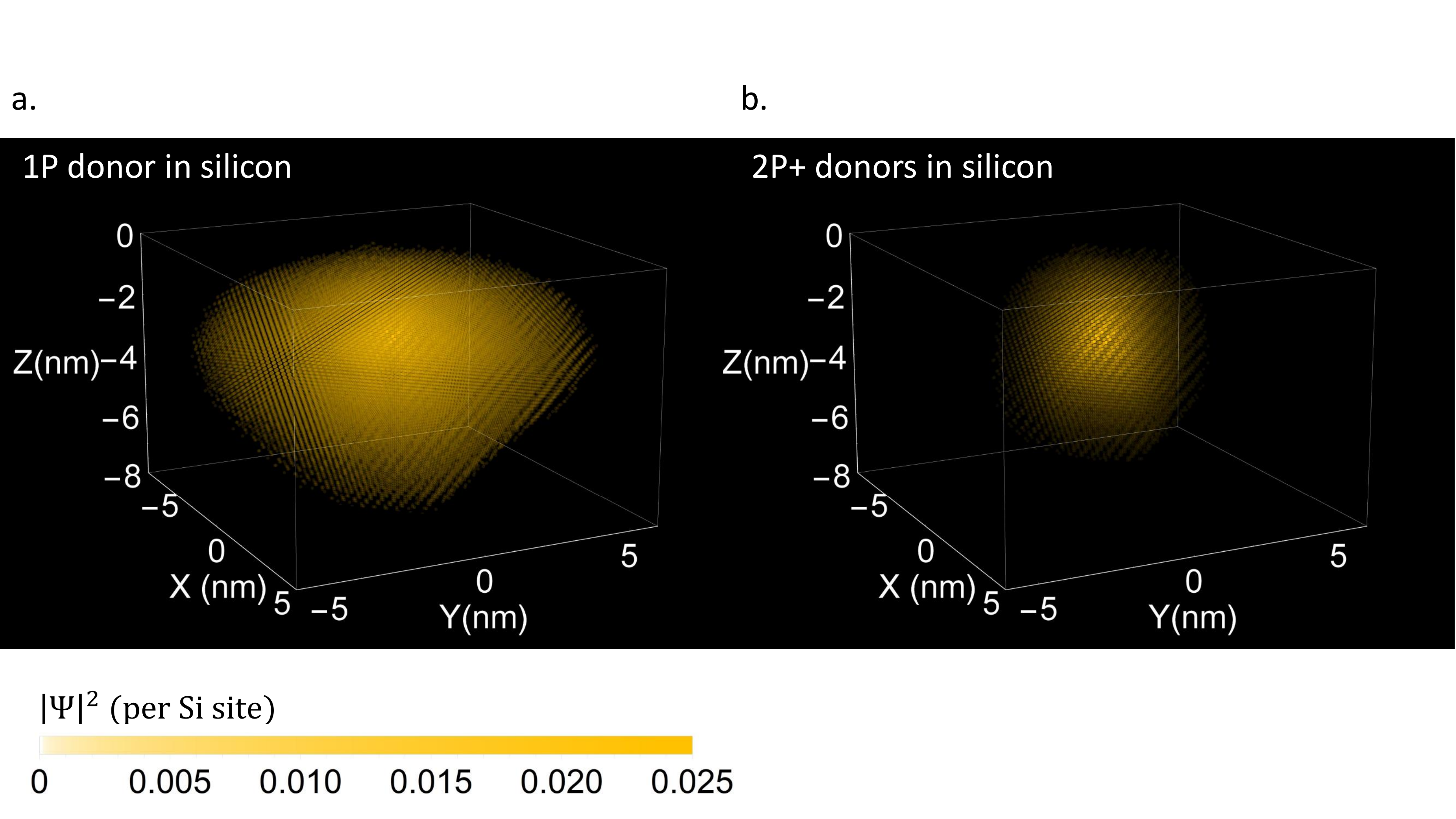}
 \vspace{-5pt}
  \caption{\textbf{The charge induced confinement of the Si:P wave function.}
  \textbf{a}) Depicted is a relatively large probability density of a single $4.5\,a_0$ deep P donor in silicon with a Bohr radius of $\sim2.5\,\rm nm$. 
  \textbf{b}) The second P donor is placed at the nearest neighbour site relative to the first in the same horizontal plane (equal depth, $\sim3.8\,\rm\AA$ lateral distance). When this system is ionised (2P+), the remaining unpaired electron is trapped by the net change of two P atoms, leading to a more confined wave function (approximately halving the Bohr radius).
  }\label{fig:SiP_Gradient_Eng_1}
   \vspace{15pt}
\end{figure}

In this section, we take a brief look at the impact of the wave function confinement on the gradient field above the surface. One way to confine the wave function is to increase the Coulomb potential of the donor's electron. This can be achieved by fabricating a device with two P donors in close proximity. We use our tight-binding model described in the main text (Methods section) to compute the electron wave function confined on 2P$^+$ system. In the 2P$^+$ system, where a pair of donor atoms is allowed to share only a single electron, the electrostatic potential effectively doubles leading to greater spatial confinement. In Fig.\,\ref{fig:SiP_Gradient_Eng_1}, we see that the 3D probability density 2P$^+$ system has approximately half the radius of 1P system.

\begin{figure}[htbp!]%
\centering
 \includegraphics[width=0.75\linewidth]{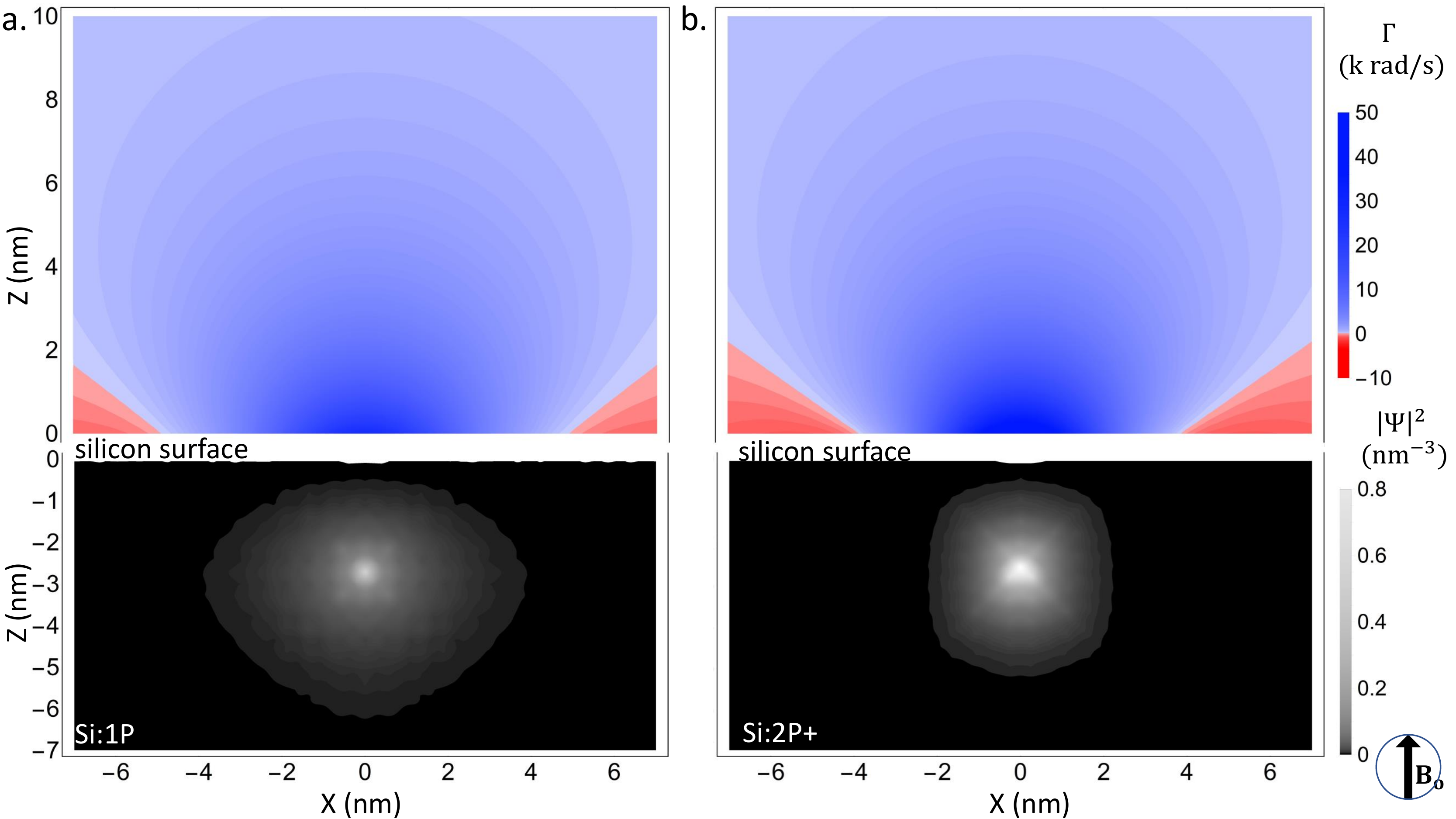}
 \vspace{-5pt}
  \caption{\textbf{The x-z cross-section of the confined wave function.}
  Compared to the 1P system (\textbf{a}) the dipole-dipole field of the 2P$^+$ system (B) has a more confined wave function (\textbf{b}, lower) with pronounced dipolar features - notice the horizontal contraction of the dipole-dipole field with an increase in magnitude (\textbf{b}, top).
  }\label{fig:SiP_Gradient_Eng_2}
   \vspace{15pt}
\end{figure}

Figure\,\ref{fig:SiP_Gradient_Eng_2} features the comparison between 1P and 2P$^+$ dipolar gradients generated above the silicon surface. The more localised wave function of the 2P$^+$ system (Fig.\,\ref{fig:SiP_Gradient_Eng_2} b. lower) produces a dipolar field with features closer to that of a point dipole (Fig.\,\ref{fig:SiP_Gradient_Eng_2} b. upper). The radial symmetry of the 2P$^+$ system is somewhat broken. Nonetheless, this effect is negligible in the context of the external dipolar field. However, it is worth noting that this is not the case for systems containing a greater number of donor atoms. Systems such as the 2P$^{++}$ have both greater physical size and a less symmetric wave function, leading to a decrease in the dipolar gradient field compared to the 2P$^+$ system.

\begin{figure}[htbp!]%
\centering
 \includegraphics[width=0.75\linewidth]{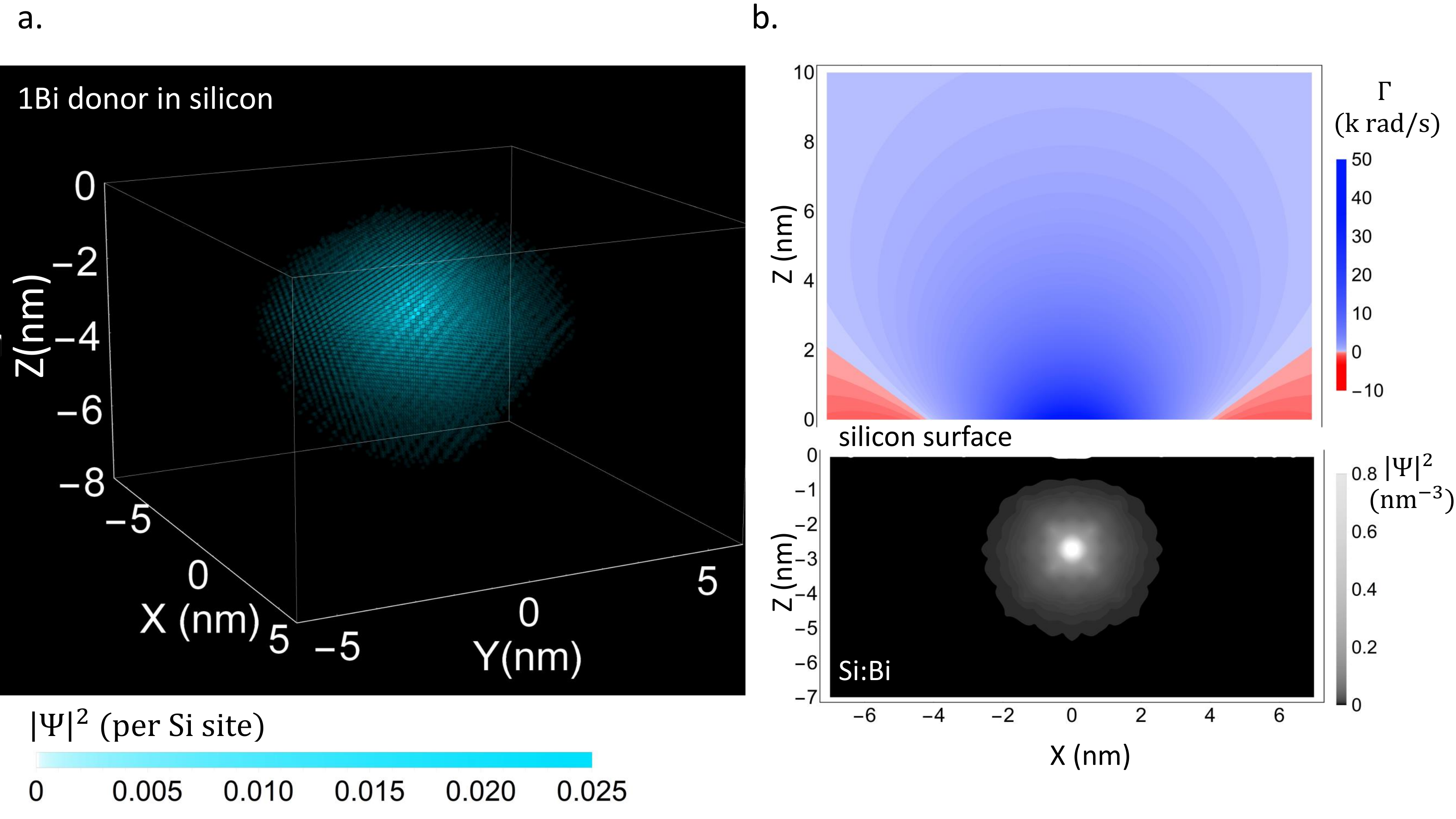}
 \vspace{-5pt}
  \caption{\textbf{Bismuth in silicon.}
  \textbf{a}) The dipolar gradient can also be increased by use of the Si:Bi donor, which wave function in silicon is more confined compared to other group-V elements. Depicted is the 3D probability density of a $4.75\,a_0$ deep Bi donor.
  \textbf{b}) The x-z cross-section of the Si:Bi system, depicting greater coupling magnitude compared to Si:P system.
  }\label{fig:SiP_Gradient_Eng_3}
   \vspace{15pt}
\end{figure}

Another method to increase the dipolar gradient is to use bismuth as a donor, depicted in Figure\,\ref{fig:SiP_Gradient_Eng_3}. Its wave function dimensions are similar to the 2P$^+$ system (Fig.\,\ref{fig:SiP_Gradient_Eng_3} a.), leaving its gradient field relatively confined (Fig.\,\ref{fig:SiP_Gradient_Eng_3} b.).

\begin{figure}[htbp!]%
\centering
 \includegraphics[width=0.75\linewidth]{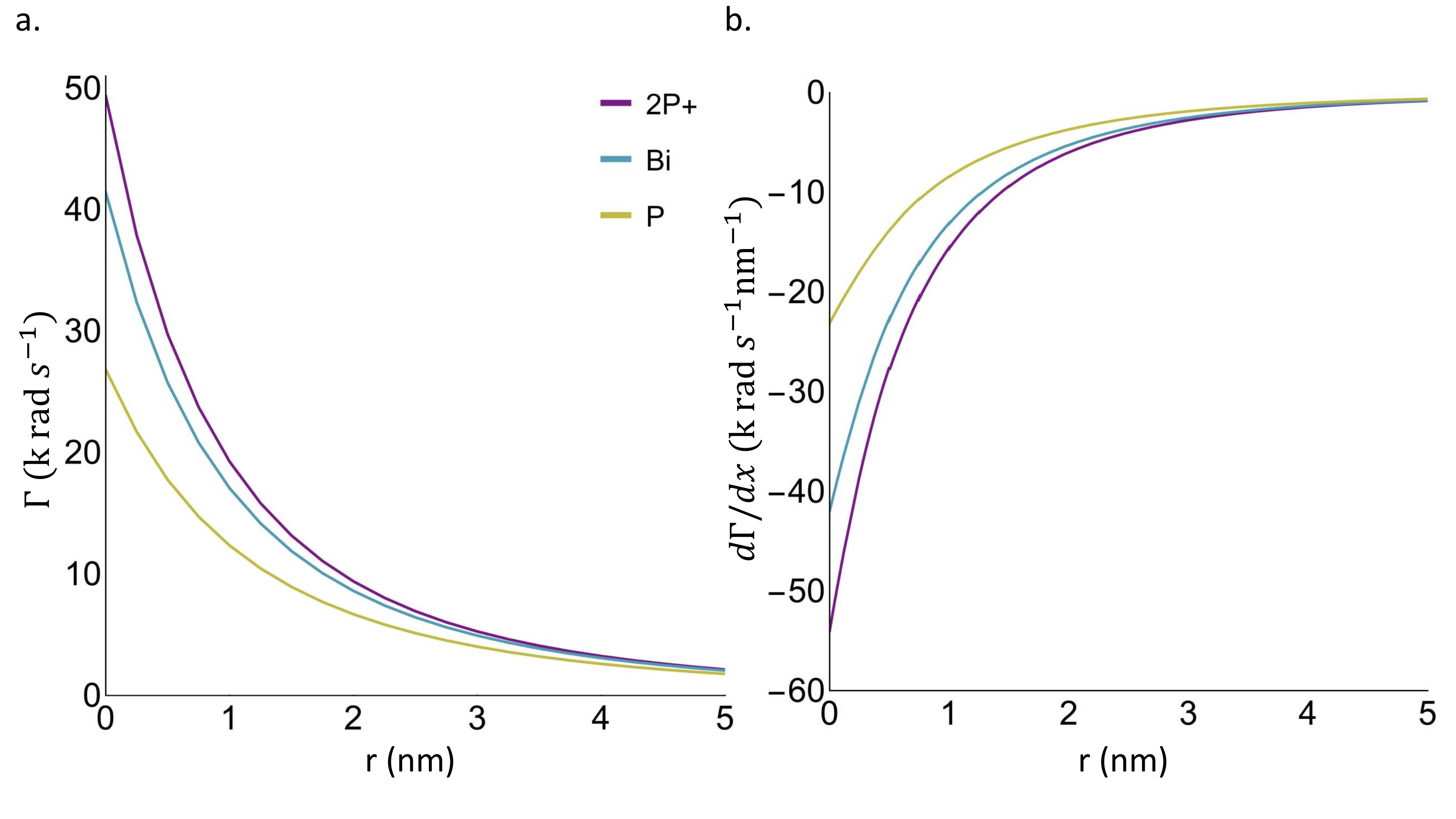}
 \vspace{-5pt}
  \caption{\textbf{Comparison of  effective dipole coupling for 1P, 2P$^+$ and 1Bi donor systems in silicon.}
  \textbf{a}) The dipole-dipole field strength of 1P, 2P+ 1Bi at the same depth of $4.75\,a_0$ in silicon. (Note: The field is calculated along the vector $r$ that starts from the silicon surface (at the point directly above the donor) and is parallel to the background magnetic field orientated vertically  $\theta=0^\circ$).
  \textbf{b}) The gradient of the associated dipole fields. The differences become less relevant with distance, indicating that the greatest improvement in resolution would be expected in the first $1-2\,\rm nm$ away from the surface.
  }\label{fig:SiP_Gradient_Eng_4}
   \vspace{15pt}
\end{figure}

Figure\,\ref{fig:SiP_Gradient_Eng_4} allows for  comparison of the dipolar fields of 1P, 2P$^+$ and 1Bi donor systems. The coupling field strengths (Fig.\,\ref{fig:SiP_Gradient_Eng_4} a.) of 2P$^+$ and Bi are similar to one another, being approximately twice of the 1P system. It is worthwhile noting that increasing the gradient through wave function confinement has a limited impact on the net sensing volume. The gradients of the three systems converge asymptotically with distance ($r\approx4-6\,\rm nm$) as the near-field effects become less relevant. The increased coupling gradient allows for decreased sampling time (for the same resolution) over the physical region of only a couple of nanometers above the surface. Therefore, the greatest advantage of the increased gradient would be in applications where fast identification of molecules is needed. In such cases, it would be possible to profile a known molecule based on its segments that fall in the near-surface detection volume.

\FloatBarrier\section{Testing spatial-frequency encoding and stability of NSS detection protocol}
\label{Si_Sc:Testing-spatial-frequency-encoding-and-stability-of-NSS-detection protocol}

\begin{figure}[htbp!]%
\centering
 \includegraphics[width=0.75\linewidth]{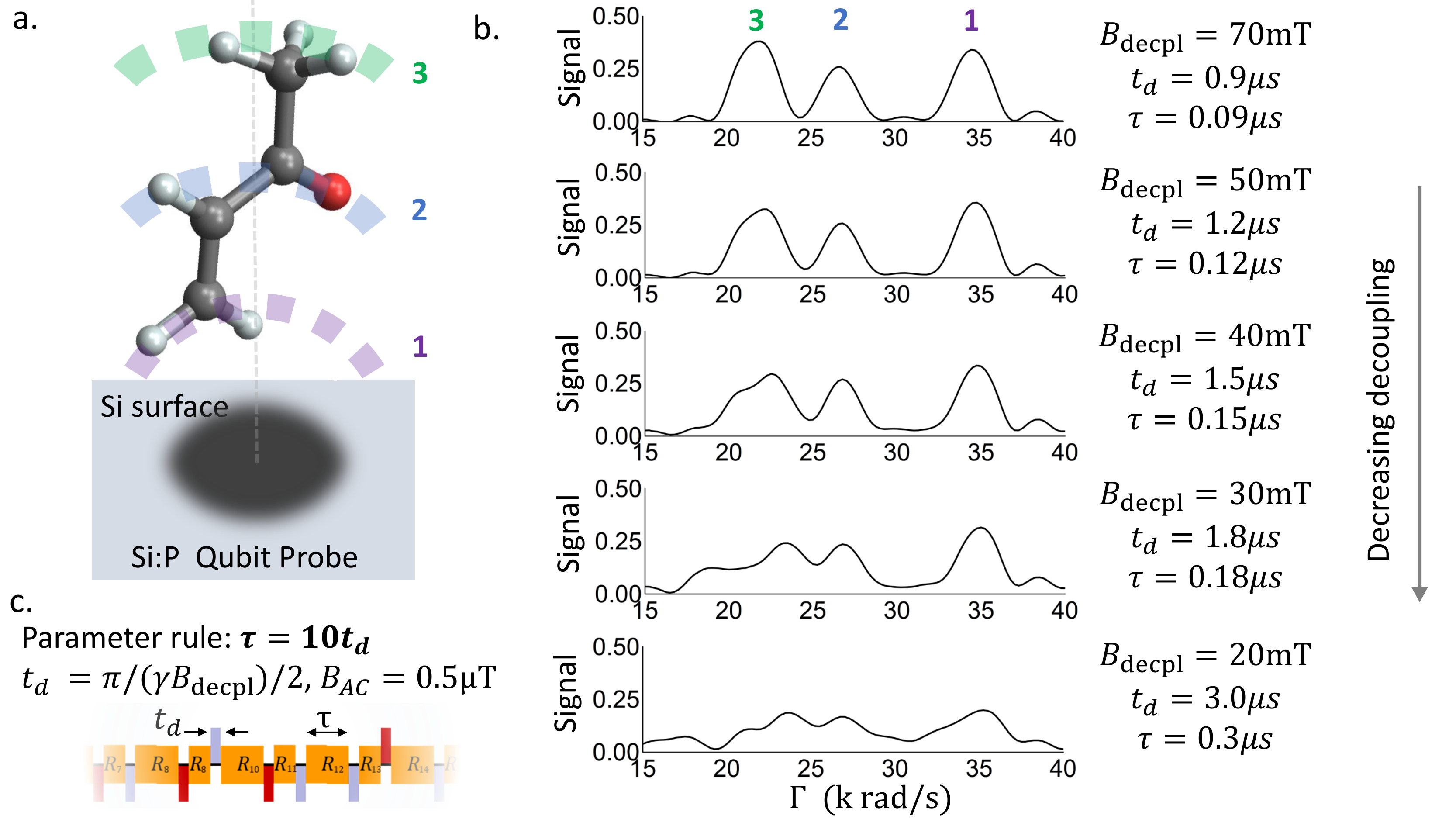}
 \vspace{-5pt}
  \caption{\textbf{Testing the  NSS protocol for spatial-frequency encoding.}
  \textbf{a}) The NSS protocol has been applied to an organic molecular segment located 2 Angstroms above the silicon surface containing a $4.75\,a_0$ $(\sim2.5\,\rm nm)$ deep Si:P qubit. 
  \textbf{b}) Using the NSS protocol to scan across the dipolar slices produces a signal that features peaks proportional to the number of target hydrogen spins located on the given slice. The peaks are distinct and well shaped (top panel), however, as the strength of the decoupling field $B_{\rm decpl}$ decreases, the dipole-dipole coupling between the target nuclei starts splitting the peaks, degrading the spatial-frequency encoding (bottom panel). 
  \textbf{c}) The particular interleaved protocol has been based on is CORY-24 sequence, with the time between decoupling pulses one order of magnitude longer than the pulse length $\tau=10t_d$, while the fine-driving field in the interleaved protocol is $B_{\rm AC}=0.5\rm\,\mu T$.
  }\label{fig:SiP_the_pulse_sequance_2}
   \vspace{15pt}
\end{figure}

Our testing methodology features the NSS protocol applied to a small cluster of tightly coupled hydrogen nuclear spins that is readily tractable. Figure\,\ref{fig:SiP_the_pulse_sequance_2} a. illustrates the set up containing a biologically inspired group of 6 hydrogens 4 carbons and one oxygen atom spatially arranged in line with their chemical bonding properties. The cluster is placed a $\sim2\,\rm\AA$ above the silicon surface featuring a P donor at a depth $4.75\,a_0$ ($\sim2.5\,\rm nm$). We target the hydrogen spins while, in order to aid visual interpretation, positioning the cluster such that each of the three illustrated dipole slices (Fig.\,\ref{fig:SiP_the_pulse_sequance_2} a. 1-3) contains a different number of nuclear spins. The scan across the slices $\Gamma$ is depicted in Fig.\,\ref{fig:SiP_the_pulse_sequance_2} b. for various decoupling parameters. In the top panel, we see three sharp peaks heights of which correspond to the number of spins present on each slice. It is convenient to read the plot from right to left, as the greatest coupling (far right) spatially corresponds to the silicon surface. The dipole-dipole decoupling sequence interleaved into the NSS protocol is CORY-24 (note, various decoupling sequence can equally be used). We observe the peaks becoming broader and eventually breaking down (bottom panel) as the decoupling field becomes weaker. The parameter rule used in this analyses (illustrated in Fig.\,\ref{fig:SiP_the_pulse_sequance_2} c.) maintains the duration of interleaved fine-driving times (orange) an order of magnitude longer than the individual decoupling pulses $\tau=10t_d$.

\begin{figure}[htbp!]%
\centering
 \includegraphics[width=0.75\linewidth]{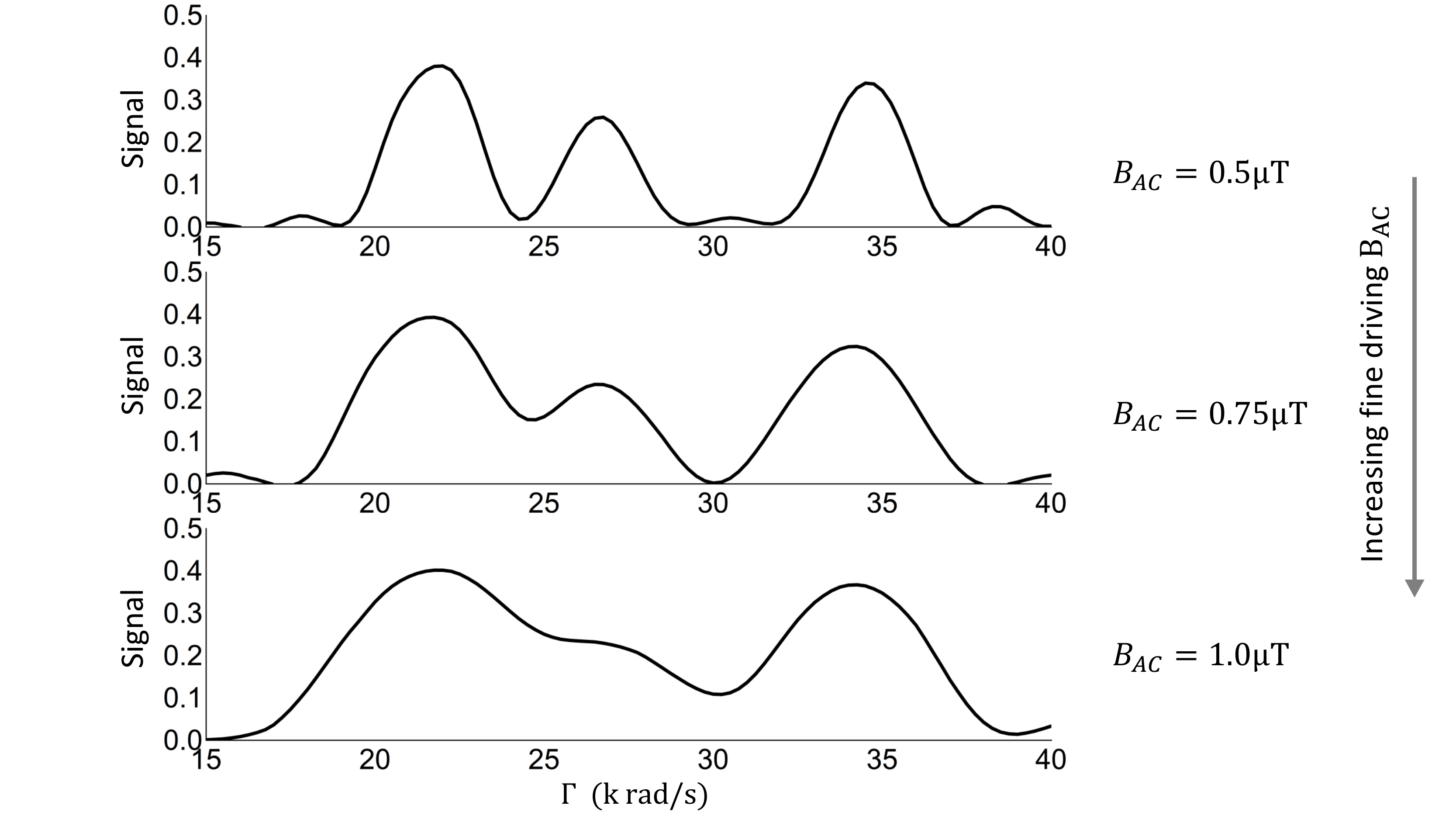}
 \vspace{-5pt}
  \caption{\textbf{Testing the NSS protocol interleaving stability.}
  The NSS protocol demonstrates typical power broadening characteristics over the fine-driving range of interest (top bottom panel), indicating stability of the NSS protocol with respect to pulse interleaving. The decoupling field is fixed to $B_{\rm decpl}=70 mT$ in order to provide well defined peaks ensuring visual clarity. Note: see Fig.\,\ref{fig:SiP_the_pulse_sequance_2} a. for the spital configuration of nucelli.
  }\label{fig:SiP_the_pulse_sequance_3}
   \vspace{15pt}
\end{figure}

We also check the stability of the quantum protocol with respect to the strength of the fine-driving field $B_{\rm AC}$. Figure\,\ref{fig:SiP_the_pulse_sequance_3} illustrates the behaviour of the probe signal as the  fine-driving field $B_{\rm AC}$ increases. To provide visual clarity, decoupling parameters for all of the panels have been set to those of the top panel in Fig.\,\ref{fig:SiP_the_pulse_sequance_2} b. We note that as $B_{\rm AC}$ increases, peaks follow the usual power broadening behaviour without splitting, indicate the stability of decoupling with respect to fine-driving interleaving. The fine-driving magnitude illustrated here, $B_{\rm AC}=0.5-1.0\rm\,\mu T$, is at the range we are interested in, corresponding to the spatial resolution on the Angstrom scale.

\FloatBarrier\section{Target coherence under NSS detection protocol}
\label{Si_Sc:Target_coherence_under_NSS_detection_protocol}

It is essential to consider the coherence time $T_{\rm \rho\_target}$ that target spins can achieve while being driven by the NSS protocol, as this is one of the critical limiting factors in the spatio-frequency encoding. We will analyse the  $T_{\rm \rho\_target}$  in the context of our detection protocol at hand, however, it is important to realise that the coherence time of the target nuclear spins is primarily dependent on the decoupling pulse sequence. Here, we work with the CORY-24 pulse sequence for the interleaving framework as it is a good compromise between a realistically complex sequence and a manually tractable one. It is worth remembering that there are more efficient decoupling protocols available, that would yield the same $T_{\rm \rho\_target}$ using smaller $B_{\rm decpl}$. Therefore, we recognise that the following analysis should not be taken as the upper limit on the $T_{\rm \rho\_target}$.

\FloatBarrier\subsubsection{Nuclear target coherence time estimation procedure}

For a particular decoupling sequence, the target coherence depends on the magnitude of $B_{\rm decpl}$ that forms the decoupling pulses and the precession time $\tau$ that separates the same pulses. To numerically analyse the NSS protocol's behaviour, we consider a highly dense cluster of hydrogen atoms, choosing a molecule of propane as an example (note, proteins on average have a lower hydrogen density and are therefore easier to decouple). Propane has a cluster of eight hydrogen spins which we interrogate by polarising one at a time and tracking its state through the NSS protocol (other spins being prepared in a mixed state).

\begin{figure}[htbp!]%
\centering
 \includegraphics[width=0.75\linewidth]{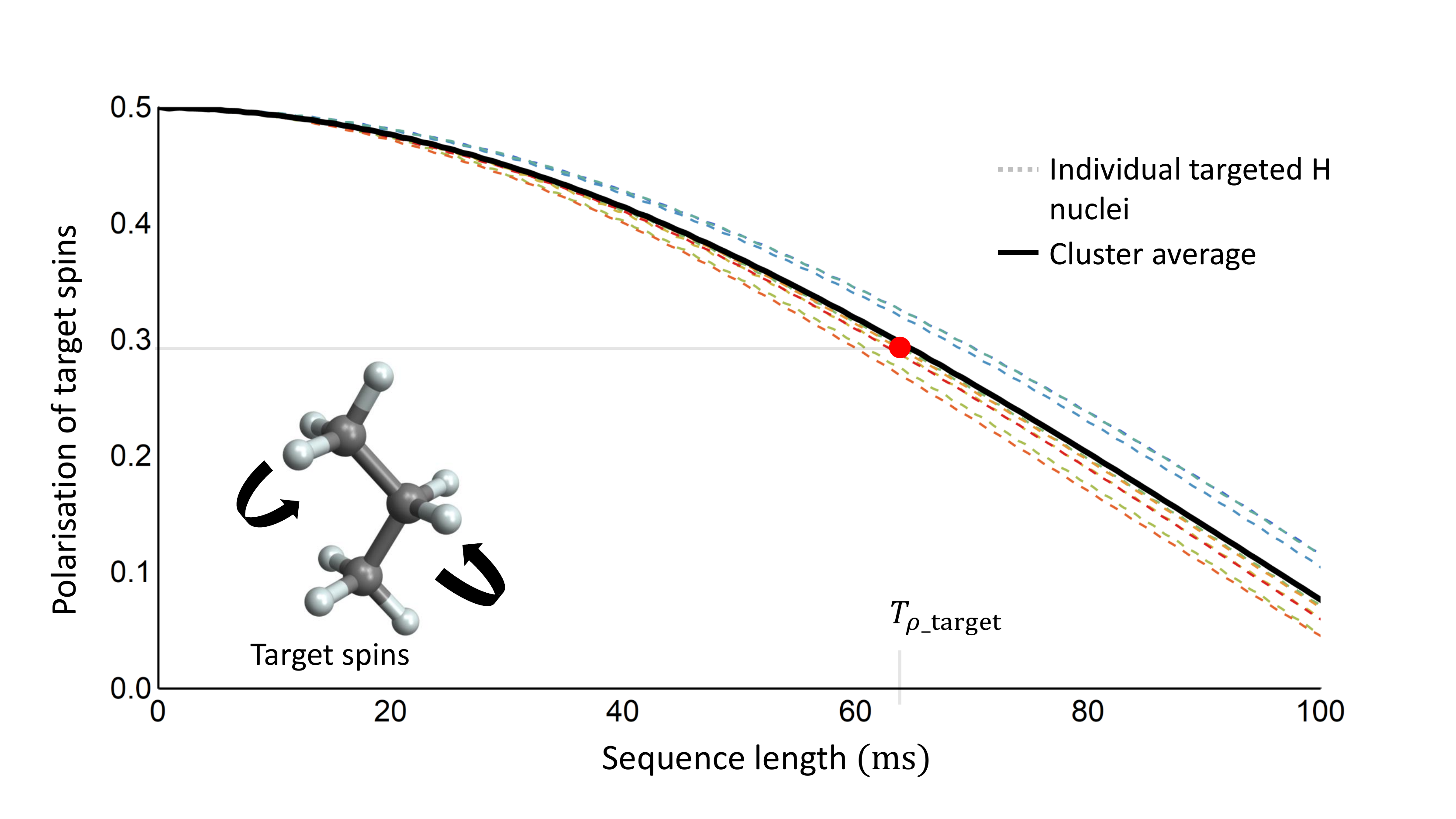}
 \vspace{-5pt}
  \caption{\textbf{The nuclear target coherence time $T_{\rm \rho\_target}$ estimation procedure.}
  The coherence time envelopes (dashed coloured lines) of each of the hydrogen nuclei in the propane molecule (chosen as a high spin density cluster) subjected to the quantum NSS detection protocol. The average envelope (solid black line) is analysed using an 1/e exponential measure to estimate $T_{\rm \rho\_target}$ (recognising that underestimation occurs for longer times where the envelop shape is not exponential, as in the case at hand). The procedure is repeated for 10-15 orientations of the nuclear cluster (lower left) to sample various instances of dipole-dipole interaction configuration.
  }\label{fig:SiP_the_pulse_sequance_4}
   \vspace{15pt}
\end{figure}

 The $T_{\rm \rho\_target}$ coherence time envelope for each of the spins is illustrated in Fig.\,\ref{fig:SiP_the_pulse_sequance_4} (coloured, dashed lines). Taking the average envelope (solid black line), we use a simple exponential measure to estimate $T_{\rm \rho\_target}$, noting that this will lead to an underestimate of longer coherence times where envelopes do not have an the exponential shape, as illustrated in Fig.\,\ref{fig:SiP_the_pulse_sequance_4}. Since the dipole-dipole interaction is orientation dependent, this procedure is repeated for many different orientations of the propane molecule, allowing us to average over various instances of the coupling between individual target nuclei. This approach enables us to estimate the coherence time parameter $T_{\rm \rho\_target}$ for the nuclei targeted by the NSS detection protocol in relatively large individual proteins which comprise of numerous nuclear clusters similar in dimension and structure to the one considered here. Note, the estimate of $T_{\rm \rho\_target}$ can be improved by the use of an ensemble of varied clusters with hydrogen densities closer to those found in the targeted protein, however this is beyond the present scope.

\FloatBarrier\subsubsection{Estimates of target coherence time}

\begin{figure}[htbp!]%
\centering
 \includegraphics[width=0.75\linewidth]{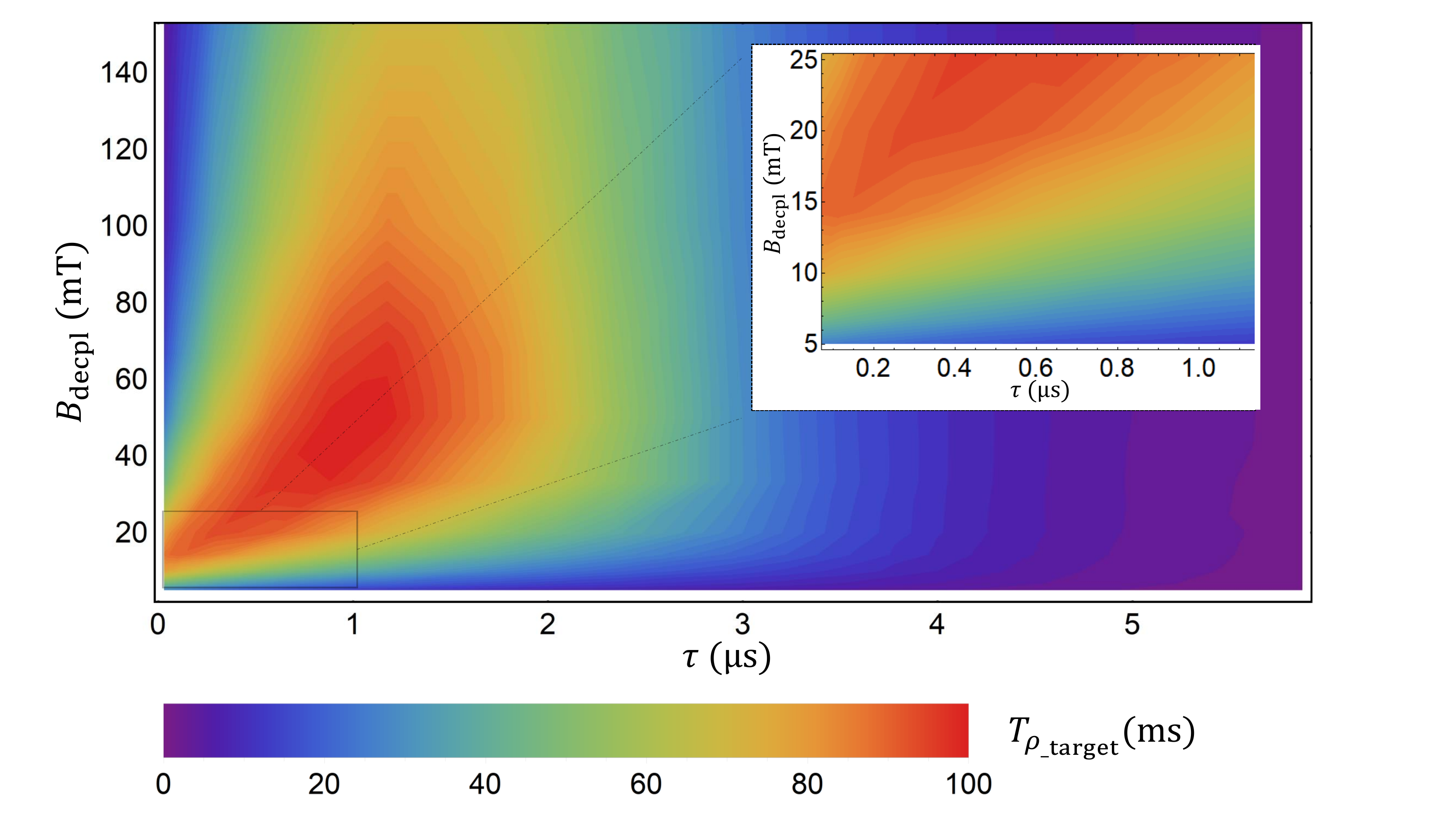}
 \vspace{-5pt}
  \caption{\textbf{The target coherence time $T_{\rm \rho\_target}$ as a function of the decoupling field $B_{\rm decpl}$ and time between the decoupling pulses $\tau$.}
  The cluster of spins used for the numerical analysis was a molecule of propane (procedure described in main text). For consistency, the NSS detection protocol utilises the CORY-24 decoupling sequence. The particular maximal coherence time depends on the particular decoupling sequence, and can be improved by the use of more efficient decoupling sequences.
  }\label{fig:SiP_the_pulse_sequance_5}
   \vspace{15pt}
\end{figure}

The results are depicted in Fig.\,\ref{fig:SiP_the_pulse_sequance_5}, where the coherence time of the target spins $T_{\rm \rho\_target}$ is estimated as a function of the decoupling field strength $B_{\rm decpl}$ and the precession time $\tau$. We note that the maximal coherence time $T_{\rm \rho\_target}$ is around $100\rm\,\mu s$. We will touch upon the imaging consequences of this later. Even when a suboptimal decoupling sequence, such as the CORY-24, is used, the coherence times reached with the decoupling fields in the range of $5-10\rm\,mT$ are suitable for 3D imaging up to $r\sim3-4\,\rm nm$ in radial distance, while $20\rm\,mT$ would push $T_{\rm \rho\_target}$ closer to its upper limit.

\FloatBarrier\section{Silicon surface and the optimal qubit depth}
\label{Silicon_surface_and_the_optimal_qubit_depth}

\begin{figure}[htbp!]%
\centering
 \includegraphics[width=0.75\linewidth]{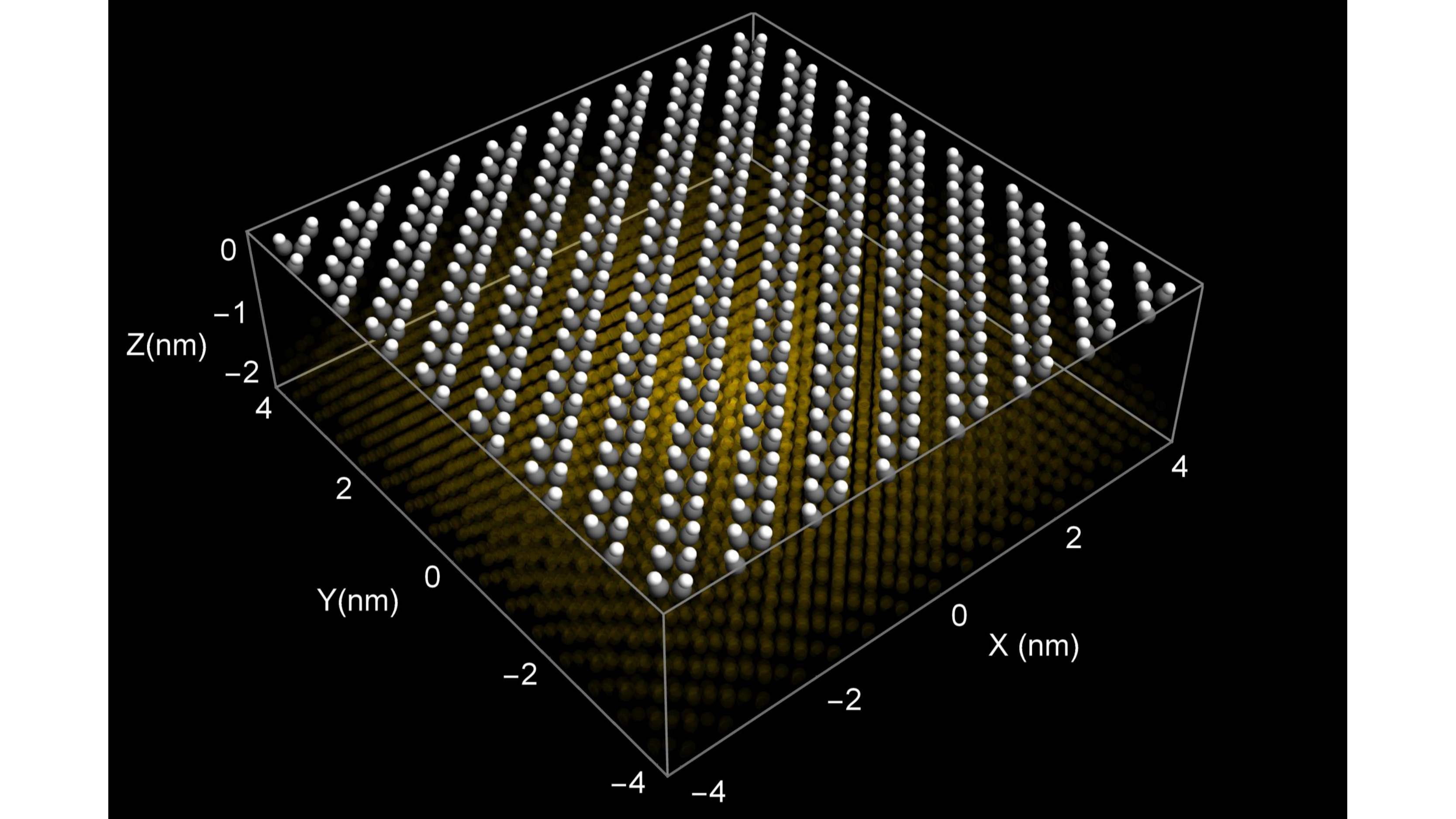}
 \vspace{-5pt}
  \caption{\textbf{The silicon (100) surface.} The externally facing layer of Si dimer rows (grey) terminated with hydrogen (white). As each Si atom is bonded to a single H atom, the distances between the neighbouring hydrogen atoms are considerably larger compared to the average nearest neighbour separation found in carbon-based organic molecules. The presence of shallow phosphorus or another group-V donor election wave function (yellow) introduces contact hyperfine coupling between its wave function and the hydrogen nuclei.
  }\label{fig:SiP_3D_H_on_top_Si_layer_1}
   \vspace{5pt}
\end{figure}

The surface of silicon is routinely terminated with oxide or hydrogen. Oxygen termination is free of nuclear spins, it is stable and well understood, however, it typically has a thickness on the nanometre scale and is prone to accumulation of charge traps. Therefore, oxide termination is suboptimal for our purposes, as it limits the physical extent of the imaging protocol by inducing substantial separation between the donor and the target molecules.
In hydrogen termination only a single, well-formed monolayer of hydrogen atoms passivates the Si surface, Fig.\,\ref{fig:SiP_3D_H_on_top_Si_layer_1}.

\FloatBarrier\subsection{Hyperfine interaction between donor electron and surface hydrogen}

Hydrogen termination indeed takes up minimal vertical space, nonetheless it presents new questions. Hydrogen atoms have a nuclear spin, forming a uniform blanket of highly coupled nuclear dipoles on the silicon surface. In principle, this should pose no difficulty to the decoupling sequence, quite the contrary. Being in the same plane removes some dipole-dipole terms that would otherwise require suppression. Hydrogen atoms bonded to silicon are further apart than in carbon-based organic molecules, again making decoupling easier.

However, we have to consider that the hydrogen surface termination is in contact with phosphorus wave function. Therefore, apart from being coupled to its neighbours via dipole-dipole interaction, each H nuclei is also subjected to hyperfine interaction. Unlike dipole-dipole, the hyperfine coupling Hamiltonian for the $i^{\rm th}$ H spin has the symmetric form $A_i{\bf \sigma}_{\rm probe}\cdot {\bf \sigma}_{i}$. Since the evolution of the probe spin throughout the NSS protocol is effectively passive free precession and the energy discrepancy between the probe's electron and target nuclear spin being large, it follows that $\sigma_x\sigma_x$ and $\sigma_y\sigma_y$ terms can be neglected, leaving the hydrogen surface spins with a dominant $\sigma_z\sigma_z$ coupling. In the context of our detection protocol, this represents an effective detuning of $\pm A_i$ away from the background Zeeman level $\gamma_t B_0$. Depending on the magnitude of this effective detuning, the decoupling pulses may become ineffective in removing the dipolar terms, in turn turning the surface into a decoherence source. Furthermore, the donor's probability distribution has a high degree of internal variation, exposing the surface H nuclei to the complex distribution of hyperfine coupling.

\FloatBarrier\subsubsection{Method for hyperfine interaction estimation}

\begin{figure}[htbp!]%
\centering
 \includegraphics[width=0.75\linewidth]{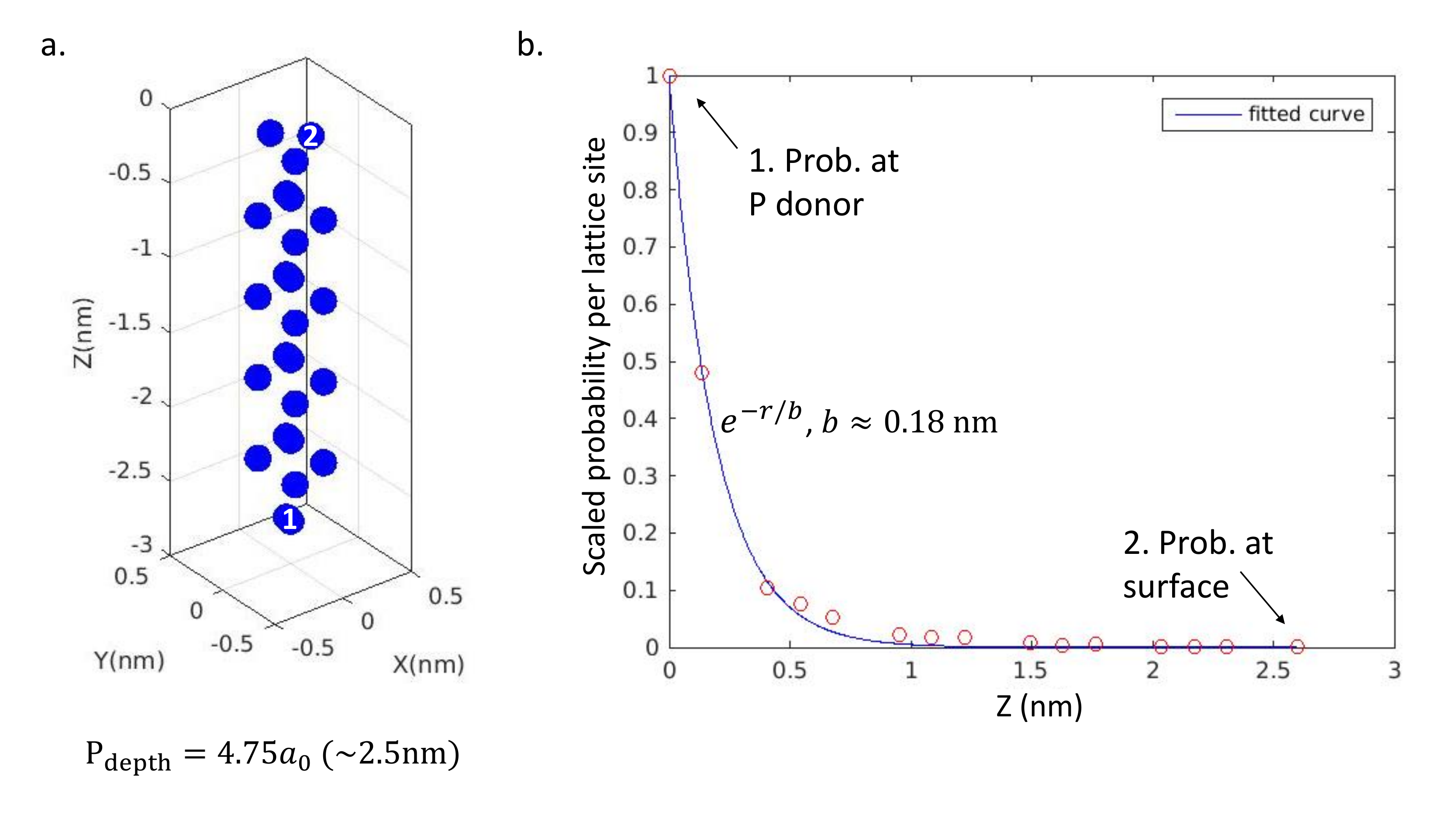}
 \vspace{-5pt}
  \caption{\textbf{The estimation procedure of the surface hydrogen hyperfine coupling.}
  \textbf{a}) A column of Si atoms inside the crystal leading upward from the donor site (P donor depth $4.75\,a_0$ $\sim2.5\,\rm nm$)  towards the surface.
  \textbf{b}) The donor's probability distribution at the sites located in the column. 
  \textbf{c}) Exponential decay of the wave function in the direction towards the surface, used to estimate the maximal presence of the wave function in the hydrogen layer.
  }\label{fig:SiP_3D_H_on_top_Si_layer_2}
   \vspace{15pt}
\end{figure}

Our set up is based on a single group-V donor near an infinite hydrogen terminated surface. Estimates of H-P hyperfine coupling in this particular configuration are scarce in both experimental and theoretical literature. An appropriate theoretical approach would involve solving the hyperfine coupling for a small cluster of surface hydrogen atoms via DFT calculations and passing the findings into the tight binding model in order to examine a wider surface area, however, this is beyond the present scope. We proceed to estimate an upper bound on the hyperfine coupling by extrapolating the phosphorus probability distribution into the surface hydrogen layer. This is achieved by considering a vertical column of Si atoms between a donor at the particular depth and the surface (Fig.\,\ref{fig:SiP_3D_H_on_top_Si_layer_2} a.), calculating the net probability density at each Si site based on a tight binding model of the wave function (Fig.\,\ref{fig:SiP_3D_H_on_top_Si_layer_2} b.) and estimating the probability density's exponential decay (Fig.\,\ref{fig:SiP_3D_H_on_top_Si_layer_2} c.). The decay envelope is used to estimate the probability present at the H atoms above the last Si layer.

\FloatBarrier\subsubsection{Surface hydrogen hyperfine interaction estimates}

\begin{figure}[hp!]%
\centering
 \includegraphics[width=0.75\linewidth]{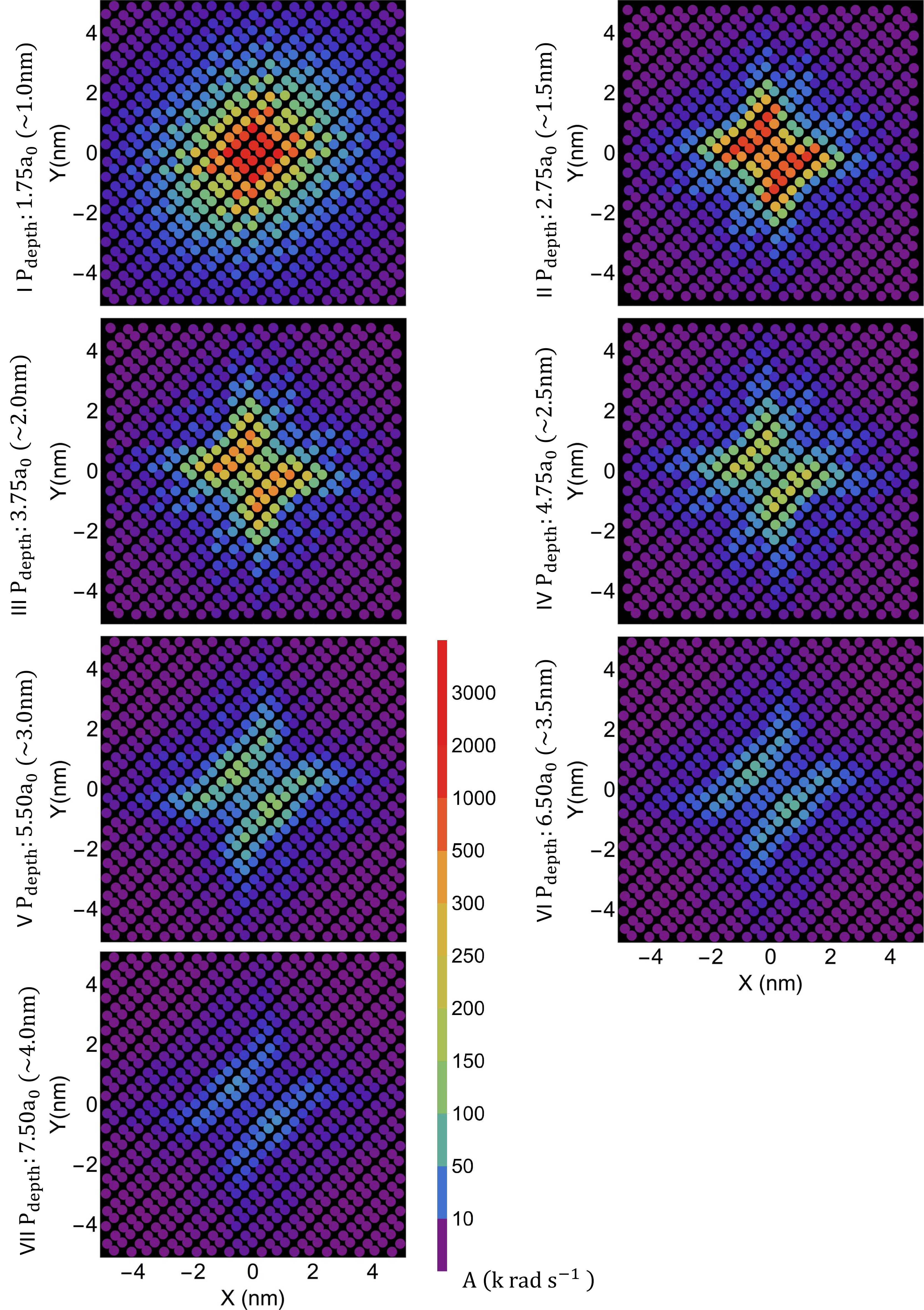}
 \vspace{-5pt}
  \caption{\textbf{Upper estimates of the hydrogen hyperfine interaction above the phosphorus donors with depths in the range of $\sim1-4\,\rm nm$ (I-VII).} Calculations assume that the hydrogen layer is electrostatically indistinguishable from the surrounding Si environment, the donor wave function is thus free to extend into it as if the H layer was part of the silicon lattice.
  }\label{fig:SiP_3D_H_on_top_Si_layer_SurfaceA_maps}
   \vspace{15pt}
\end{figure}

Once the Si:P probability density at the surface termination has been estimated, the hyperfine interaction can be calculated based on the following relationship \cite{Lloyd_book_2006}:
\begin{eqnarray} \label{eq:A_from_prob}
A_i=\frac{8\pi}{3\hbar}\gamma_{\rm probe}\gamma_{\rm t}|\Psi({\bf r}_i)|^2,
\end{eqnarray}
where $\Psi$ is the estimate of the donor's wave function at the location ${\bf r}_i$ of the $i^{\rm th}$ surface hydrogen atom. Figure\,\ref{fig:SiP_3D_H_on_top_Si_layer_SurfaceA_maps} depicts the 2D surface hyperfine distribution for various depths of the phosphorus donor. The lateral drop off is fast - on average creating a $1-2\,\rm nm$ wide pocket of H spins that are experiencing the bulk of the interaction. Our decoupling protocols are designed to cope with the interaction strengths found in neighbouring hydrogen atoms in organic molecules, $\sim250\,\rm k\,rad\,s^{-1}$, corresponding to the green shade in Fig.\,\ref{fig:SiP_3D_H_on_top_Si_layer_SurfaceA_maps}.

\begin{figure}[htbp!]%
\centering
 \includegraphics[width=0.75\linewidth]{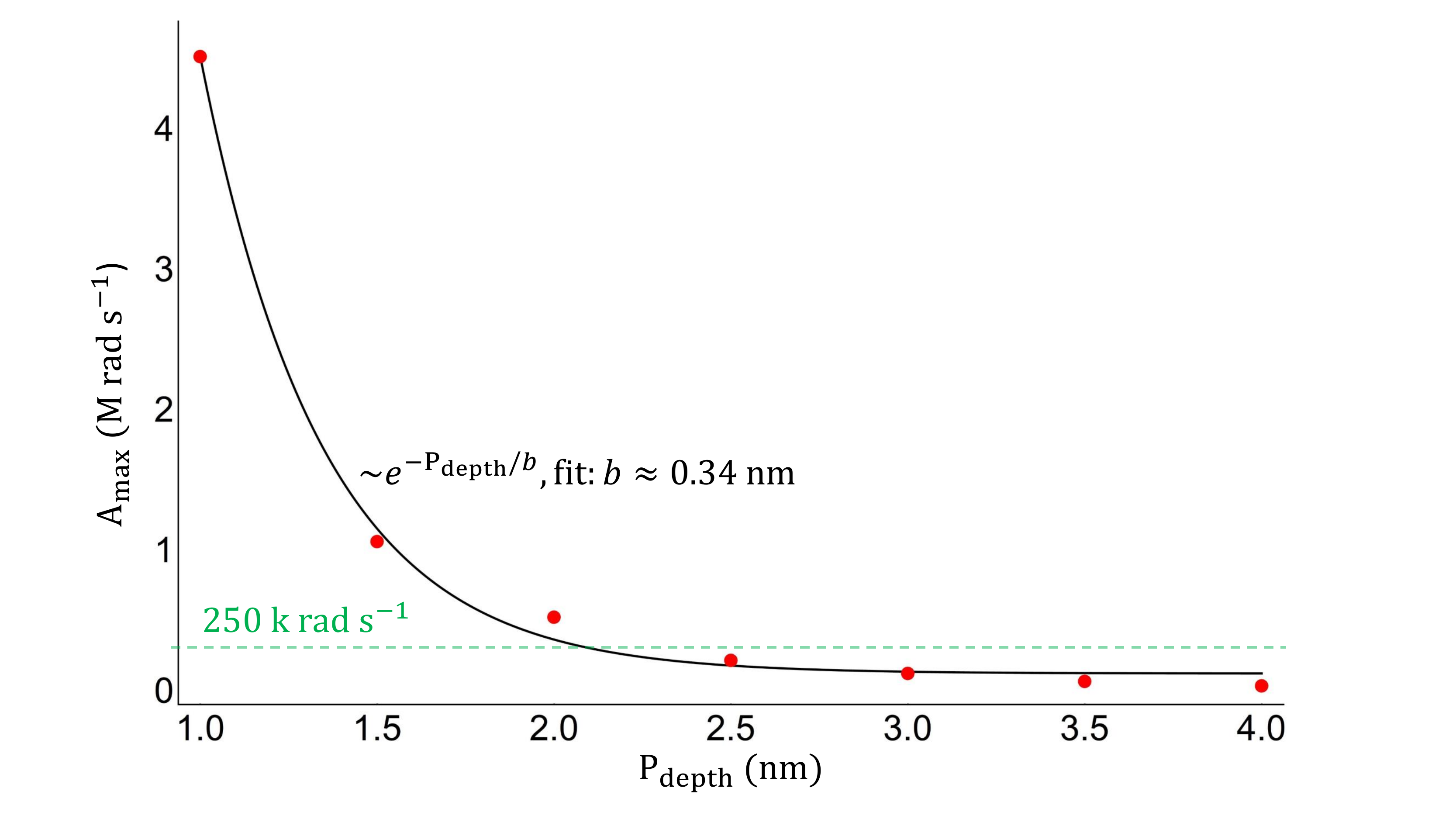}
 \vspace{-5pt}
  \caption{\textbf{The maximum hyperfine coupling experienced by the surface hydrogen as a function of donor depth.}
  The magnitude guideline of $\sim250\,\rm k\,rad\,s^{-1}$ (green dashed) represents the coupling strengths associated with the dipole-dipole interaction between neighbouring H atoms in organic molecules.
  }\label{fig:SiP_3D_H_on_top_Si_layer_3}
   \vspace{15pt}
\end{figure}

Fig.\,\ref{fig:SiP_3D_H_on_top_Si_layer_3} depicts the maxima $A_{\rm max}$ of each distribution as a function of donor depth. It is worth noting that $A_{\rm max}$ reduces to the level of dipole-dipole coupling between the molecular H nuclei for donor depths below $\sim2.5\,\rm nm$, corresponding to the cases IV to VII in Fig.\,\ref{fig:SiP_3D_H_on_top_Si_layer_SurfaceA_maps}.

\FloatBarrier\subsection{Finding the optimal donor depth}

These results raise the question about the optimal donor depth. We would like the donor to be as close to the surface as possible,  in order to ensure a strong dipolar gradient. However, the surface hydrogen may cause problems if its coherence times become similar or smaller than the maximal running time of the NSS detection protocol, which we assume is $\sim100\,\rm ms$ based on maximal $T_{\rm \rho\_target}^{\rm max}$ (ignoring limits of the probe's coherence time for the moment). Fig.\,\ref{fig:SiP_3D_H_on_top_Si_layer_3} indicates that doubling of the decoupling pulse power may be able to overcome the hyperfine decoupling, however, this alone does not provide any information about the hydrogen nuclear spin coherence times.

\begin{figure}[htbp!]%
\centering
 \includegraphics[width=0.75\linewidth]{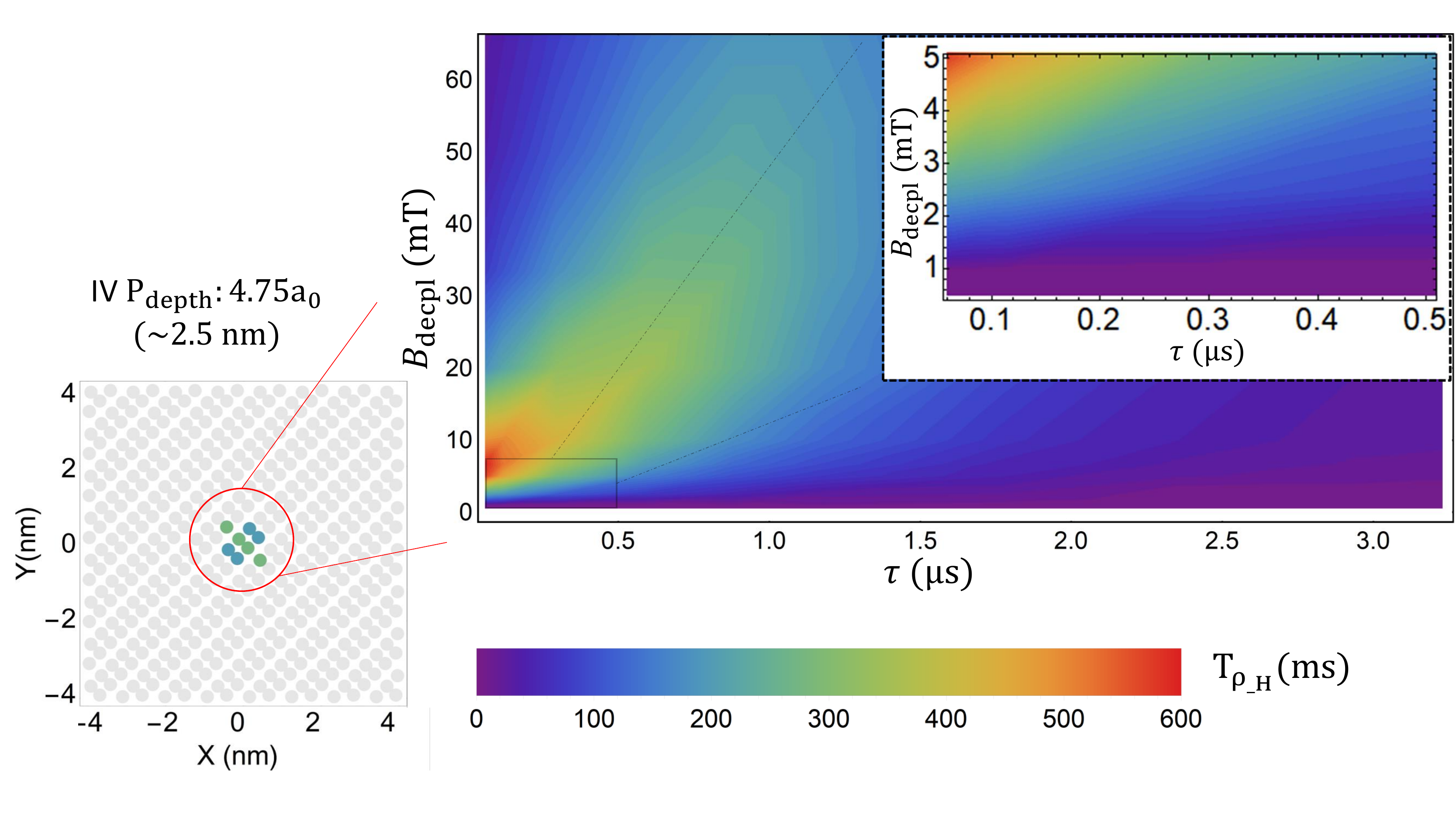}
 \vspace{-5pt}
  \caption{\textbf{The $T_{\rm \rho\_H}$ profile of the central cluster of surface hydrogen spins subjected to the NSS detection protocol.} 
  This cluster of spins exhibits the highest hyperfine coupling (bottom left) to a phosphorus donor at depth $4.75\,a_0$ ($\sim 2.5\,\rm nm$).
  }\label{fig:SiP_the_pulse_sequance_SurfaceH_1}
   \vspace{15pt}
\end{figure}

\FloatBarrier\subsubsection{Impact of hyperfine interaction on surface H coherence}

To address this question in detail, we apply the NSS detection protocol to a cluster of surface hydrogen spins experiencing the strongest hyperfine coupling. The same $T_{\rm \rho}$ procedure discussed earlier  (excluding averaging for various spatial orientations) is used to estimate H coherence, in the presence of both the full hyperfine terms and the mutual dipole-dipole coupling.

Figure\,\ref{fig:SiP_the_pulse_sequance_SurfaceH_1} depicts the cluster of eight (size restricted by numerical resources) hydrogen surface spins, for which the hyperfine coupling is the greatest. As before, the graph depicts the coherence time of the hydrogen spins $T_{\rm \rho\_H}$ as the function of $B_{\rm decpl}$ and $\tau$. For the instance where the phosphorus donor depth is $4.75\,a_0$, the maximal coherence time $T_{\rm \rho\_H}^{\rm max}$ is around $600\,\rm ms$, significantly higher compared to the molecular hydrogen targets. Also, the pattern of the $T_{\rm \rho\_H}$ graph in Fig.\,\ref{fig:SiP_the_pulse_sequance_SurfaceH_1} is different compared to $T_{\rm \rho\_target}$ in Fig.\,\ref{fig:SiP_the_pulse_sequance_5}. The presence of hyperfine detuning ZZ terms causes the maximum of $T_{\rm \rho\_H}$ to shift towards the left, in the direction of reduced $\tau$, indicating that the decoupling sequence is managing the hyperfine coupling by brute force averaging of the surface spin states. The effect is analogous to motional narrowing induced by fast, on-resonance Rabi driving, in contrast to exploiting the off-resonance magic axis driving.

\begin{figure}[htbp!]%
\centering
 \includegraphics[width=0.75\linewidth]{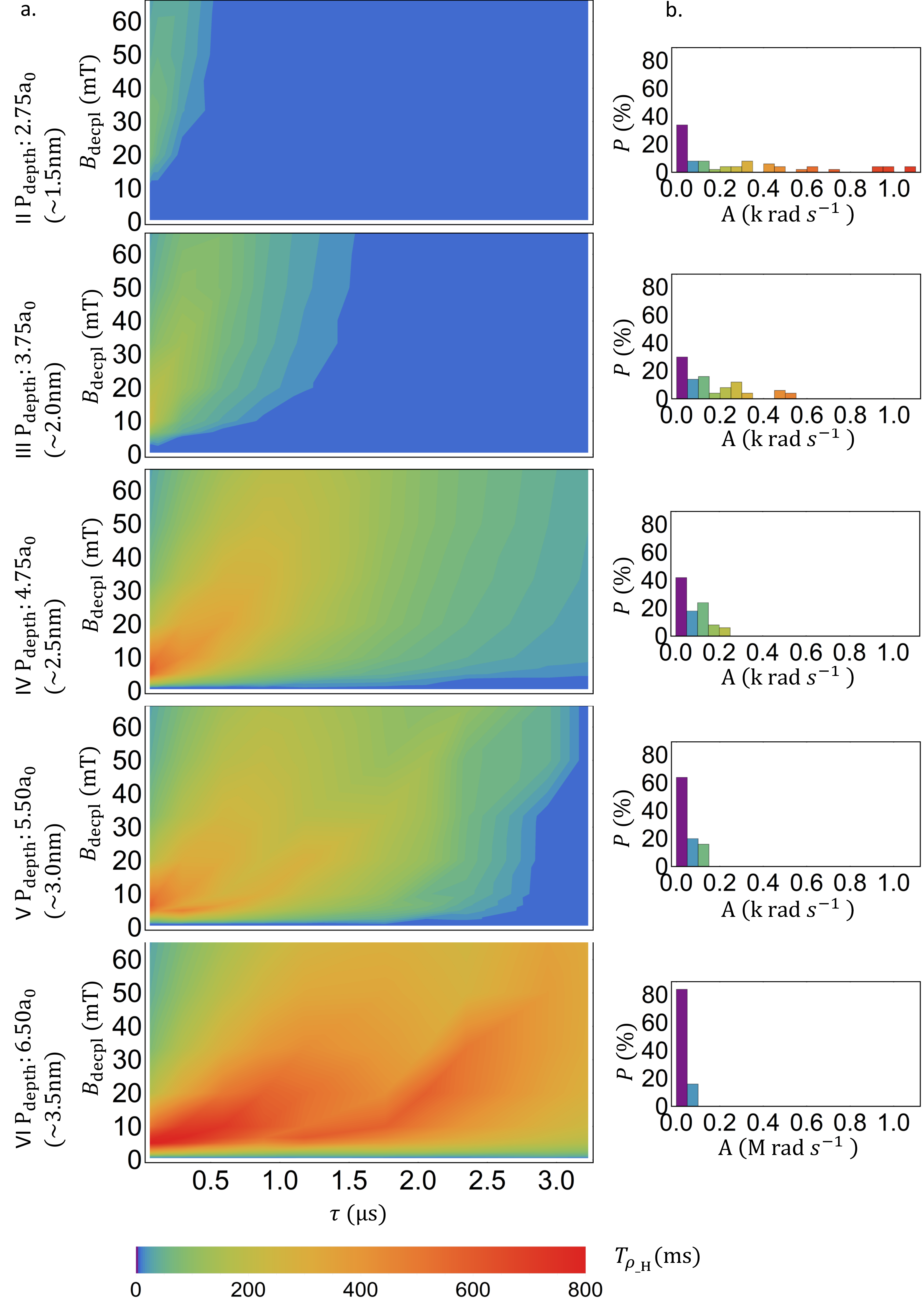}
 \vspace{-5pt}
  \caption{\textbf{The analysis of surface hydrogen coherence for a range of phosphorus donor depths (instances II to VI).}
  \textbf{a}) $T_{\rm \rho\_H}$ profiles obtained by application of the NSS detection protocol as a function of $B_{\rm decpl}$ and $\tau$. 
  \textbf{b}) The distribution of hyperfine coupling in the $2\rm\,nm$ by $2\rm\,nm$ surface area above the donor.
  }\label{fig:SiP_the_pulse_sequance_SurfaceH_2}
   \vspace{15pt}
\end{figure}

\FloatBarrier\subsection{Conclusions on donor depth}

Further conclusions are drawn by analysing a series of donor depths. Figure\,\ref{fig:SiP_the_pulse_sequance_SurfaceH_2} a. depicts a range of instances from $1.5\,\rm nm$ to $3.5\,\rm nm$, while Fig.\,\ref{fig:SiP_the_pulse_sequance_SurfaceH_2} b. depicts the distribution of hyperfine surface coupling. We note that in the first panel (case II), the effect of hyperfine coupling is severe, causing the coherence times of the H surface spins to be poor in the presence of the NSS detection protocol. However, these effects rapidly decrease with the donor distance. In conclusion, we note that from the perspective of hydrogen termination spin coherence, the optimal Si:P donor depth should be $2\,\rm nm$ or greater. In particular, instances III \& IV provide an optimal compromise between the dipole gradient strength above the surface and the impact of hydrogen surface termination.

\FloatBarrier\section{Optimal detection segment length}
\label{Si_Sc:Optimal_detection_segment_length}

\begin{figure}[htbp!]%
\centering
 \includegraphics[width=0.75\linewidth]{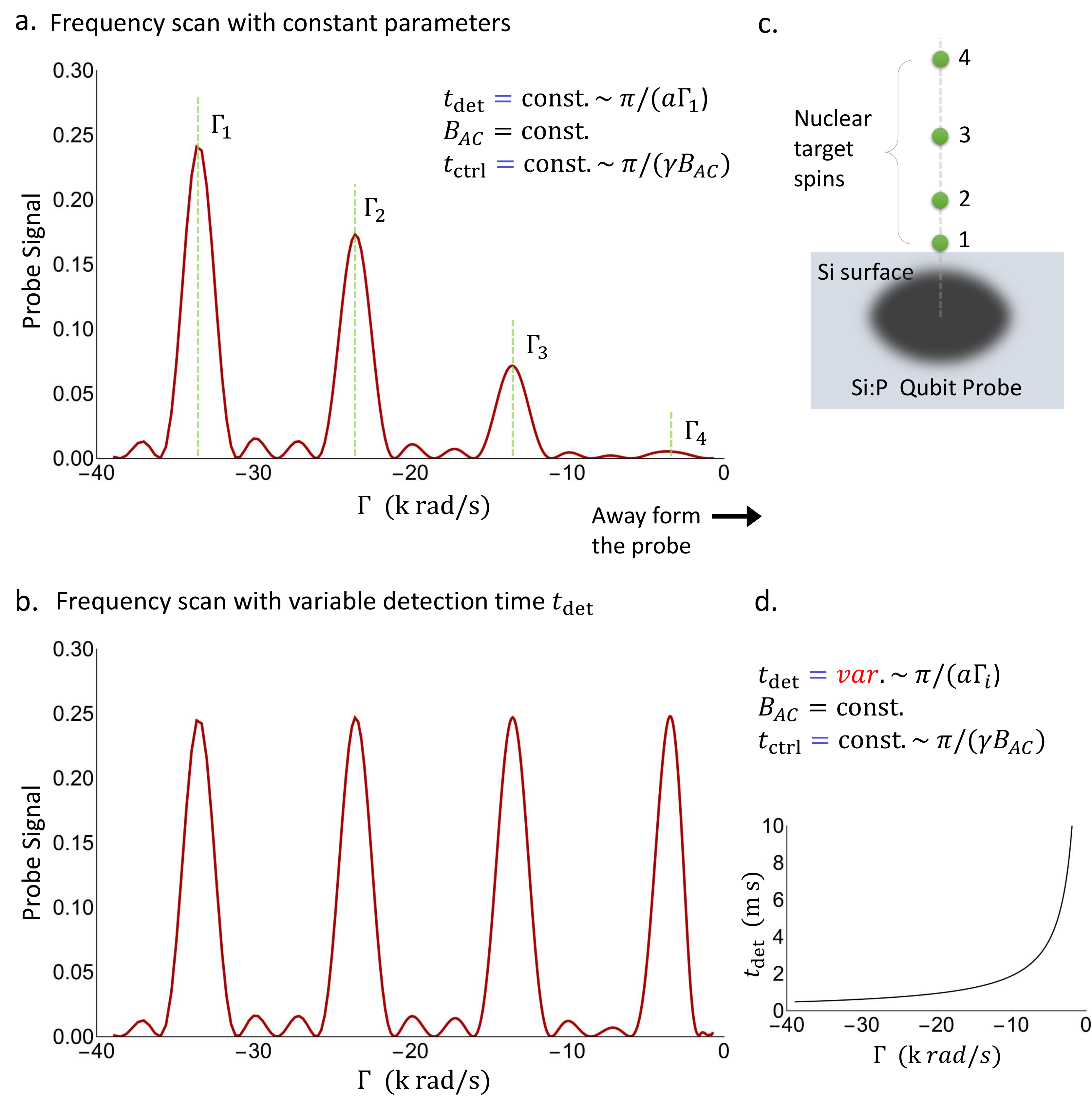}
 \vspace{-5pt}
  \caption{\textbf{The NSS detection protocol frequency sweep.}
  \textbf{a}) Holding the sampling parameters $(B_{\rm AC},t_{\rm det},t_{\rm ctrl})$ constant during the sweep leads to a reduction of the signal ($S^*_{\rm net}$) contrast as the detection segment time $t_{\rm det}$ becomes inadequate for recording of the coupling strength of the slice addressed by $B_{\rm AC}$. 
  \textbf{b}) Varying $t_{\rm det}$ with the slice frequency as per Eq. \ref{eq:Signal_from_sigle_nucleus}, leads to a constant signal response. 
  \textbf{c}) A depiction of hydrogen nuclear spins above the probe's wave function corresponding to a. and b., well separated so as to allow for individual signal response to be observed. Note, the signal has been plotted against the negative frequency for visual convenience, in order to keep the ordering of peaks (left to right) consistent with increasing distance between the target and the probe.
  }\label{fig:SiP_k2r_slice_sampling_plots_1}
   \vspace{15pt}
\end{figure}

We take a detailed look at the signal produced by a single target nucleus, as given in the main text of the paper:

\begin{align} \label{eq:Signal_from_sigle_nucleus}
S(\Gamma,k)=\left(1-\cos\left(ak/2\,t_{\rm det}\right)\right) \frac{(\gamma_{\rm t}B_{\rm AC})^2}{\Omega_{\rm k}^2 + (2\pi/T_{\rm \rho\_target})^2 } \sin^2(\Omega_{\rm k}/2\,t_{\rm ctrl}),
\end{align}
\begin{align} \label{eq:fine_Rabi_rate}
\Omega_{\rm k}&=(a(\Gamma-k))^2 + (\gamma_{\rm t}B_{\rm AC})^2,
\end{align}

Figure\,\ref{fig:SiP_k2r_slice_sampling_plots_1} a. depicts the signal received by performing a frequency sweep across the slices while keeping the other parameters $(B_{\rm AC},t_{\rm det},t_{\rm ctrl})$ constant. Each of the peaks corresponds to a target spin as depicted in Fig.\,\ref{fig:SiP_k2r_slice_sampling_plots_1} c. The first term in Eq.\,\ref{eq:Signal_from_sigle_nucleus} containing $t_{\rm det}$ corresponds to the rate by which the information is transferred from the targets to the probe. The consequences of keeping $t_{\rm det}$ constant can be seen in the figure - as the slice frequency  $\Gamma$ increases the detection time becomes less optimal, reducing the signal's amplitude. The answer is to dynamically modulate the detection time as a function of the slice frequency $\Gamma$ that we wish to probe.  This can be achieved by varying $t_{\rm det}$ as follows:
\begin{eqnarray} \label{eq: varing_t_det}
 t_{\rm det}= \frac{\pi\rho_{\rm det}}{a\Gamma}.
\end{eqnarray}
Equation\,\ref{eq: varing_t_det} provides the detection time that allows a spin on the given slice $\Gamma$ to induce a relative phase of $\rho_{\rm ctrl}\pi$ on the probe, where $\rho_{\rm det}$ is the fraction of the coupling period $\pi/a\Gamma$ covered by the detection segment (note: for relatively large molecules, where there are many target spins per slice, $\rho_{\rm det}\ll1$ prevents signal saturation). Figure\,\ref{fig:SiP_k2r_slice_sampling_plots_1} b. depicts the signal with the control established over $t_{\rm det}$, where each of the target spins induces a uniform signal response.

\FloatBarrier\subsection{Pathways to greater resolution}
Equation\,\ref{eq:Signal_from_sigle_nucleus} produces a $sinc$ function signal shape (Fig.\,\ref{fig:SiP_k2r_slice_sampling_plots_1}), as the interleaved pulses are square in shape. That translates to the cross-sectional profile of 3D dipolar lobe slices (note, the transformation is nonlinear). Pulse engineering can improve the physical cross-sectional profile. For example, use of Gaussian shaped interleaved pulses would lead to a net Gaussian shape of the slices' signal profiles, effectively increasing the resolution of the spatial-frequency encoding. Spatial differentiability can be further improved by use of $sinc$ shaped interleaved pulses, resulting in a rectangular real space profile of the slices.

\FloatBarrier\section{Controlling the cross-sectional slice profile in real space} 
\label{Si_Sc:Controlling_the_cross-sectional_slice_profile_in_real_space}

\FloatBarrier\subsection{Relationship between the spectral width and 3D space slice cross-section}
\begin{figure}[htbp!]%
\centering
 \includegraphics[width=0.75\linewidth]{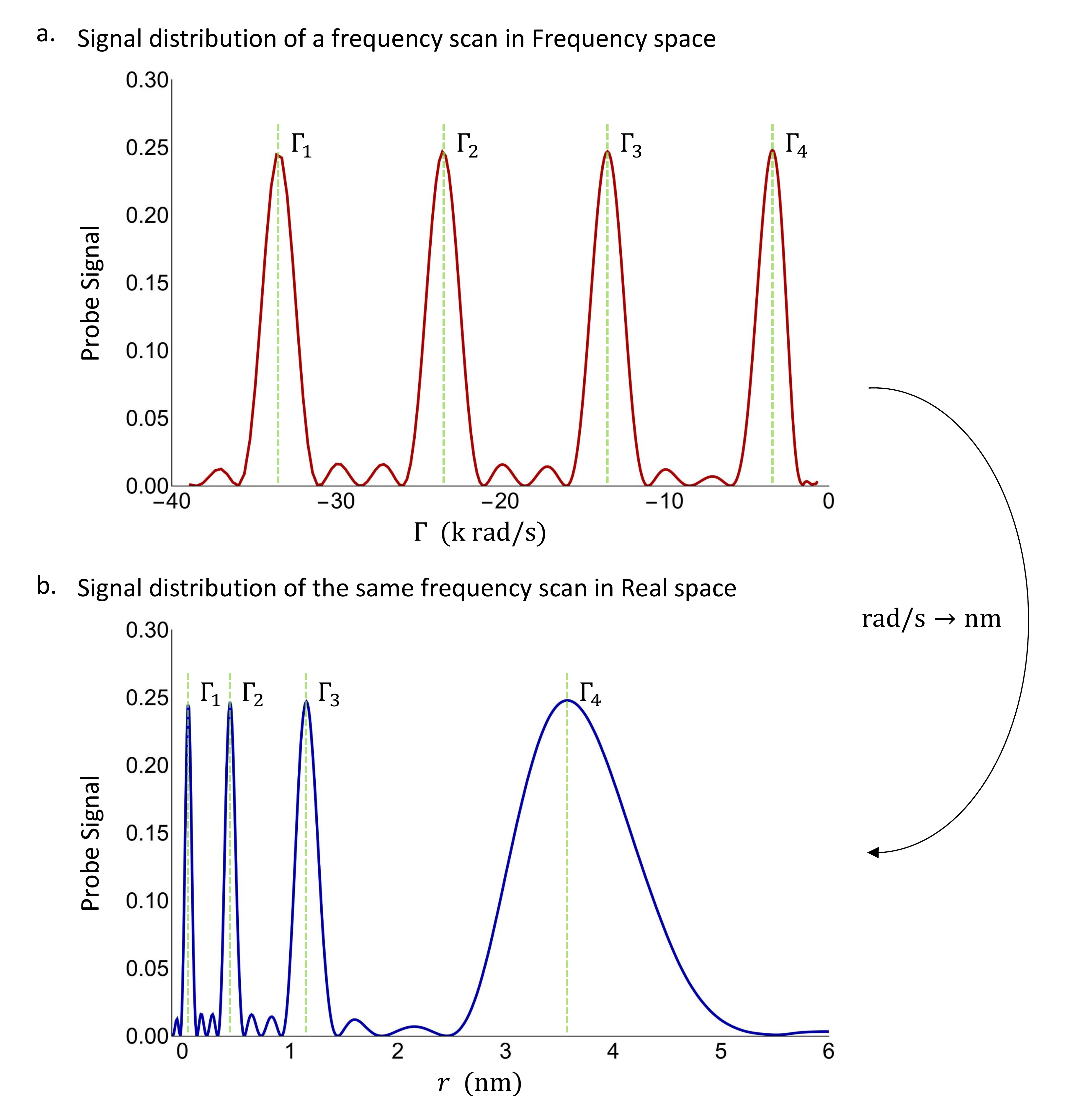}
 \vspace{-5pt}
  \caption{\textbf{The width of the dipolar slices in 3D space.} 
  When the uniform signal in the frequency space (\textbf{a}) is represented in the 3D space (\textbf{b}), the cross-sectional profile of the peaks becomes nonuniform, increasing in width with the targets radial distance away from the probe. The signal associated with a single nuclear target also represents the cross-sectional profile of the dipole-dipole slice in which the given target nucleus is located. Therefore, the spatial width of the slices is increasing with distance away from the probe. This poses an issue for the resolution as the wider slices carry less spatially discriminating information. Note: target spin configuration is the same as in Fig.\,\ref{fig:SiP_k2r_slice_sampling_plots_1}.
  }\label{fig:SiP_k2r_slice_sampling_plots_2}
   \vspace{15pt}
\end{figure}

To characterise the slices' cross-sections in real space, we represent the signal as a function of the radial vector $\bf r$ for the background field $\bf B_0$ in perpendicular direction $\theta=0$. Figure\,\ref{fig:SiP_k2r_slice_sampling_plots_2} depicts how a set of identical peaks in frequency space (Fig.\,\ref{fig:SiP_k2r_slice_sampling_plots_2} a.) appears in real space (Fig.\,\ref{fig:SiP_k2r_slice_sampling_plots_2} b.). In this simple example, individual nuclei produce separate peaks, which therefore also represent the cross-section of slices in the real space. As seen in Fig.\,\ref{fig:SiP_k2r_slice_sampling_plots_2} b., unlike in frequency space, the peaks are not uniform. They expand significantly as the slices move away from the probe. This poses a challenge for the 3D resolution as the image of the upper (radially more distant form the probe) parts of the molecule become increasingly blurred. Broader slice widths have a secondary consequence for the overall resolution, as wider slices carry less spatially discriminating information per slice. In the context of the inversion methodology (moving from the slice-space to 3D Cartesian space), this effect is detrimental in areas where varying slice widths intersect, which also encompass much of the closer parts of the molecule as high tilt ($\theta$) $\bf B_0$ orientations tend to position broader slices in the area vertically above the probe.

\FloatBarrier\subsection{The control methodology for cross-sectional slice profile in real space}

\begin{figure}[htbp!]%
\centering
 \includegraphics[width=0.75\linewidth]{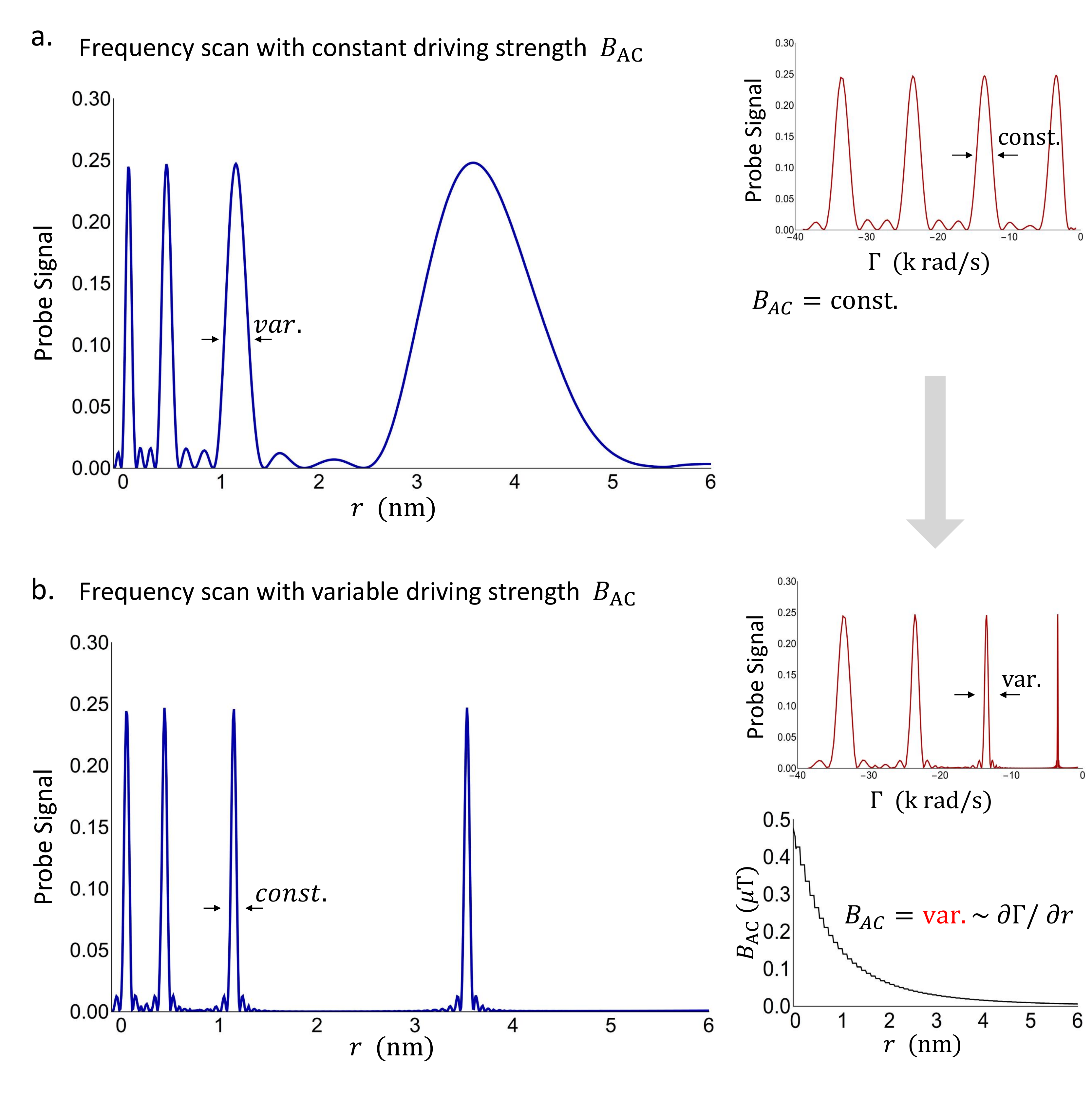}
 \vspace{-5pt}
  \caption{\textbf{Controlling the physical width of the gradient slices.}
  Modulating  $B_{\rm AC}$ and $t_{\rm ctrl}$ in accordance with the dipole-dipole field gradient transforms the constant $B_{\rm AC}$  scan, with expanding slice widths (\textbf{a}), to the scan width constant slice width in real space (\textbf{b}). Side note, the effect of the factor $p_{\rm ctrl}$ on the signal is illustrated by setting $p_{\rm ctrl}=.025$. Note: target spin configuration is the same as in Fig.\,\ref{fig:SiP_k2r_slice_sampling_plots_1}.
  }\label{fig:SiP_k2r_slice_sampling_plots_3}
   \vspace{15pt}
\end{figure}

The overall resolution and the imaging efficiency can be improved by making the real space cross-section of slices uniform. Here, this is achieved by dynamically controlling the power broadening across the slices in accordance with the gradient of the dipole-dipole coupling field. Therefore, the fine-driving field $B_{\rm AC}$ becomes a dependent parameter that satisfies the following condition for a given slice $\Gamma$:
\begin{eqnarray} \label{eq: varing_B_AC}
 B_{\rm AC}(\Gamma)= \frac{B^{\rm AC}_0}{\partial \Gamma/\partial r(r=0)} \frac{\partial \Gamma}{\partial r},
\end{eqnarray}
where, $B^{\rm AC}_0$ is the starting fine-driving field scaled by the dipolar gradient magnitude at the surface  $\partial \Gamma/\partial r(r=0)$, while the coupling gradient $\frac{\partial \Gamma}{\partial r}$ modulates the fine-driving field for every slice $\Gamma$ in order to ensure that its width remains constant. It is also important to note that the control segment time $t_{\rm cntrl}$ should also be adjusted in order to keep the signal magnitude fixed:
\begin{eqnarray} \label{eq: varing_t_cntl}
 t_{\rm ctrl}(\Gamma)= \frac{\pi \rho_{\rm ctrl}}{\gamma_t B_{\rm AC}(\Gamma)},
\end{eqnarray}
where the factor $\rho_{\rm ctrl}$ indicates by how much the nuclei in the given slice $\Gamma$ are rotated, as a fraction of a full $\pi$ flip. This factor should be set in accordance with the molecule's average nuclear density, in order to ensure the combined signal $S_{\rm net}$ does not become saturated, by asymptotically approaching 0 for any of the slices being probed.

The parameters $\rho_{\rm ctrl}$ and $\rho_{\rm det}$ both affect the phase collected by the probe, however, their effect on the signal is different. The ratio $\rho_{\rm det}$  controls the detection segment length, what effects the amplitude of any signal the probe may detect. However, $\rho_{\rm ctrl}$ controls the amount of control rotation a nuclear target is subjected to. Therefore, it affects not only the signal's amplitude but also the relative spectral width of each slice. The closer $\rho_{\rm det}$ is to 1 the higher the spectral resolution of the probed slice.

Figure\,\ref{fig:SiP_k2r_slice_sampling_plots_3} illustrates the methodology. In the constant amplitude fine-driving case, the peaks are uniform in frequency space (Fig.\,\ref{fig:SiP_k2r_slice_sampling_plots_3} a. top-right) and translate into nonuniform 3D space slice widths. Controlling the driving field $B_{\rm AC}$ and the segment $t_{\rm ctrl}$ allows the width of the frequency space peaks (Fig.\,\ref{fig:SiP_k2r_slice_sampling_plots_3} b. top-right), to be modulated in such a way as to produce slices in 3D space that have the uniform width.

\FloatBarrier\section{Carbon(13) nuclear density imaging}
\label{Si_Sc:Carbon(13)_nuclear_density_imaging}

We simulate signal acquisition and reconstruction assuming that the molecule has been $^{13}C$ saturated, that the fine-driving strength is $B^{\rm AC}_0=0.5\rm\,\mu T$. In line with achievable coherence time estimates, the coherence time of the target nuclear spins is set to $T_{\rm \rho\_target}=100\,\rm ms$. The coherence time of the P donor election spin is also set to the same length $T_2^{\rm probe}=100\,\rm ms$, however, note that experimentally demonstrated $T_2$ times of Si:P donors have been measured on the second timescale \cite{Tyryshkin2011}. The slice parameters were as follows: $r_{\rm max}=1.4\,\rm\AA$, $dr=0.25\,\rm\AA$, with $\theta_{\rm max}=63^\circ$, $d\theta=1.5^\circ$ $d\phi=6^\circ$, and $n_m=1000$ projective measurements per slice.

\begin{figure}[htbp!]%
\centering
 \includegraphics[width=0.75\linewidth]{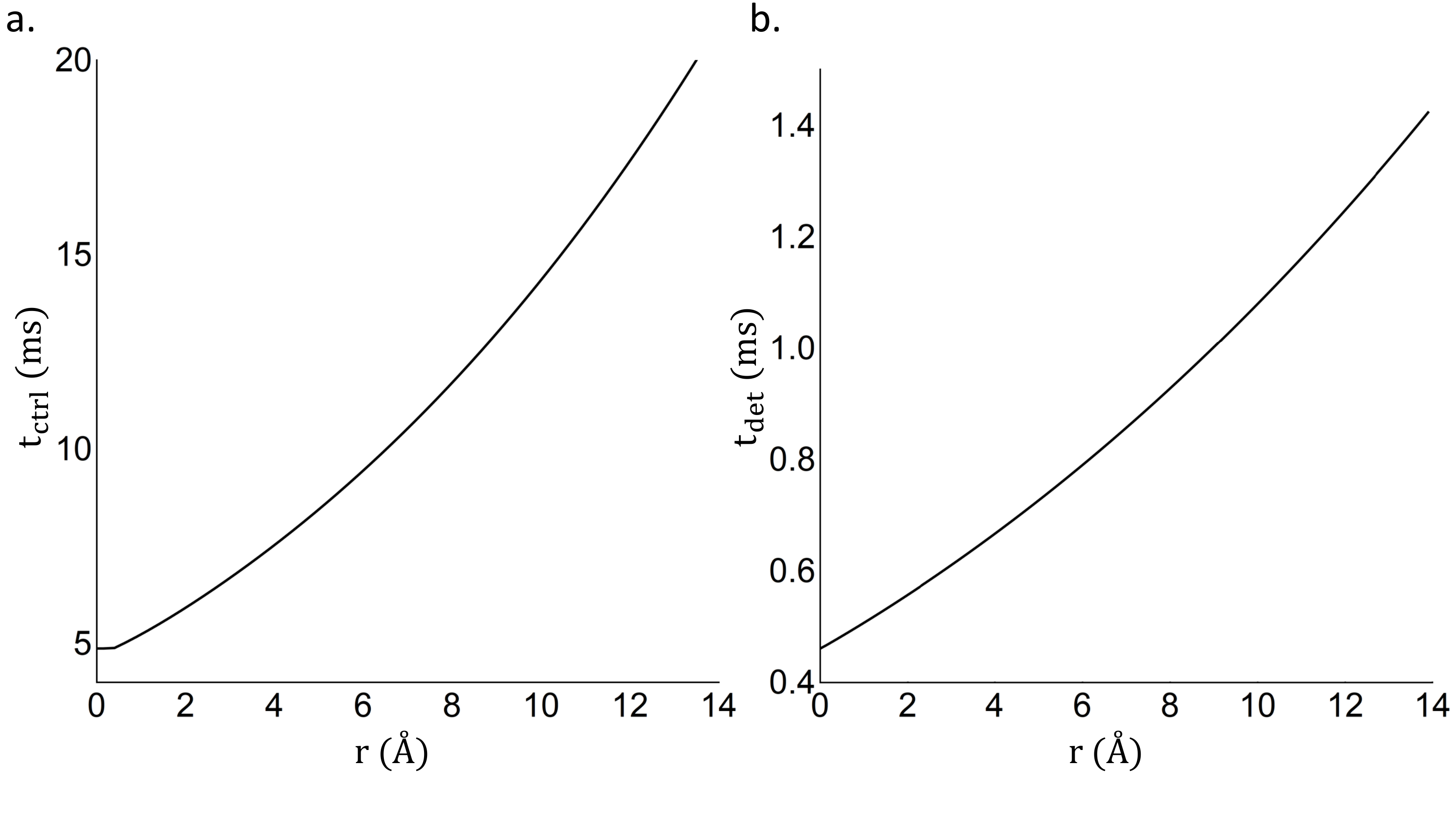}
 \vspace{-5pt}
  \caption{\textbf{The NSS detection protocol parameters for $^{13}C$ imaging in sectional mode.}
  \textbf{a}) The leading contributor to the length of the protocol is the control segment $t_{\rm ctrl}$, which has to be approximately an order of magnitude longer compare to the detection segment in order to provide a sufficient spectral resolution. 
  \textbf{b}) The detection segment time $t_{\rm det}$ which unlike $t_{\rm ctrl}$ is limited by the $T_2^{\rm probe}$.
  }\label{fig:SiP_demonstration_4_1}
   \vspace{15pt}
\end{figure}

The nuclear density inside the selected volume has led us to choose the detection and control segment parameters as follows $\rho_{\rm det}=0.2$ and $\rho_{\rm ctrl}=0.2$. Figure\,\ref{fig:SiP_demonstration_4_1} depicts the average control and detection times for the protocol as a function of the lobe radius $r$.  When it comes to the coherence of the probe, we note that the short detection segment is exposed to $T_2^{\rm probe}$, while the longer control segment is protected by a relatively longer $T_1^{\rm probe}$, while the coherence of the target nuclear spins needs to be maintained throughout the entire length of the NSS detection protocol.

It is useful to estimate the total time that would be required to perform all the measurements needed to generate the 3D density in the sectiona image mode for $^{13}C$ subdensity. Behind every slice is the NSS detection protocol whose length is the sum of the control, detection and measurement times $t_{\rm net}=t_{\rm det} + t_{\rm ctrl} + t_{\rm m}$. The sampling consists of $\approx2500$ $(\theta,\phi)$ orientations each harbouring a sweep over $r$ that counts $\approx60$ slices. The time it takes to complete one sweep over $r$ (one orientation) is broken up into  $\approx2.5s$ of control time, $\approx0.02s$ of detection time and $\approx0.3s$ of measurement time assuming $n_m=1000$ measurements per slice at a conservative $5\rm\,\mu s$ per readout. Collectively this amounts to $\approx2.82s$ per orientation. It follows that the complete course of $\approx2500$ orientations would take around $2$ hours.

\FloatBarrier\section{Impacts of signal noise in density imaging}
\label{Si_Sc:Impacts_of_signal_noise_in_density_imaging}

\begin{figure}[!ph]%
\centering
 \includegraphics[width=0.75\linewidth]{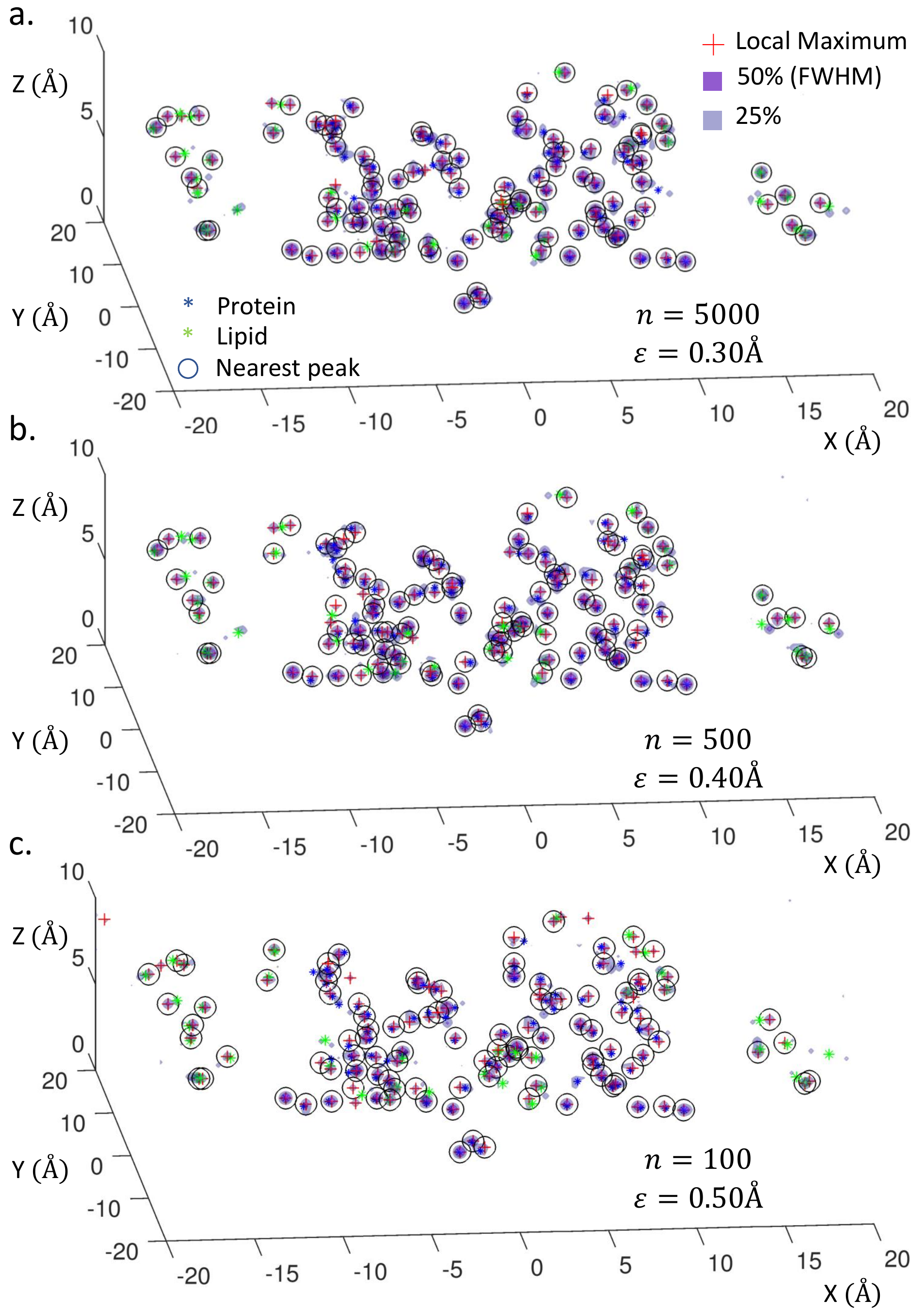}
 \vspace{-5pt}
  \caption{\textbf{The $^{13}C$ density for various shot noise conditions.}
  \textbf{a}) $n_m=5000$.
  \textbf{b}) $n_m=500$.
  \textbf{c}) $n_m=100$. The apparent resilience to the high levels of shot noise is a consequence of a large number of slices simultaneously contributing to each voxel. It indicates that the sampling and reconstruction procedure also have the capacity to tolerate signals associated with quantum errors during the runtime of the protocol.
  }\label{fig:SiP_demonstration_6}
   \vspace{15pt}
\end{figure}

Testing the density reconstruction for various amplitudes of shot noise is also useful for a couple of reasons.  Figure\,\ref{fig:SiP_demonstration_6} depicts the same carbon(13) distribution for various numbers of projective measurements per slice $n_m=(5000,500,100)$. Despite an order of magnitude reduction in the noise level between Figs.\,\ref{fig:SiP_demonstration_6} a. and c., the reconstruction remains relatively stable, with the average deviation staying within the boundaries of the voxel separation $\epsilon=0.5\,\rm\AA$. This can be explained by recognising that each point in space (voxel) is an intersection of a multitude of slices, approximate to the number of orientations (no. of $(\theta,\phi)$ combinations). In our case, this effect is analogous to an increase in the number of measurements of around 3 orders of magnitude per voxel. Apart from leading to a shorter net experimental imaging time, it also demonstrates relative robustness to decoherence events that may fatally disrupt the NSS detection protocol, rendering the particular slice measurement defective.

\FloatBarrier\section{Nitrogen(14) density imaging}
\label{Si_Sc:Nitrogen(14)_density_imaging}

\begin{figure}[hp!]%
\centering
 \includegraphics[width=0.75\textwidth]{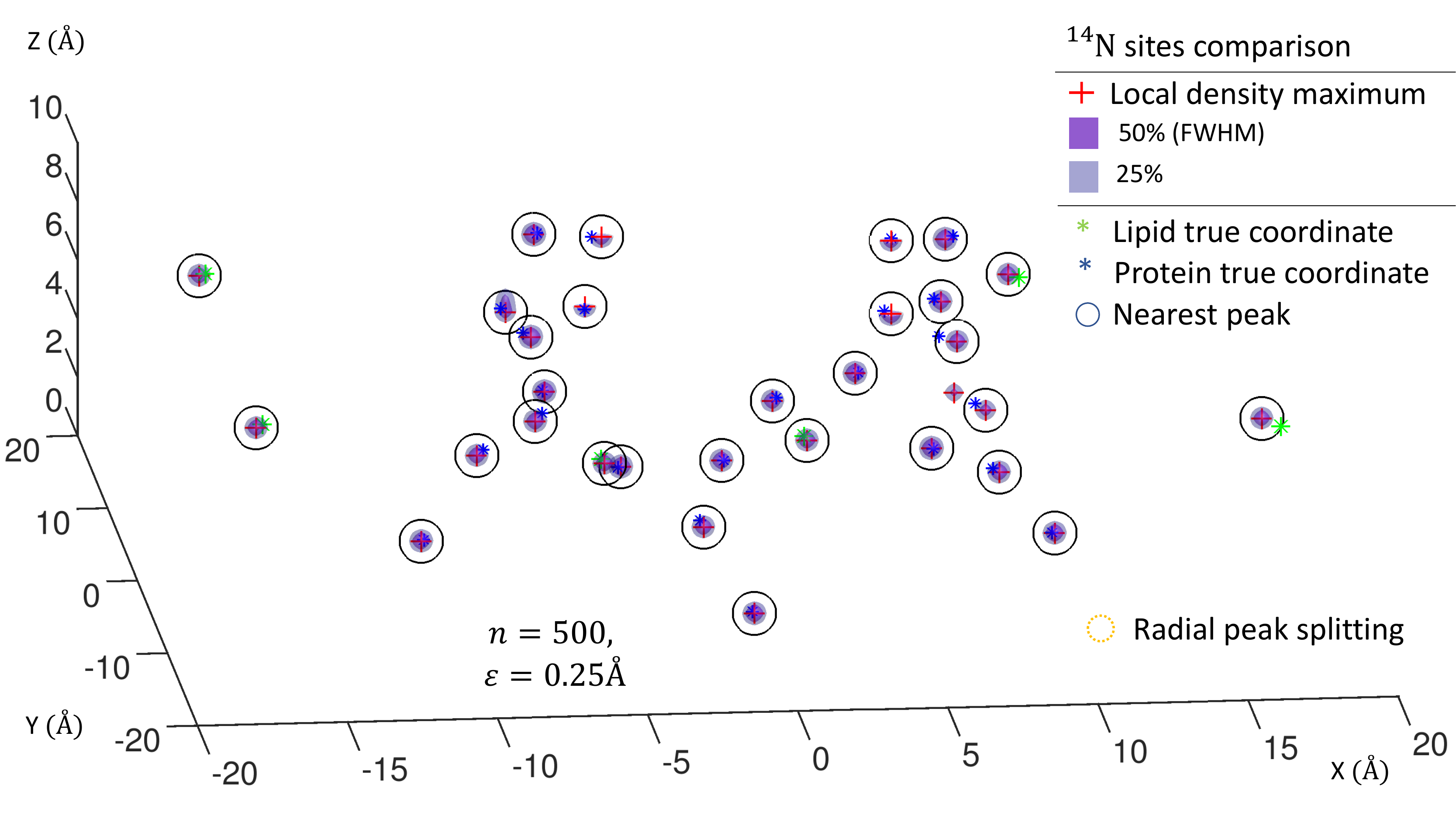}
  \caption{\textbf{The image of $^{14}N$ nuclear sub-density of the M2 proton channel-lipid system and the atomic site reconstruction for the noise case $n_m=500$.}
  The sampling parameters used were identical to that of $^{13}C$ imaging, however, a lower density of N sites has led to a more constrained and accurate image.
  }\label{fig:SiP_demonstration_7_1}
\end{figure}

\begin{figure}[htbp!]%
\centering
 \includegraphics[width=0.75\linewidth]{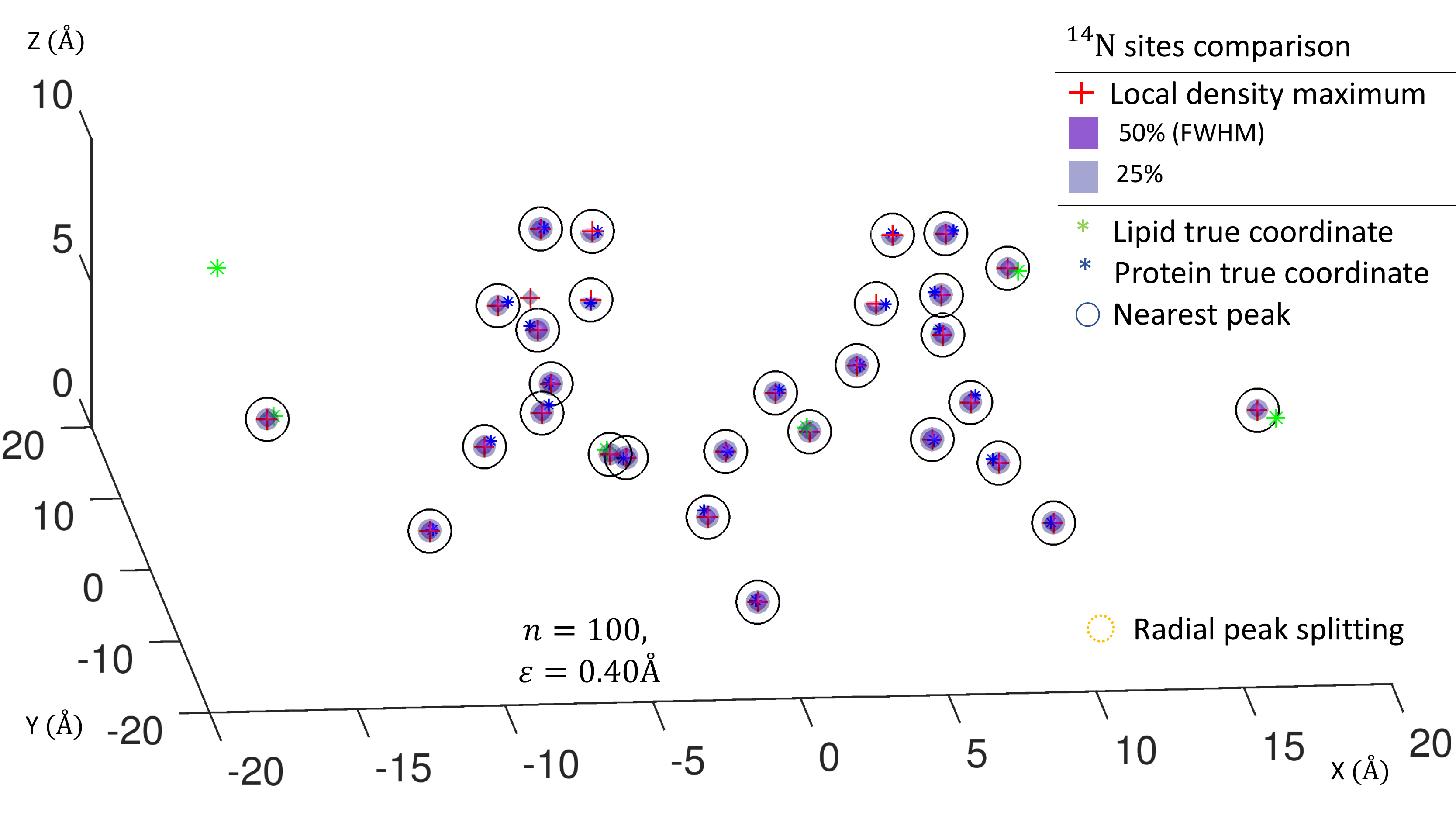}
 \vspace{-5pt}
  \caption{\textbf{The image of $^{14}N$  3D nuclear sub-density for the shot noise cases $n_m=100$.} The sampling parameters used were identical to that of $^{13}C$ imaging.
  }\label{fig:SiP_demonstration_7_2}
   \vspace{15pt}
\end{figure}

We apply the same imaging parameters (form the Appendix \ref{Si_Sc:Carbon(13)_nuclear_density_imaging}:Carbon(13) nuclear density imaging) the nitrogen nuclear(14) sub-density, Fig.\,\ref{fig:SiP_demonstration_7_1}. The nitrogen density analysis reveals a highly confined image robust to noise. The number of unknowns (N sites) in the system is small and well separated allowing a high degree of over-constraint allowing for nitrogen atomic site reconstruction even under high shot noise conditions (Fig.\,\ref{fig:SiP_demonstration_7_2}).

The greater distances between the nitrogen atoms also improve the ability of the voxel grid to sample the space. By reducing the number of slices and making them broader, it would be possible to optimise the N density imaging further. We take this approach in order to image the entire transmembrane protein.

\FloatBarrier\section{Hydrogen density imaging}
\label{Si_Sc:Hydrogen_density_imaging}

\begin{figure}[hp!]%
\centering
 \includegraphics[width=0.75\textwidth]{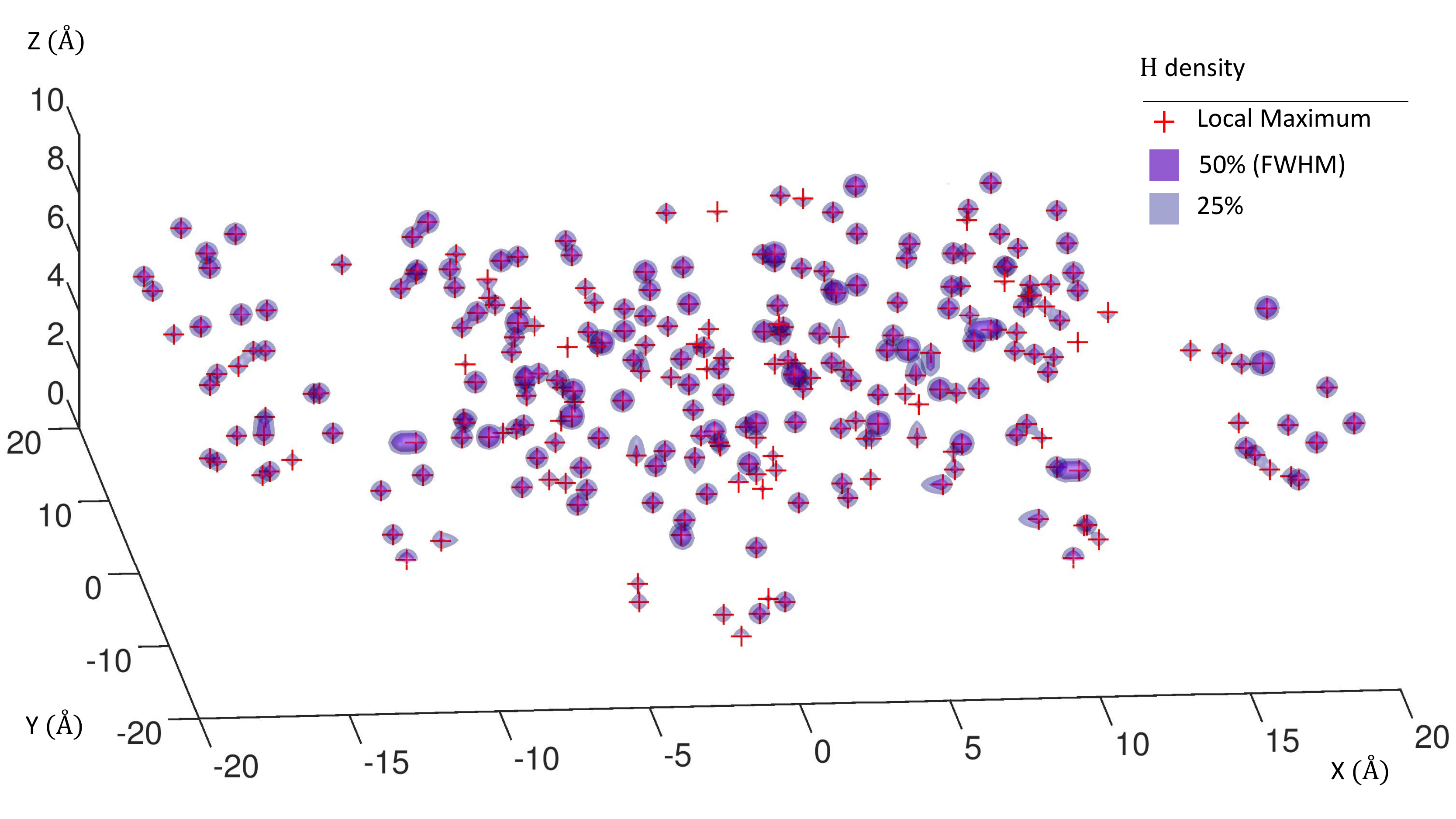}
  \caption{\textbf{The image of the single M2 proton channel-lipid system $H$ nuclear 3D density (simulation).}
  For comparison, the acquisition parameters applied to hydrogen are the same as for the $^{13}C$ density.
  }\label{fig:SiP_demonstration_8_1}
\end{figure}

\begin{figure}[!htb]%
\centering
 \includegraphics[width=0.75\linewidth]{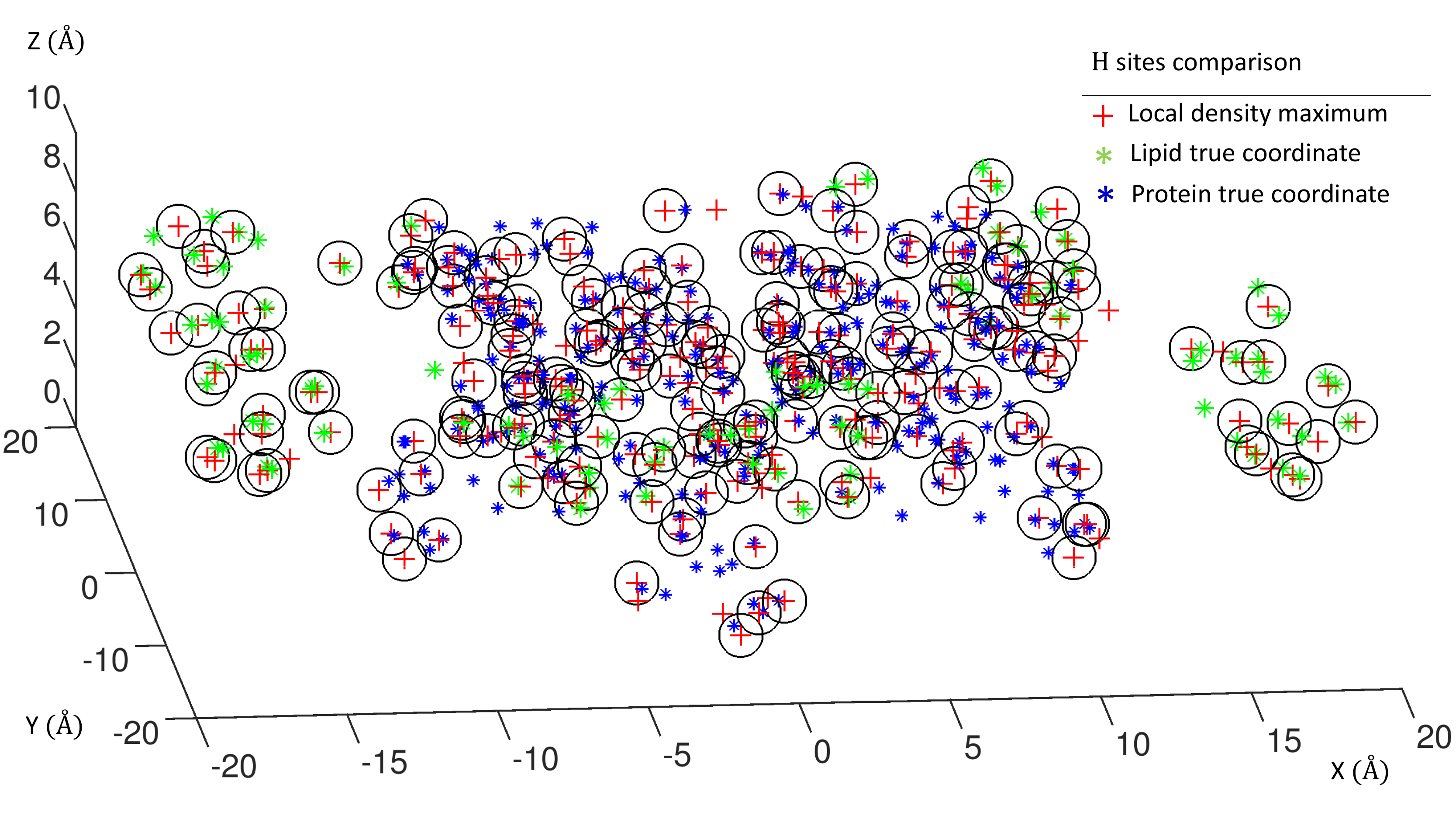}
 \vspace{-5pt}
  \caption{\textbf{The reconstructed $H$ sites of the M2 proton channel-lipid system in comparison with the true atomic coordinates.}
  Deviation from the true coordinates (marked with *) is calculated automatically by determining the radial distances between the true atomic site and its nearest density peak (pairs circled). The average deviation is $\epsilon=0.5\,\rm\AA$. The peaks show significant radial splitting while a fraction of the sites ($\sim 20\%$) are unmatched with true coordinates in the higher density regions. Note, the simple peak-detection procedure used here also has a higher rate of failure as some of the peaks are relatively close together compared to the voxel grid spacing.
  }\label{fig:SiP_demonstration_8_2}
   \vspace{15pt}
\end{figure}

\begin{figure}[!htb]%
\centering
 \includegraphics[width=0.75\linewidth]{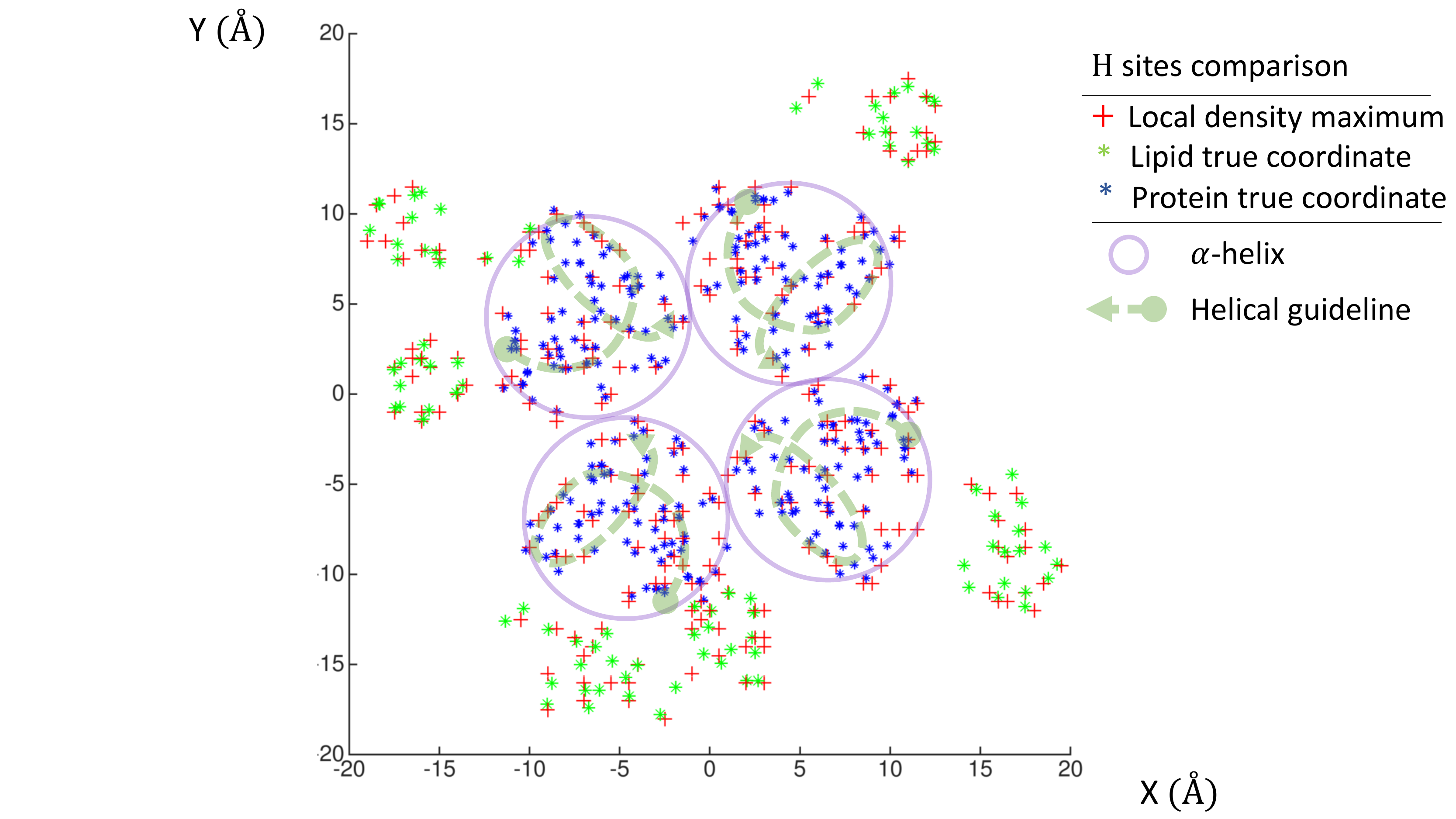}
 \vspace{-5pt}
  \caption{\textbf{The reconstructed $H$ sites of the M2 proton channel-lipid system top-down view.}
  The z-projection of the image in Fig.\,\ref{fig:SiP_demonstration_8_2} provides a more intuitive insight into the protein's symmetric structure. Notice the projections of 4 $\alpha$-helix segments (circled in purple), with helical guidelines superimposed to aid visual interpretation (green dashed).
    }\label{fig:SiP_demonstration_8_3}
   \vspace{15pt}
\end{figure}

For consistency, we simulate the hydrogen density imaging using the same parameters as for the carbon(13) density (Fig.\,\ref{fig:SiP_demonstration_8_1}). In contrast to nitrogen, the hydrogen density image appears to have suffered a reduction in quality. However, the reconstructed image shows a relatively low deviation $\epsilon=0.5\,\rm\AA$ considering that hydrogen density is $\sim2.5$ times greater than that of carbon and therefore would benefit from a finer voxel grid. Higher nuclear density has also accentuated the effects of nonuniform radial sampling, leading to increase in noise and missing sites ($\sim20\%$) at the edges of the sample and in higher density areas. Also, note that the neighbours of missing sites tend to have relatively greater amplitudes, this is related to the iterative nature of the numerical inversion process converging to the nearer voxel. Due to the smaller separations, H atoms would benefit from a finer voxel grid, however, this is inconsequential for the measurement protocol - generating and inverting larger maps $\underline{\underline{M}}$ is numerically well understood, yet resource intensive in the context of the work at hand.

\FloatBarrier\section{Imaging the entire transmembrane protein-lipid system}
\label{Si_Sc:Imaging_the_entire_transmembrane_protein-lipid system}

We apply the imaging protocol to the atomic structure of the entire protein-lipid system.  We focus on the relatively scarce nitrogen nuclear sub-density, noting that higher nuclear densities (i.e. H and C) would require greater numerical recourses to simulate imaging over the volume of the system ($40\times40\times40\rm\AA$).

\begin{figure}[htbp!]%
\centering
 \includegraphics[width=0.75\linewidth]{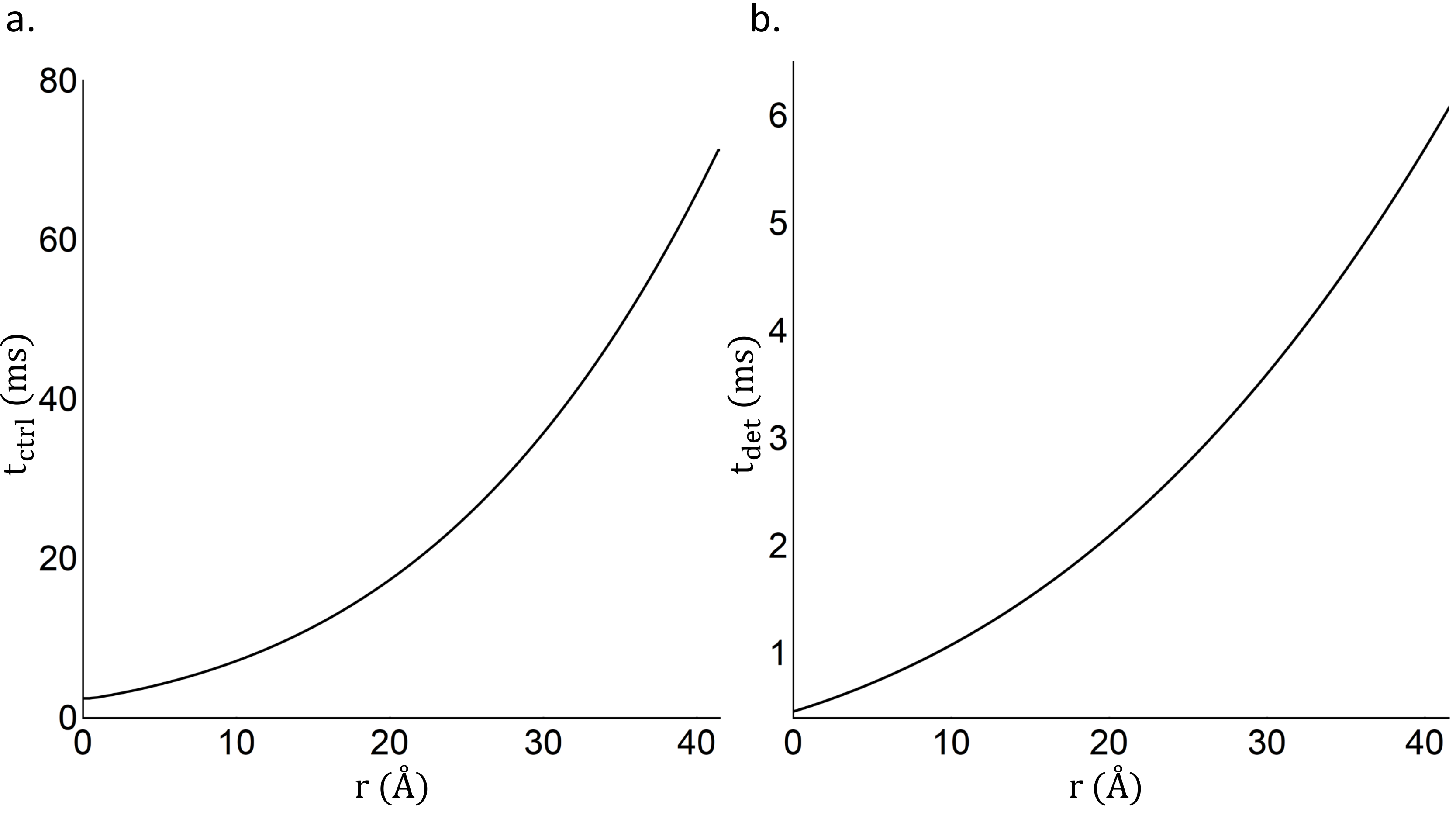}
 \vspace{-0pt}
  \caption{\textbf{Variable NSS protocol parameters required for a larger imaging volume.}
\textbf{a}) In case of a P donor at depth $\sim2.5\rm\,nm$, the control segment $t_{\rm ctrl}$ times required for imaging up to $40\,\rm\AA$ away from the Si surface approach the target's assumed coherence limit of $\sim100\rm\,ms$. However, this provides enough room for imaging across the lipid bilayer.  
\textbf{b}) The behaviour of the detection segment time $t_{\rm ctrl}$ under the same assumptions.
  }\label{fig:SiP_demonstration_9_td_tc}
   \vspace{15pt}
\end{figure}

Adapting the parameters to a sparser voxel grid $d=0.75\,\rm\AA$ and taking broader slices $B_0^{\rm AC}=1.25\rm\,\mu T$, $dr=0.5\,\rm\AA$, $r_{\rm max}=4\,\rm\AA$ $d\phi=5^\circ$, $\theta_{\rm max}=45^\circ$, allows us to increase the height of the sampling volume from $1\rm\,nm$ to $4\rm\,nm$ and remain within our numerical resources. The protocol's control and detection times remain within the decoherence timescales, as depicted in Fig.\,\ref{fig:SiP_demonstration_9_td_tc}.

These conditions are sufficient to demonstrate imaging of the entire molecule, particularly the nitrogen sub-density. The $r$ sweep counts 85 slices,  over longer periods of time, leading to $\approx9s$ per orientation. As there are 2100 orientations, the net imaging time comes to $\approx6$ hours. Several effects become visible  mainly in the upper section of the volume. Aside from the sub-optimal radial sampling, the higher parts of the sample are experiencing a reduction in the signal contrast due to the NSS protocol length approaching the $100\,\rm ms$ presumed decoherence limit of the target spins. As a result, the average deviation of the reconstruction is higher than the voxel grid spacing at $\epsilon=1.3\,\rm\AA$.

\section*{References}
\bibliography{Perunicic_arxiv}

\end{document}